\documentstyle[epsfig]{elsart}
\newcommand{\newc}{\newcommand}
\newc{\lra}{\leftrightarrow}
\newc{\beq}{\begin{equation}}
\newc{\eeq}{\end{equation}}
\newc{\barr}{\begin{eqnarray}}
\newc{\earr}{\end{eqnarray}}

\newcommand{\beqa}{\begin{eqnarray}}
\newcommand{\eeqa}{\end{eqnarray}}
\newcommand{\bdm}{\begin{displaymath}}
\newcommand{\edm}{\end{displaymath}}

\renewcommand{\thefootnote}{\#\arabic{footnote}}
\renewcommand{\theequation}{\thesection.\arabic{equation}}
\renewcommand{\thefigure}{\thesection.\arabic{figure}}
\renewcommand{\thetable}{\thesection.\arabic{table}}

\begin{document}
%\begin{frontmatter}
\date{\today}
%\title* { Neutrinos in a Spherical Box}
\title {EXPLORING NEW FEATURES OF NEUTRINO OSCILLATIONS WITH A TRITON SOURCE
AND A LARGE SPHERICAL TPC}
%
%
%\toctitle{ Neutrinos \protect\newline in a Spherical Box}
% allows explicit linebreak for the table of content
%
%
%\titlerunning{NEUTRINOS IN A SPHERICAL BOX}
% allows abbreviation of title, if the full title is too long
% to fit in the running head
%
\author{Y. Giomataris$^{1}$ and J.D. Vergados$^{2}$ }
%
%\authorrunning{Giomataris and Vergados}
% if there are more than two authors,
% please abbreviate author list for running head
%

\address{
1 CEA, Saclay, DAPNIA, Gif-sur-Yvette, Cedex,France}
\address{
2 University of Ioannina, Ioannina, GR 45110, Greece
\\E-mail:Vergados@cc.uoi.gr}
%\maketitle              % typesets the title of the contribution
\begin{frontmatter}
\begin{abstract}
The purpose of the present paper is to study the neutrino
oscillations as they may appear in the low energy neutrinos emitted
in triton decay:
$$^3_1 H \rightarrow ^3_2He +e^-+\tilde{\nu}_e$$
with maximum neutrino energy of $18.6~KeV$. Such low energy
neutrino oscillations can only be detected via neutrino electron
scattering. The novel feature is the fact that the cross section
is sensitive to both charged and neutral currents. Thus one can
 simultaneously probe
the electron neutrino disappearance as well as the appearance of
the other two flavors. By folding the differential cross section
with the neutrino spectrum one can study the dependence of the
oscillation parameters as a function of the electron energy. The
technical challenges to this end can be summarized as building a
very large TPC capable of detecting low energy recoils, down to a
few 100 eV, within the required low background constraints have
been previously described \cite{NOSTOS1}. We only mention here
that the oscillation length involving the small angle
$s_{13}=\sin{\theta_{13}}$, directly measured in this experiment,
is fully contained inside the detector. A sensitivity of a few
percent for the measurement of the above angle will be achieved.

\end{abstract}

\end{frontmatter}
\section{Introduction.}
The discover of neutrino oscillations can be considered as one of the greatest triumphs of modern physics.
It began with atmospheric neutrino oscillations \cite{SUPERKAMIOKANDE}interpreted as
 $\nu_{\mu} \rightarrow \nu_{\tau}$ oscillations, as well as
 $\nu_e$ disappearance in solar neutrinos \cite{SOLAROSC}. These
 results have been recently confirmed by the KamLAND experiment \cite{KAMLAND},
 which exhibits evidence for reactor antineutrino disappearance.
  As a result of these experiments we have a pretty good idea of the neutrino
mixing matrix and the two independent quantities $\Delta m^2$, e.g $m_2^2-m^2_1$ ans $m^2_3-m^2_2|$.
 Fortunately these
two  $\Delta m^2$ are vastly different $$\Delta m^2(21)
m^2_{21}=|m_2^2-m_1^2|=(5.0-7.5)\times 10^{-5}(eV)^2$$ and
$$\Delta m^2_{32}=|m_3^2-m_2^2|=2.5\times 10^{-3}(eV)^2.$$
 This means that the associated oscillation
lengths are very different and the experimental results can effectively be described as two generation oscillations.
This has been followed by an avalanche of interesting three generation analyses  \cite{BAHCALL02}-\cite{BARGER02}.

 In all of these analyses the
oscillation length is much larger than the size of the detector. So one is able to see the effect, if the detector is
placed in the right distance from the source. It is, however, possible to design an experiment with an oscillation
length of the order of the size of the detector. In this case one can start with zero oscillation near the
source, proceed to maximum oscillation near the middle of the detector and end up again with no oscillation on
the other end. This is achieved, if one considers a neutrino source with the lowest possible average neutrino energy
such as a triton source with a maximum energy of $18.6$ keV. Thus the average oscillation length is $6.5$m, which is
smaller than the radius of $10$m of a spherical TPC detector (for a description of the apparatus see our
earlier work \cite{NOSTOS1}.

 Since the neutrino energies are so small, the only possible detector sensitive to oscillations is one, which
is capable of detecting electrons. With this detector can
 detect:
\begin{itemize}
\item electrons which are produced by electron neutrinos via the charged current interaction,
 which are depleted by {\bf electron neutrino disappearance}
\item electrons which may arise from the other two neutrino flavors due to the neutral
current interaction. These two flavors are produced by the {\bf muon and tau neutrino appearance,
which operates simultaneously}.
\end{itemize}
 Since these two mechanisms are competing with each other and the subsequent electron
cross section depends on the electron energy, one has a novel feature, i.e. the dependence of the effective
 oscillation probability
on the electron energy. Thus the results may appear as disappearance oscillation in some kinematical regime
and as appearance oscillation in some other regime. In this paper we will examine how these features can best be
exploited in determining the neutrino oscillation parameters.
\section{ Elastic Neutrino Electron Scattering}
\begin{figure}[!ht]
 \begin{center}
 \epsfxsize=.5
 \textwidth
 \epsffile{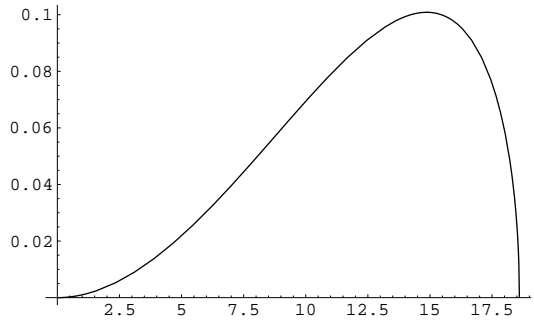}
 \caption{The neutrino spectrum distribution function for the triton source (normalized to unity).}
 \label{spectrum}
  \end{center}
  \end{figure}

 For low energy neutrinos the historic process neutrino-electron scattering \cite{HOOFT} \cite{REINES}
 is very useful.
The differential cross section \cite{VogEng} takes the form
%('t Hooft and Vogel $\&$ Engel)
\begin{equation}
\frac{d\sigma}{dT}=\left(\frac{d\sigma}{dT}\right)_{weak}+
\left(\frac{d\sigma}{dT}\right)_{EM} \label{elas1a}
\end{equation}
The cross section due to weak interaction alone becomes:
 \begin{eqnarray}
 \left(\frac{d\sigma}{dT}\right)_{weak}&=&\frac{G^2_F m_e}{2 \pi}
 [(g_V+g_A)^2\\
\nonumber
&+& (g_V-g_A)^2 [1-\frac{T}{E_{\nu}}]^2
+ (g_A^2-g_V^2)\frac{m_eT}{E^2_{\nu}}]
 \label{elasw}
  \end{eqnarray}
 $$g_V=2\sin^2\theta_W+1/2~~ (\nu_e)~,~g_V=2\sin^2\theta_W-1/2~~ (\nu_{\mu},\nu_{\tau})$$
 $$g_A=1/2~~~~ (\nu_e)~~~~,~~~~g_A=-1/2~~~~ (\nu_{\mu},\nu_{\tau})$$
 For antineutrinos $g_A\rightarrow-g_A$. To set the scale 
\beq \frac{G^2_F m_e}{2 \pi}=4.45\times 10^{-48}~\frac{cm^2}{keV}
\label{weekval} 
\eeq
% In the above expressions for the $\nu_{\mu},\nu_{\tau}$ only the
% neutral current has been included, while for $\nu_e$ both the
% neutral and the charged current contribute.
The electron energy depends on the  neutrino energy and the
scattering angle and is given by:

$$T= \frac{2~m_e (E_{\nu}\cos{\theta})^2}{(m_e+E_\nu)^2-(E_{\nu}
\cos{\theta})^2}$$
%$$T=\frac{X^2}{2 m_e}~~,~~X=2E_{\nu} \frac{m_e(m_e+E_{\nu})\cos{\theta}}
%{(m_e+E_\nu)^2-(E_{\nu} \cos{\theta})^2}$$
The last equation can be simplified as follows:
 $$T \approx \frac{ 2(E_\nu \cos{\theta})^2}{m_e}\Rightarrow$$
 The maximum electron
energy depends on the neutrino energy. For $E_{\nu}=18.6~KeV$
one finds that the maximum electron kinetic energy is:
 $$ T_{max} \approx 1.27~KeV$$
I what follows, whenever appropriate, we are going to average our results with the neutrino spectrum 
shown in Fig. \ref{spectrum}.
\section{ NEUTRINO OSCILLATIONS IN THE NOSTOS EXPERIMENT}
 
\begin{figure}[!ht]
 \begin{center}
 \epsfxsize=.3
 \textwidth
  \epsffile{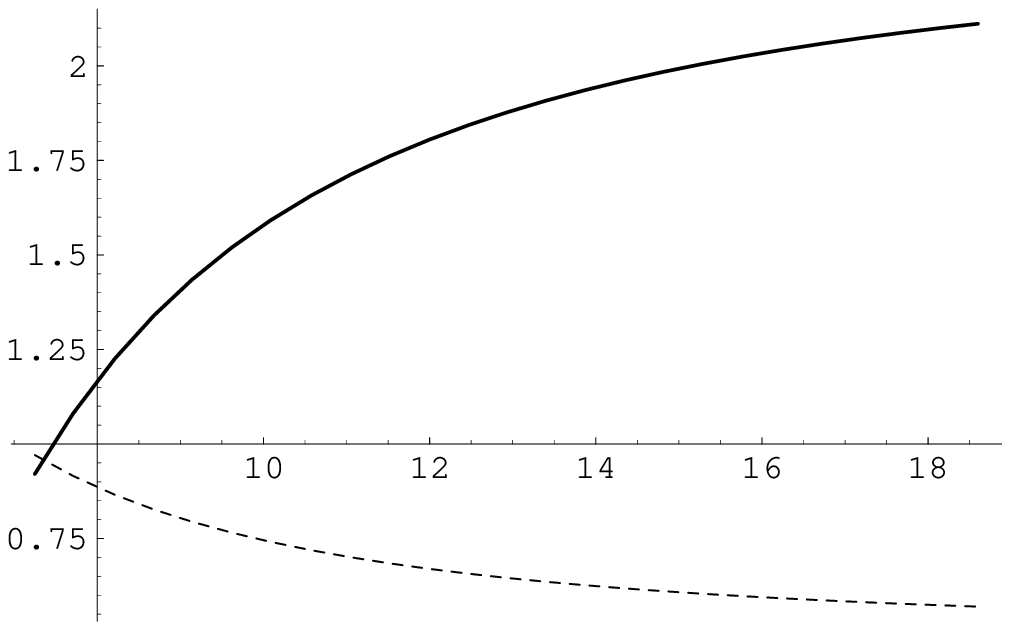}
  \epsfxsize=.3
 \textwidth
  \epsffile{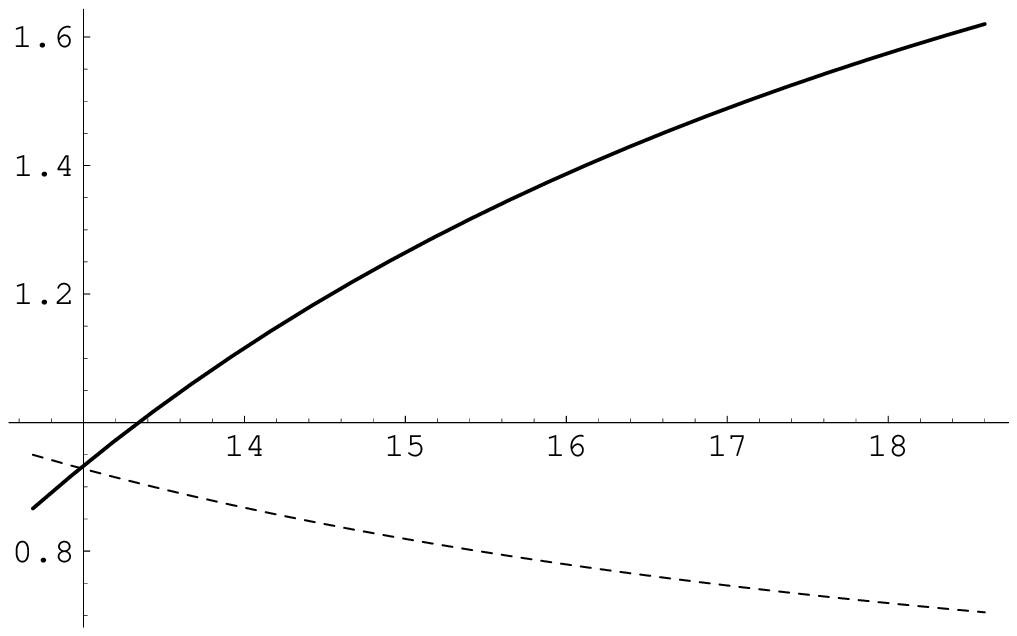}
  \epsfxsize=.3
 \textwidth
  \epsffile{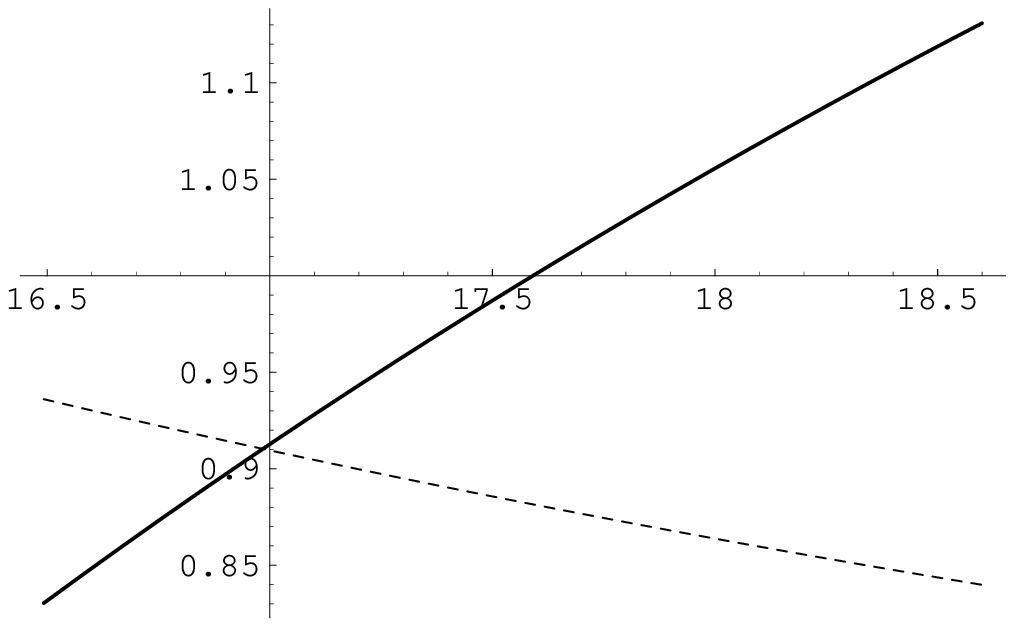}
 \caption{The differential cross $(\nu_e,e^-)$ section, in arbitrary units, as a function of the 
neutrino energy for electron 
energies from left to right 0.2, 0.6 and 1.0 keV. The the dashed curve corresponds to the cross
section due to $\nu_{\alpha}$, $\alpha=\mu$ or $\tau$.}
 %\end{center}
 \label{fig:1}
  \end{center}
  \end{figure}

\begin{figure}[!ht]
 \begin{center}
 \epsfxsize=.3
 \textwidth
  \epsffile{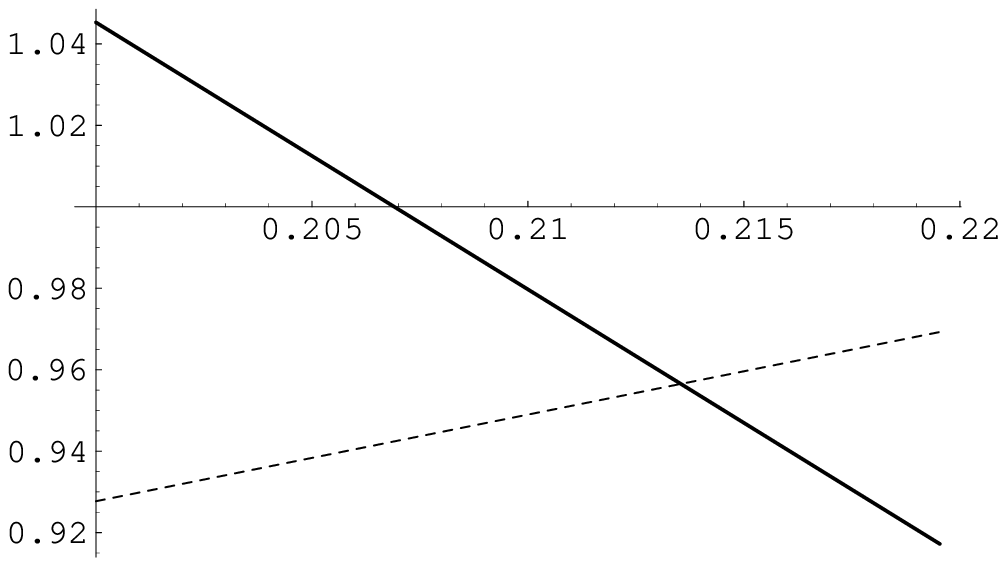}
  \epsfxsize=.3
 \textwidth
  \epsffile{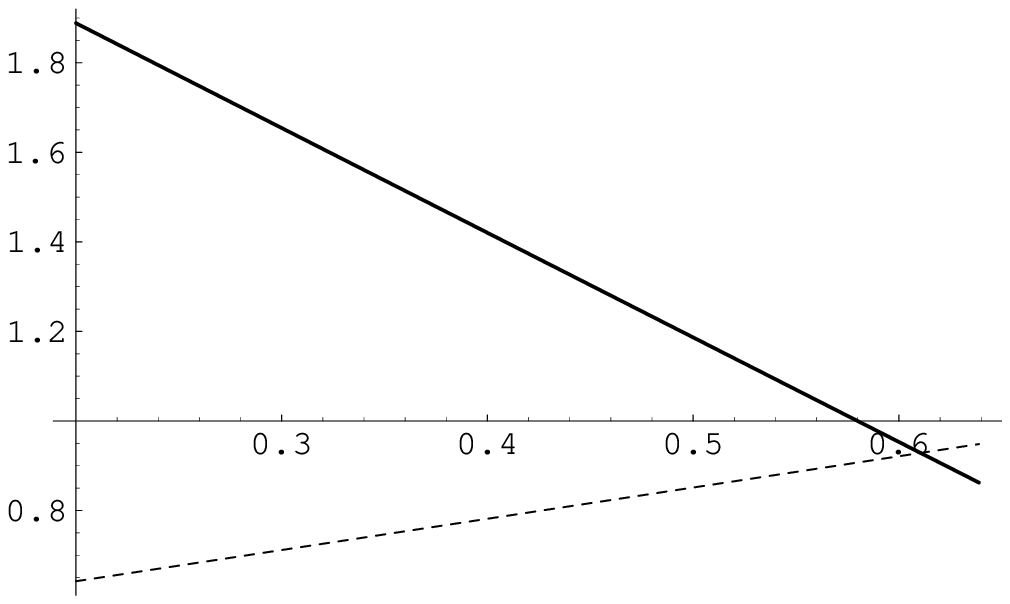}
  \epsfxsize=.3
 \textwidth
  \epsffile{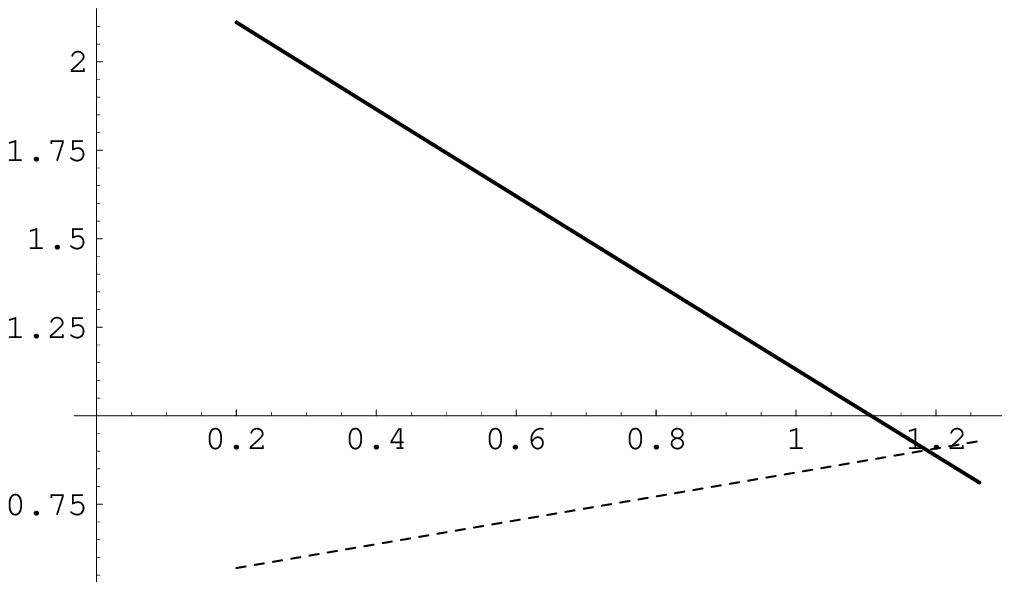}
 \caption{The differential cross $(\nu_e,e^-)$ section , in arbitrary units,
as a function of the electron energy for neutrino 
energies from left to right 7.6, 13.1 and 18.6 keV. The the dashed curve 
corresponds to to the cross
section due to $\nu_{\alpha}$, $\alpha=\mu$ or $\tau$.}
 %\end{center}
 \label{fig:2}
  \end{center}
  \end{figure}

\begin{figure}[!ht]
 \begin{center}
 \epsfxsize=.4
 \textwidth
  \epsffile{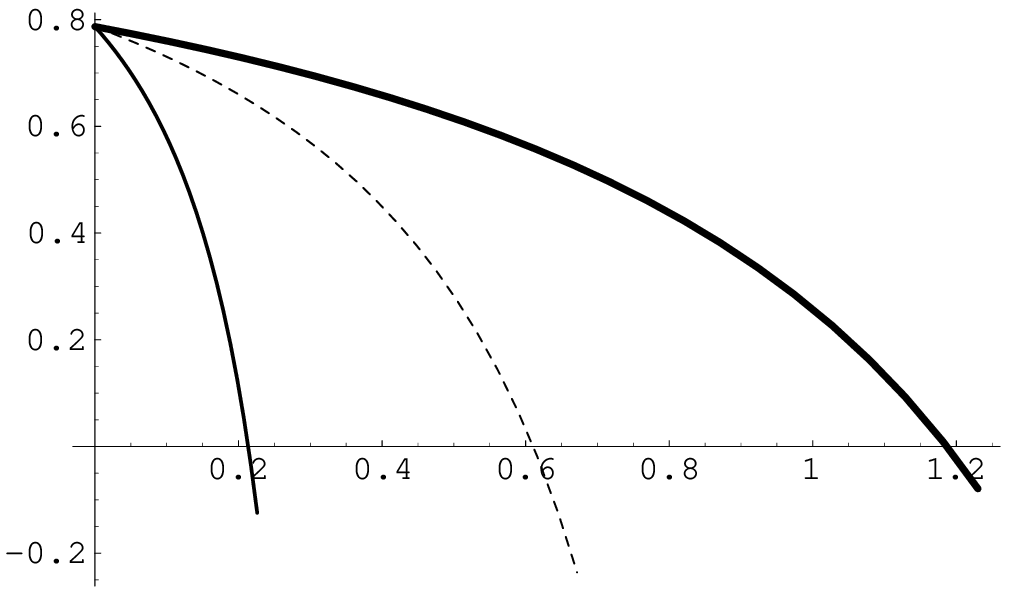}
  \epsfxsize=.4
 \textwidth
  \epsffile{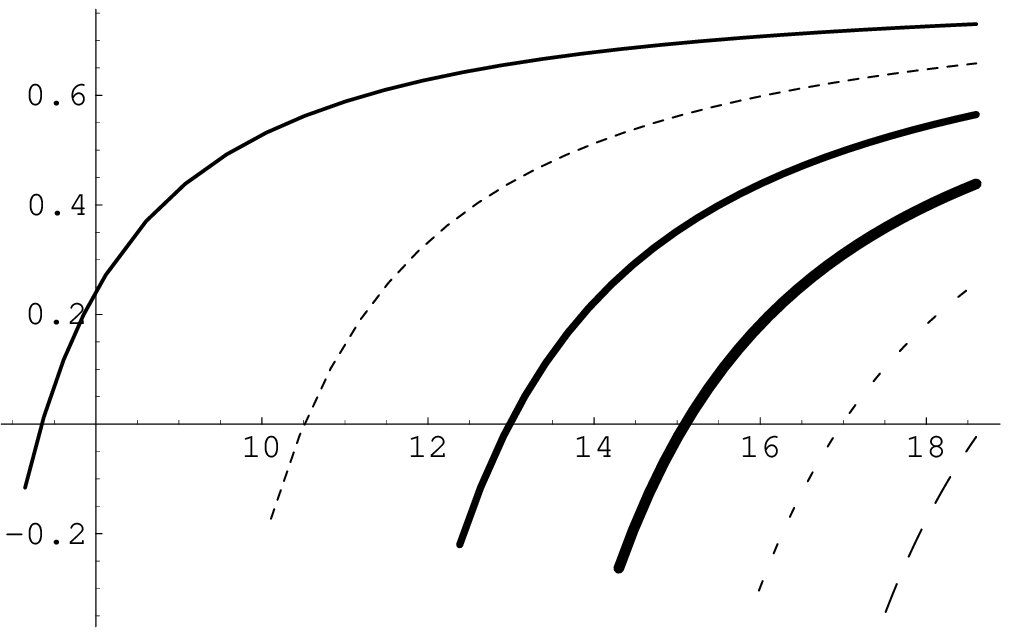}
 \caption{On left the function $\chi(E{\nu},T)$ is shown as a function of the electron energy for neutrino energies 
7.6 keV (thin line), 13.1 keV (dashed line) and 18.6  KeV (thick line). On the right we show the same quantity as a
of the neutrino energy for electron energies 0.2, 0.4, 0.6, 0.8, 1.0, 1.2  keV increasing to the right.}
 %\end{center}
 \label{fig:3}
  \end{center}
  \end{figure}

The electron neutrino is given \cite{NOSTOS1} by:
$$\nu_e=\frac{c~\nu_1+s~\nu_2+s_{13}~ \nu_3}{1+s_{13}}$$
where $s\approx sin \theta_{solar}$ $c\approx cos \theta_{solar}$, $\theta_{solar}$ determined from the 
solar neutrino data, and $s_{13}=sin \theta_{13}$ is a small quantity constrained by the
CHOOZE experiment to be small. The electron disappearance oscillation probability is
given by:
$$P(\nu_e \rightarrow \nu_e)=$$
$$1-4 \frac{s^2c^2 \sin^2(\pi \frac{L}{L_{21}})+\left(s^2 \sin^2(\pi \frac{L}{L_{31}})+
 c^2 \sin^2(\pi \frac{L}{L_{32}})\right)s^2_{13}}{(1+s_{13}^2)^2}$$
 Assuming $\Delta m^2_{31} \approx \Delta m^2_{32}$ we find:
$$P(\nu_e \rightarrow \nu_e)\approx $$
$$1-\frac{(\sin 2 \theta_{solar})^2 \sin^2(\pi \frac{L}{L_{21}})+
 4 s^2_{13} \sin^2(\pi \frac{L}{L_{32}})}{(1+s_{13}^2)^2}$$
where the term  proportional to $s^2_{13}$ is relevant for NOSTOS ($L_{21}=50L_{32}$).
The NOSTOS experiment, in addition to the electronic neutrinos, which are depleted by the oscillations,
 is sensitive to the other two neutrino flavors, which can also scatter electrons
 via the neutral current interactions. These flavors are
  generated via the appearance oscillation:
$$P(\nu_e \rightarrow \sum_{\alpha \#e} \nu_{\alpha})\approx $$
$$\frac{(\sin 2 \theta_{solar})^2 \sin^2(\pi \frac{L}{L_{21}})+
 4 s^2_{13} \sin^2(\pi \frac{L}{L_{32}}}{(1+s_{13}^2)^2}$$
 Thus the number of the scattered electrons, which bear this rather unusual
oscillation pattern, is proportional to 
 the $(\nu_e,e^-)$ scattering cross section
with a proportionality constant given by:
\beqa
P(\nu_e \rightarrow \nu_e)\approx &1& -\chi(E_{\nu},T)
\nonumber\\
& & \frac{(\sin 2 \theta_{solar})^2 \sin^2(\pi \frac{L}{L_{21}})+
 4 s^2_{13} \sin^2(\pi \frac{L}{L_{32}})}{(1+s_{13}^2)^2}
\nonumber\\
\label{oscprob}
\eeqa
where $\chi(E_{\nu},T)$ is one minus the fraction of the cross section of flavor $\nu_{\mu}$ 
or $\nu_{\tau}$ divided by that of $\nu_e$, i.e.
\begin{eqnarray}
& & \chi(E_{\nu},T)=4sin^2 \theta_w\left[2E^2_{\nu}-4T E_{\nu}+T (2T-m_e)\right]/
\nonumber\\
& &[ E^2_{\nu}(1+4\sin^2{\theta_W}+8\sin^4{\theta_W})
- 2 E_{\nu}T(1+2\sin^2{\theta_W})^2
\nonumber\\
&+& (1+2\sin^2{\theta_W})T(2(T-m_e) \sin^2{\theta_W}+T)]
\nonumber\\
\label{chinew}
\end{eqnarray}
%In very low energies, however,  $E_{\nu}\ll m_e$ and $T<E_{\nu}$.
%Thus the last term in
%the numerator  of Eq. (\ref{chinew}) becomes important. In fact if $(1-T/E_{\nu})\simeq 1$
% the function $\chi(E_{\nu},T)$ is positive. In general, however, in the region 
% \end{slide*}
% \end{document}
The function $\chi(E_{\nu},T)$ in the region:
$$E_{\nu}+\frac{m_e}{4}-\frac{1}{4}\sqrt{8 m_e E_{\nu}+
m^2_e}\leq T \leq E_{\nu}+\frac{m_e}{4}+\frac{1}{4}\sqrt{8 m_e E_{\nu}+m_e^2}$$ 
 becomes negative. This means that  the cross section due to $\nu_{\mu}$ 
or $\nu_{\tau}$ becomes larger than that of $\nu_e$. 
This can also be seen by plotting the relevant cross sections, 
see Figs \ref{fig:1} and  \ref{fig:2}. We see 
that the cross section due to flavor $\nu_{\mu}$ or $\nu_{\tau}$ is rising, while the one due to
$\nu_e$ is dropping,
see Fig. \ref{fig:2}. 
 % \newpage
%  \newpage
  This means that 
 the function $\chi(E_{\nu},T)$ eventually
 changes sign, see Fig. \ref{fig:3}. If this occurs in a kinematically interesting region, it will lead to 
neutrino oscillation results different than those naively expected. 
\section{VARIOUS PATTERNS OF OSCILLATION PROBABILITIES}
\begin{figure}[!ht]
 \begin{center}
  \epsfxsize=.3
 \textwidth
  \epsffile{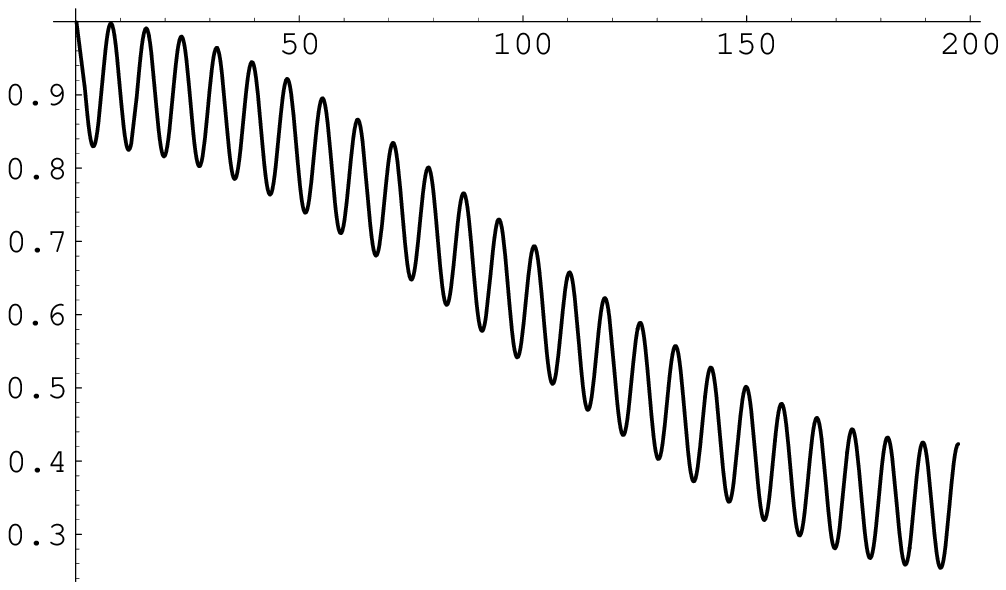}
  \epsfxsize=.3
 \textwidth
  \epsffile{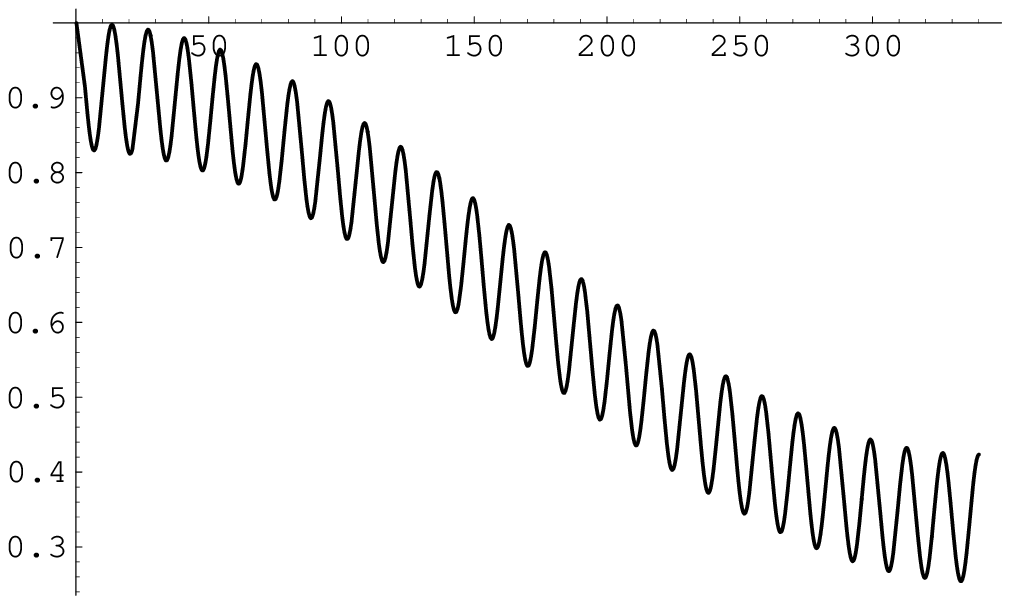}
  \epsfxsize=.3
 \textwidth
  \epsffile{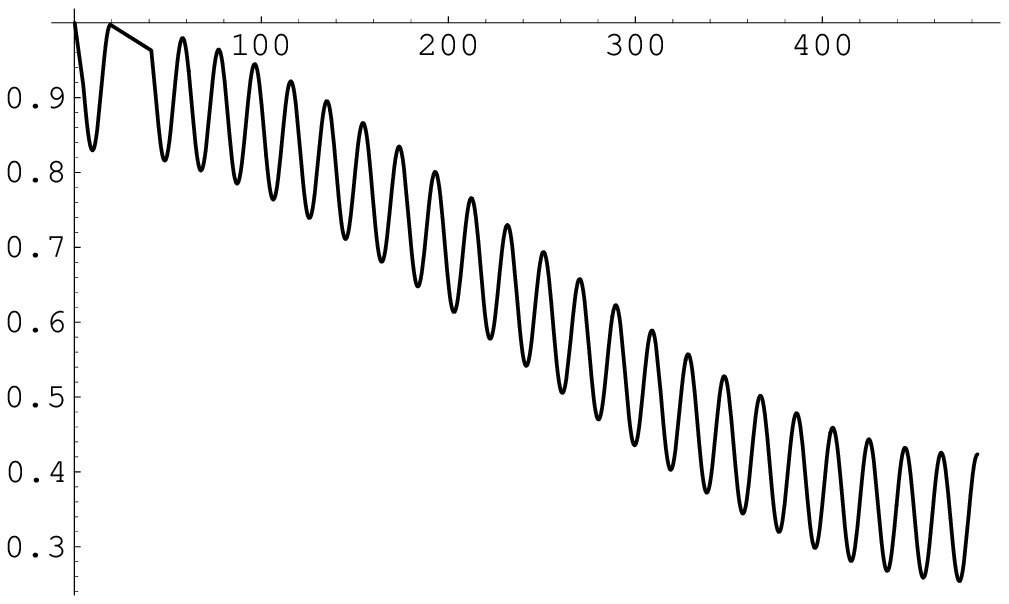}
 \caption{The oscillation probability  at distances larger than the NOSTOS size  
without the function $\chi(E_{\nu},T)$. One can see
the oscillation associated with $\Delta m^2_{12}$. If sensitive to
 to $sin^2 2\theta_{13}=0.17$ one will also see on top of this the small oscillation associated
with $\Delta m^2_{32}$. From left to right,
 $E_{\nu}=7.6,13.1$ and $18.6~keV$.}
 %\end{center}
 \label{fig:1a}
  \end{center}
  \end{figure}

\begin{figure}[!ht]
 \begin{center}
   \epsfxsize=.3
 \textwidth
  \epsffile{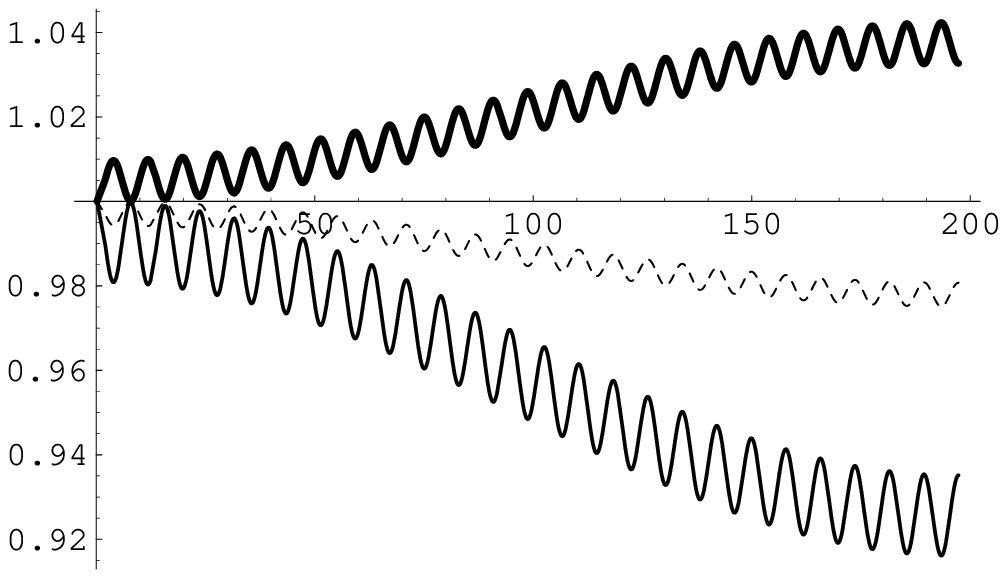}
  \epsfxsize=.3
 \textwidth
  \epsffile{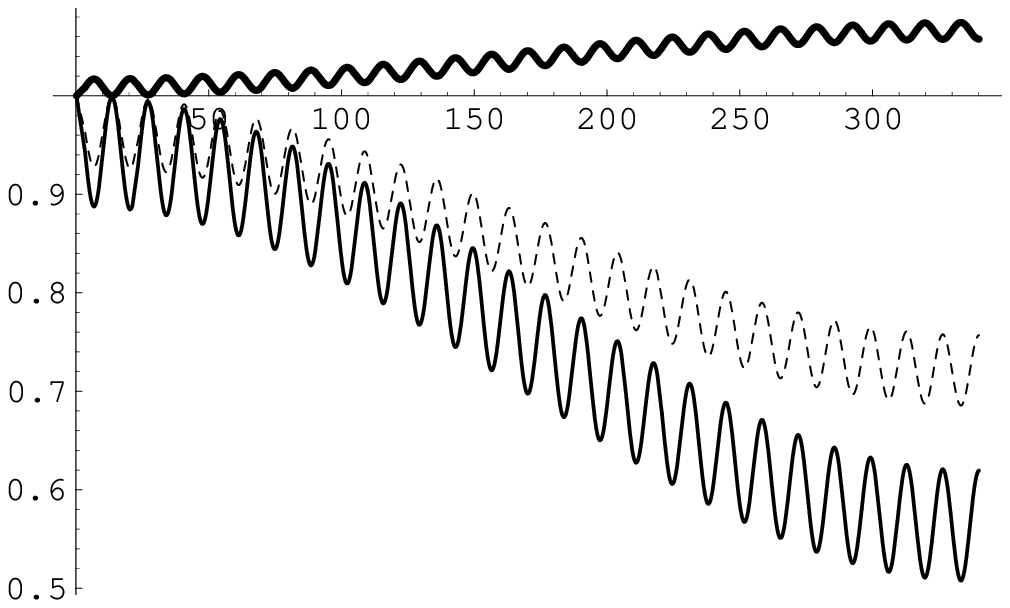}
  \epsfxsize=.3
 \textwidth
  \epsffile{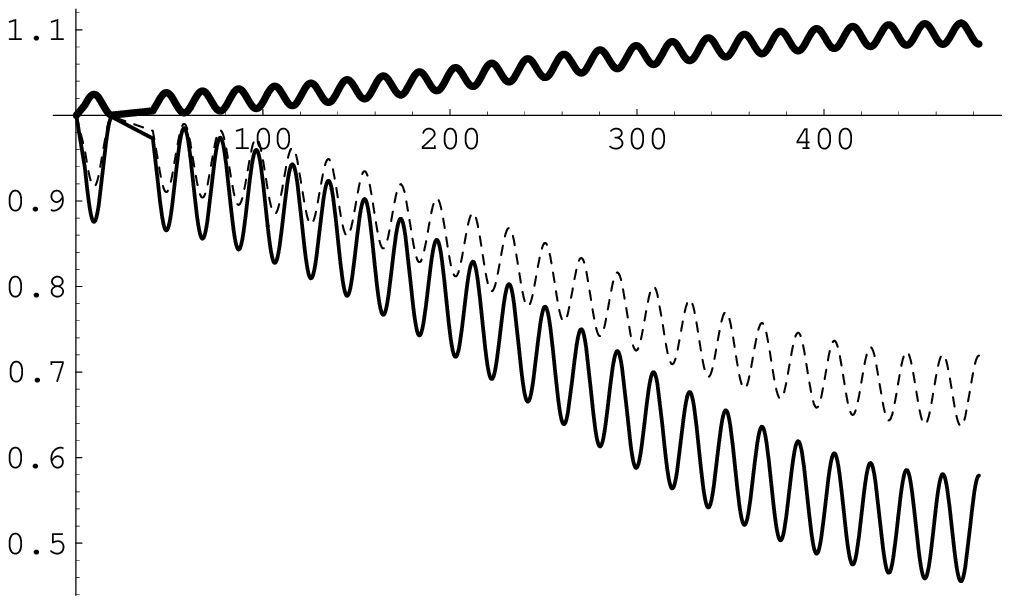}
 \caption{The same as in Fig. \ref{fig:1a} after including the function $\chi(E_{\nu},T)$. The results now
 depend on the electron energy. For  $E_{\nu}=7.6$ (left) we show the results for $T=0.200,0.213$ and $0.220$ keV 
indicated by
thin, dashed and thick line respectively. For  $E_{\nu}=13.1$ (middle), $T=0.200,0.419$ and $0.639$ keV, while
 for $E_{\nu}=18.6$ (right), $T=0.200,0.731$ and $1.262$ keV.}
 %\end{center}
 \label{fig:1b}
  \end{center}
  \end{figure}

\begin{figure}[!ht]
 \begin{center}
   \epsfxsize=.3
 \textwidth
  \epsffile{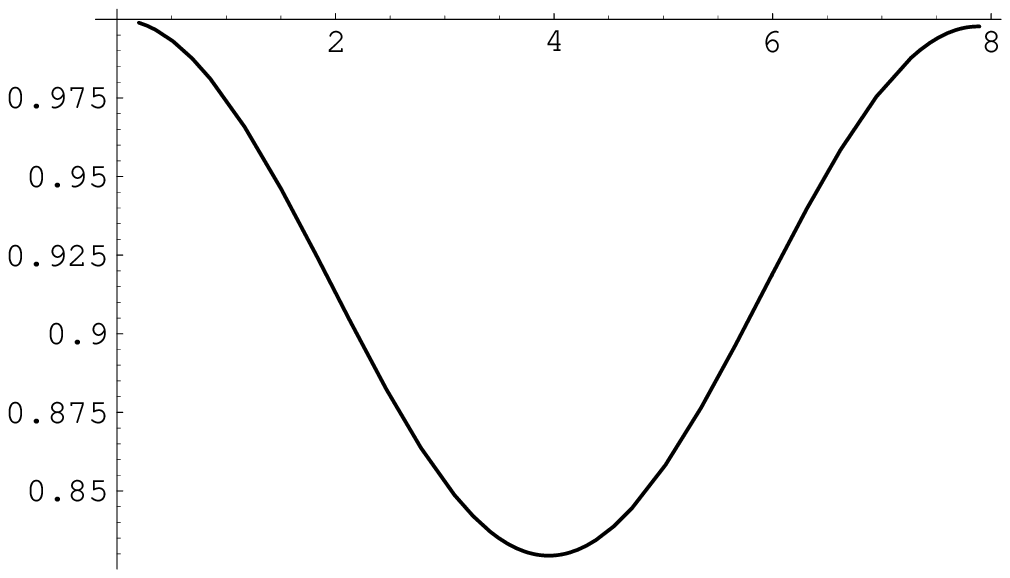}
  \epsfxsize=.3
 \textwidth
  \epsffile{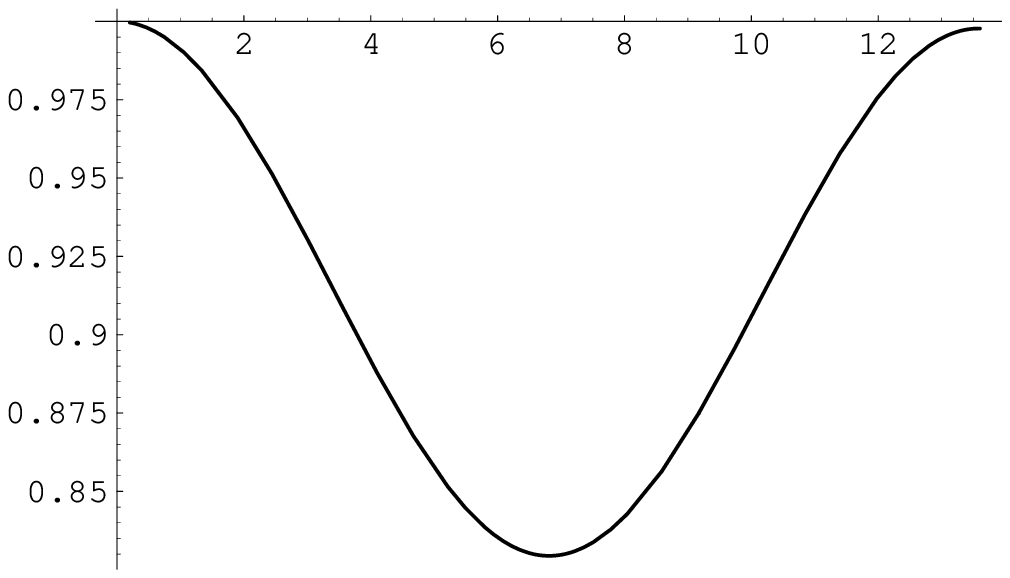}
  \epsfxsize=.3
 \textwidth
  \epsffile{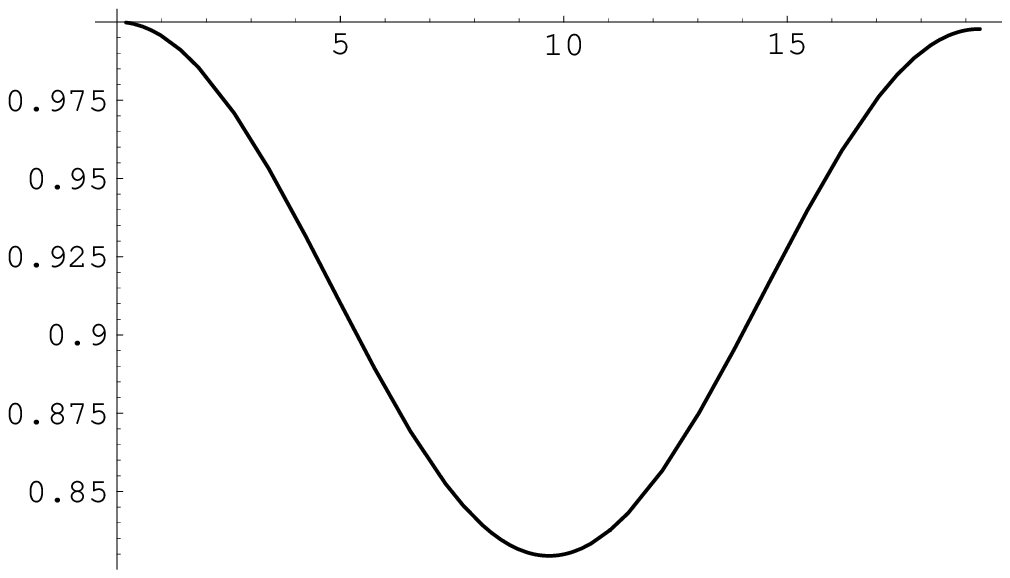}
 \caption{The oscillation probability  expected in NOSTOS with
 $sin^2 2\theta_{13}=0.17$ corresponding to, from left to right,
 $E_{\nu}=7.6,13.1$ and $18.6~keV$, without the function $\chi(E_{\nu},T)$.}
 %\end{center}
 \label{fig:1c}
  \end{center}
  \end{figure}

\begin{figure}[!ht]
 \begin{center}
   \epsfxsize=.3
 \textwidth
  \epsffile{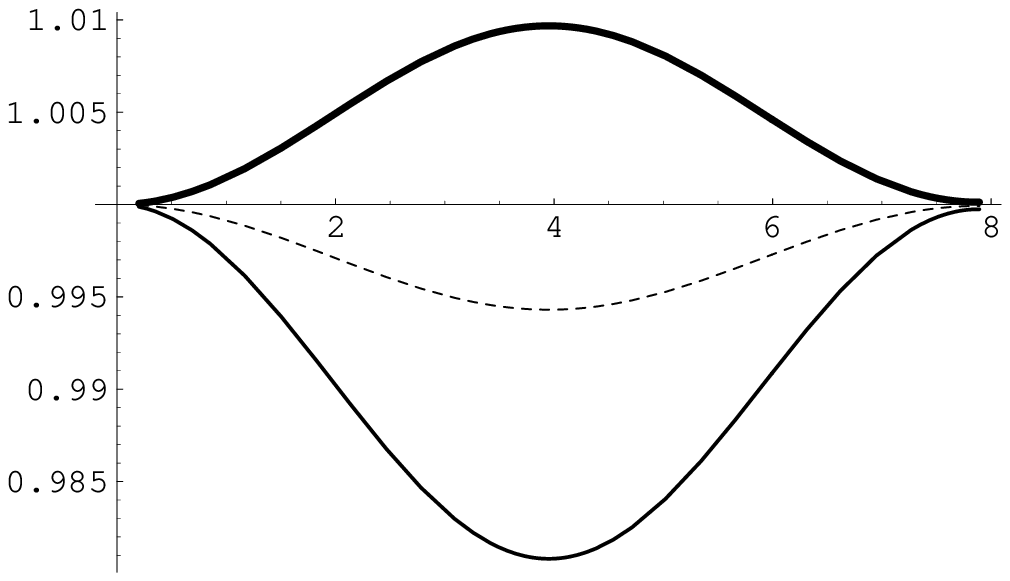}
  \epsfxsize=.3
 \textwidth
  \epsffile{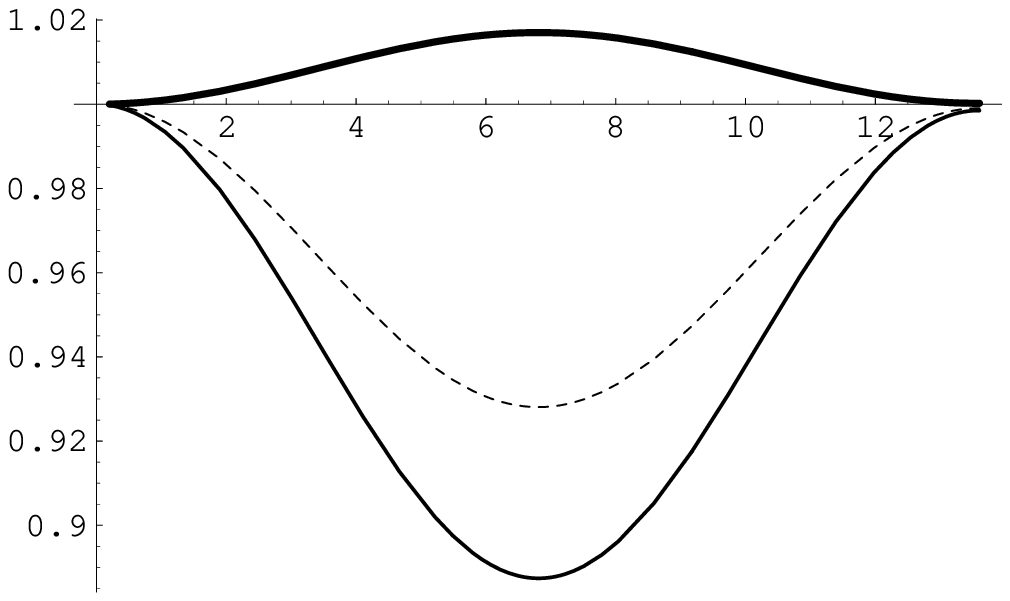}
  \epsfxsize=.3
 \textwidth
  \epsffile{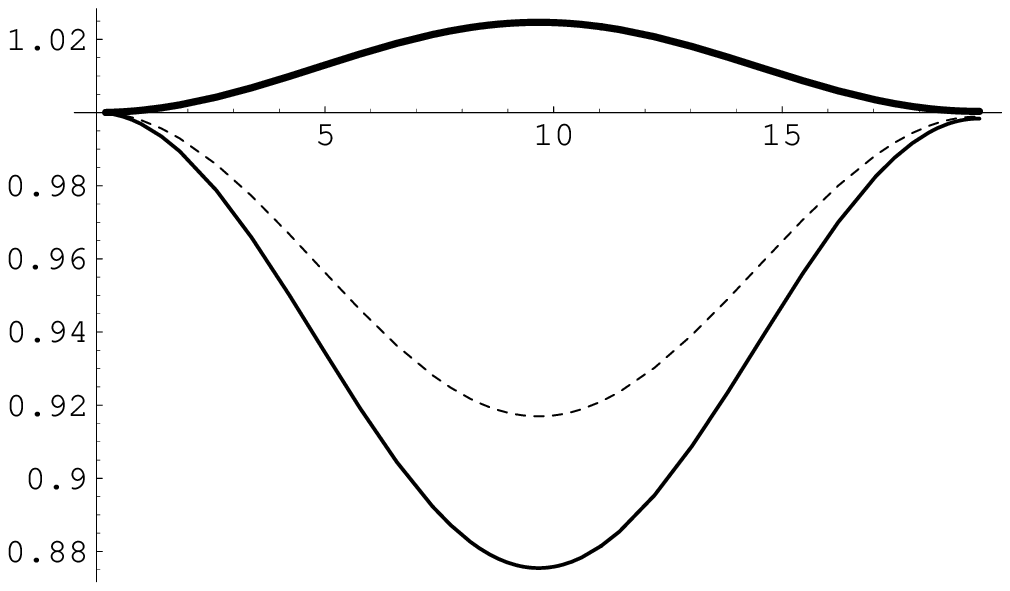}
 \caption{The same as in Fig. \ref{fig:1c} after including the function $\chi(E_{\nu},T)$ and the neutrino and
electron energies as indicated in Fig \ref{fig:1b}.}
 %\end{center}
 \label{fig:1d}
  \end{center}
  \end{figure}

\begin{figure}[!ht]
 \begin{center}
   \epsfxsize=.3
 \textwidth
  \epsffile{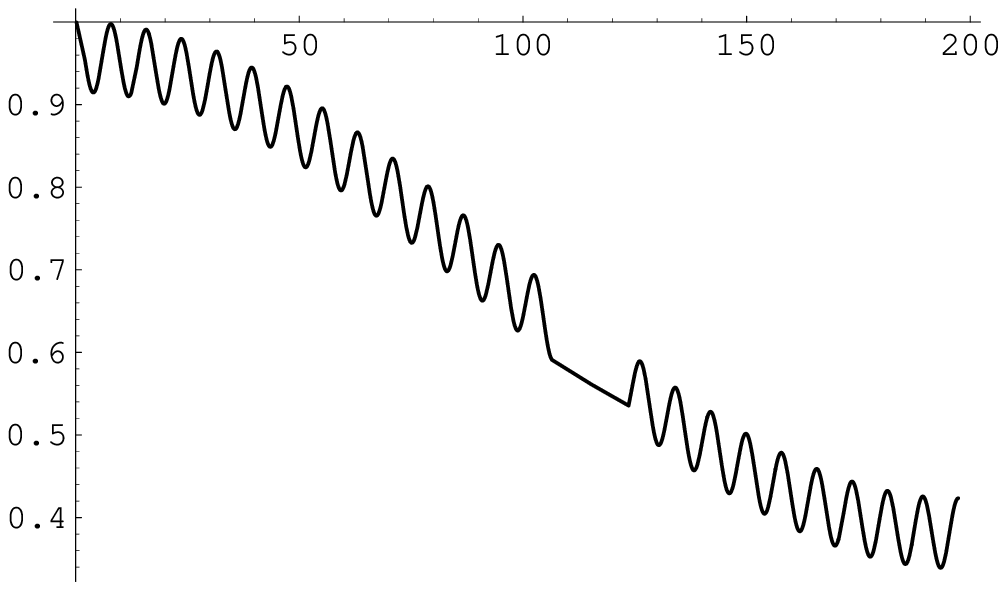}
  \epsfxsize=.3
 \textwidth
  \epsffile{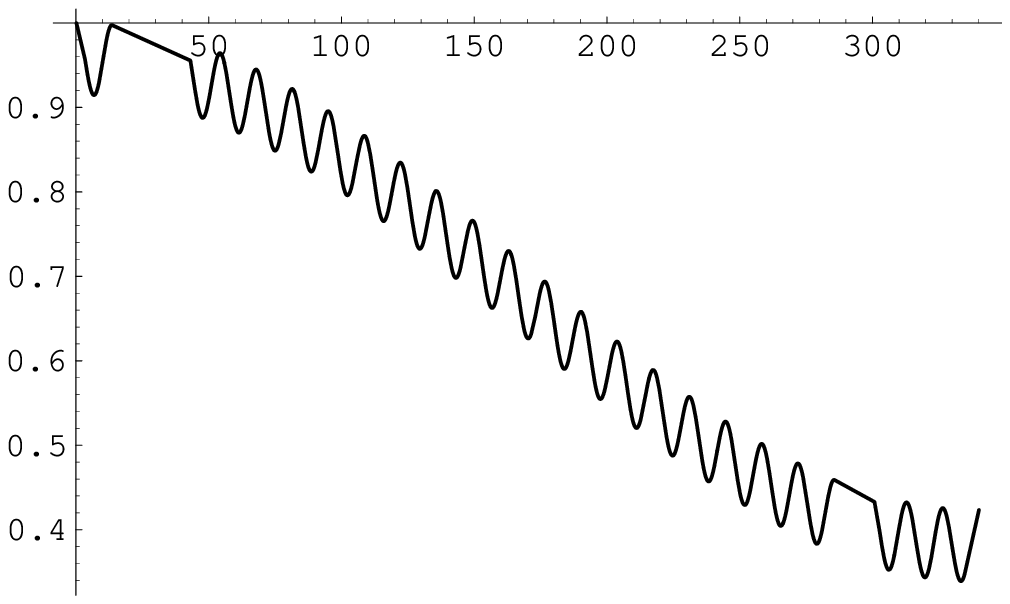}
  \epsfxsize=.3
 \textwidth
  \epsffile{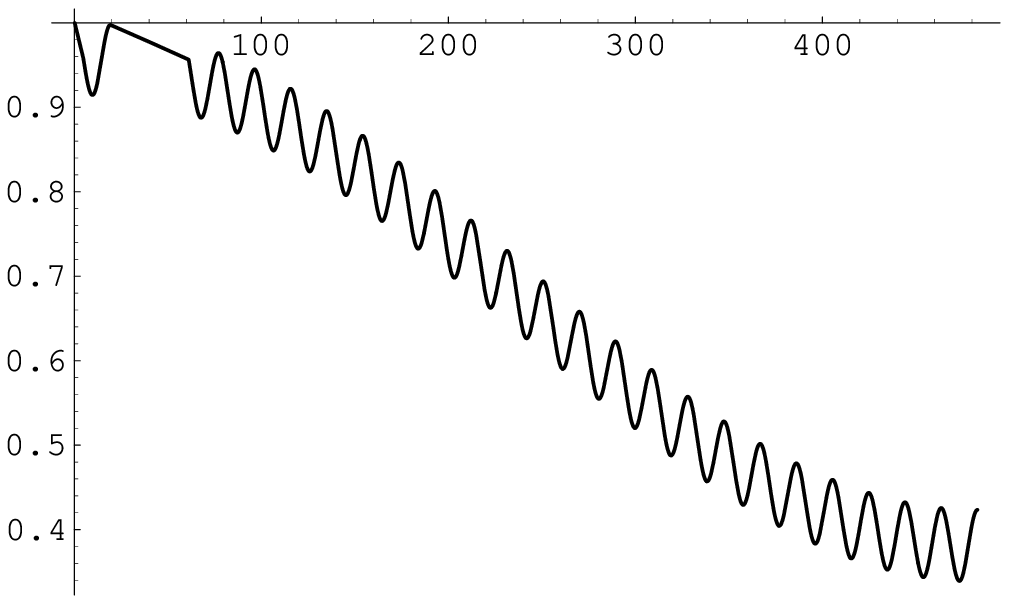}
\caption{The same as in Fig. \ref{fig:1a} with $sin^2 2\theta_{13}=0.085$.}
 %\end{center}
 \label{fig:2a}
  \end{center}
  \end{figure}

\begin{figure}[!ht]
 \begin{center}
    \epsfxsize=.3
 \textwidth
  \epsffile{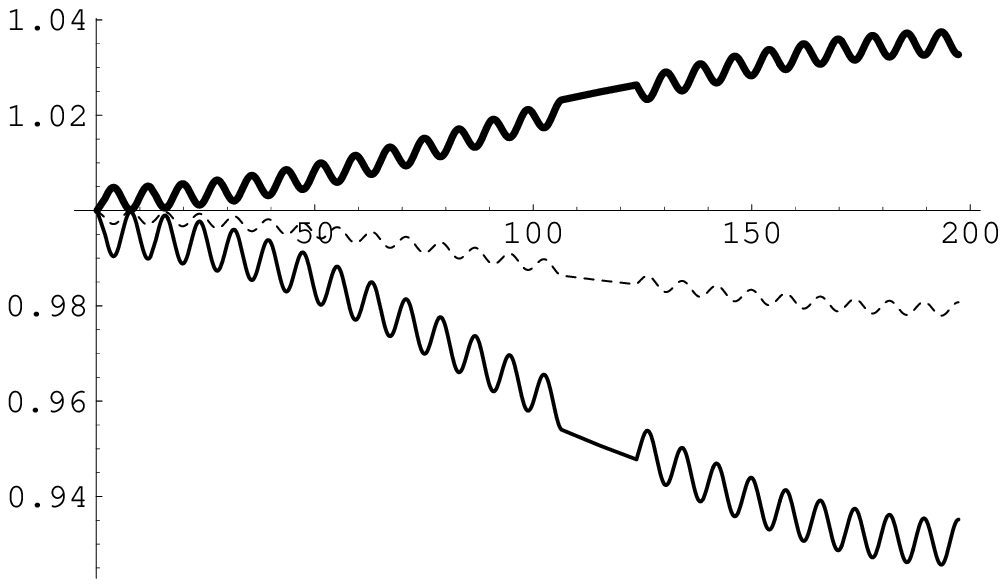}
  \epsfxsize=.3
 \textwidth
  \epsffile{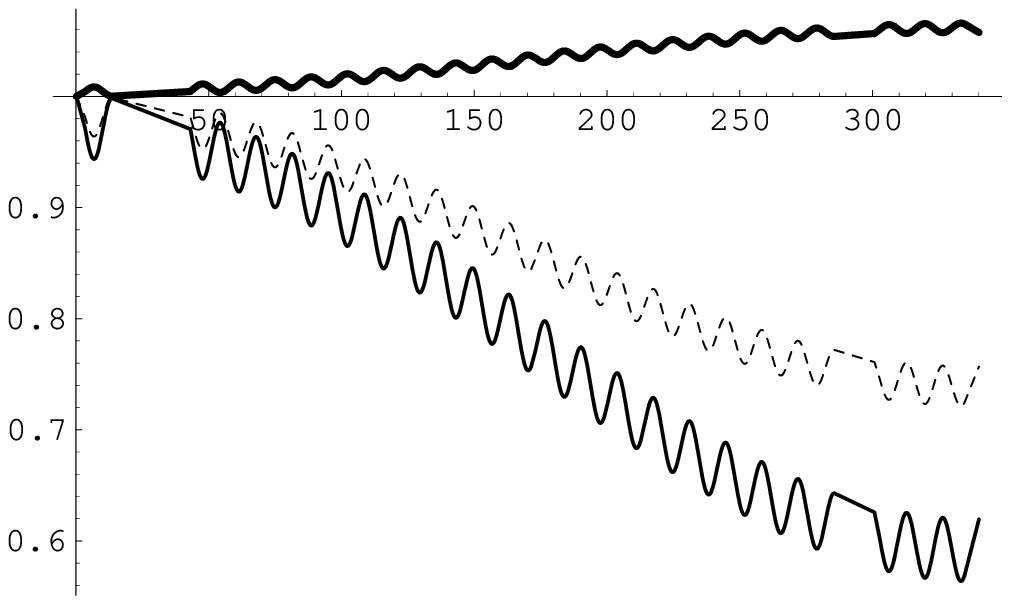}
  \epsfxsize=.3
 \textwidth
  \epsffile{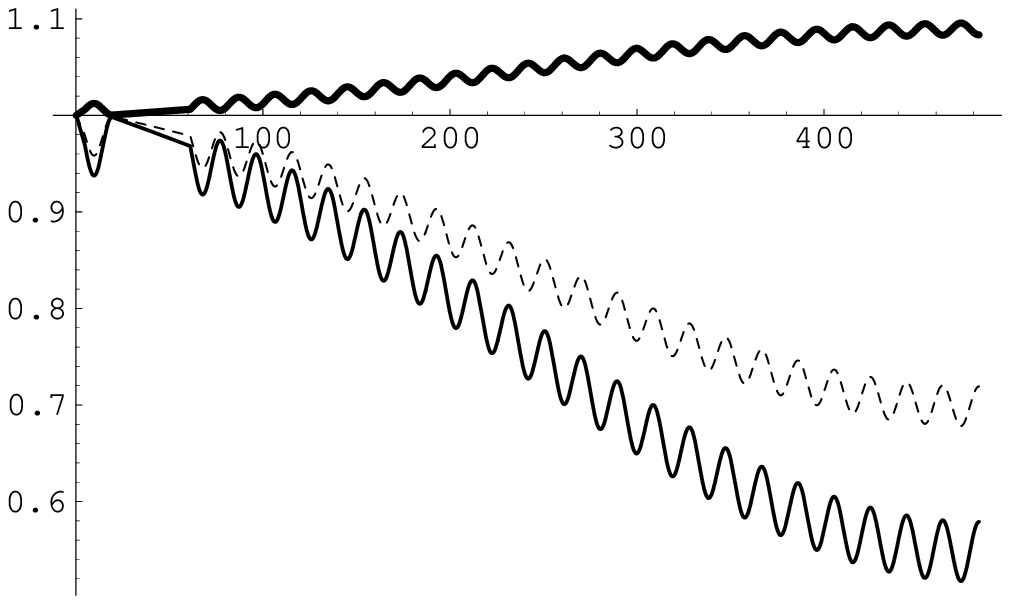}
 \caption{The same as in Fig. \ref{fig:1b} with $sin^2 2\theta_{13}=0.085$.}
 %\end{center}
 \label{fig:2b}
  \end{center}
 \end{figure}

\begin{figure}[!ht]
 \begin{center}
    \epsfxsize=.3
 \textwidth
  \epsffile{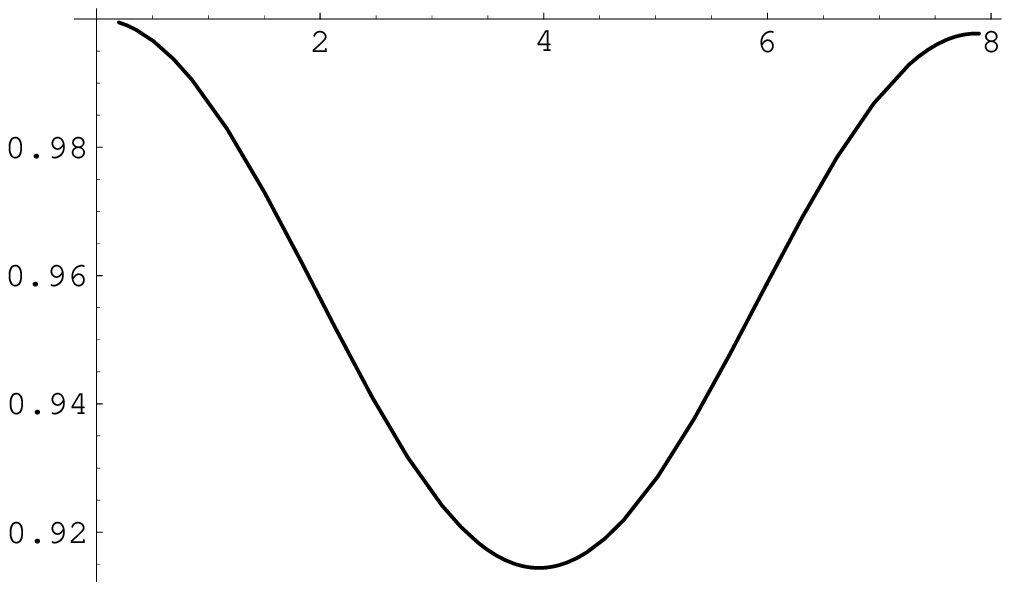}
  \epsfxsize=.3
 \textwidth
  \epsffile{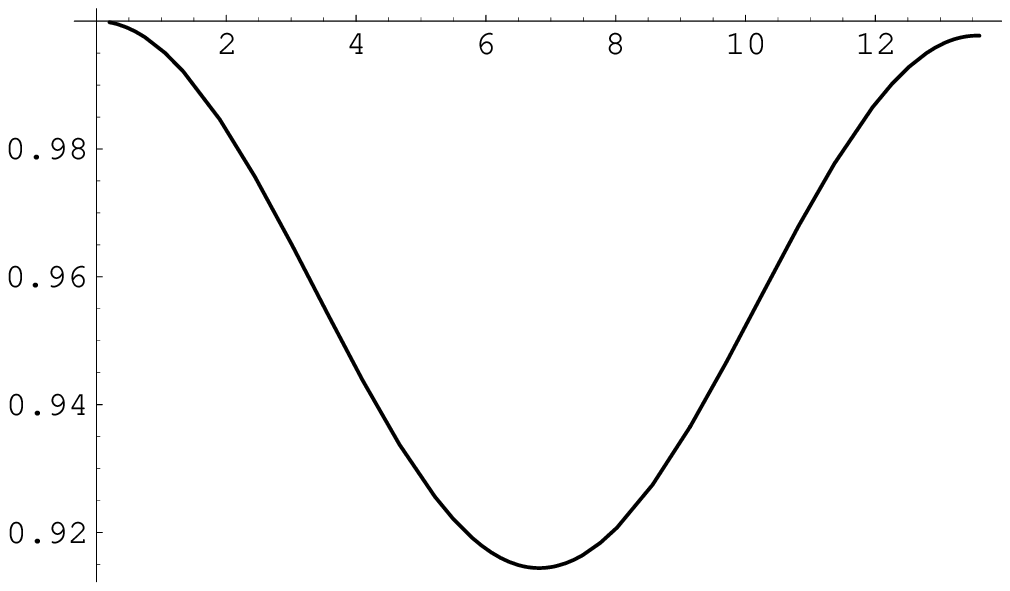}
  \epsfxsize=.3
 \textwidth
  \epsffile{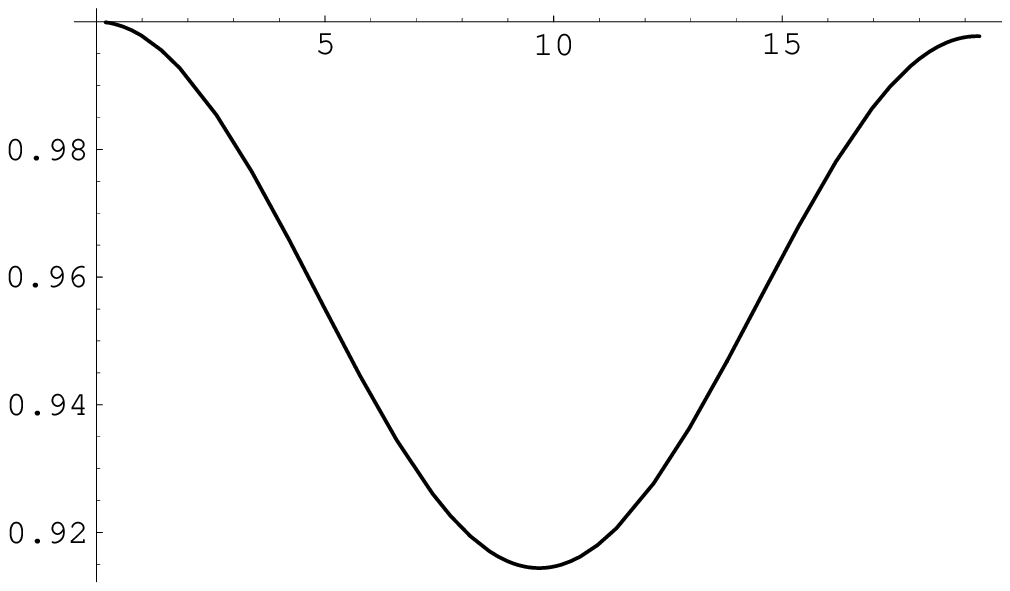}
 \caption{The same as in Fig. \ref{fig:1c} with $sin^2 2\theta_{13}=0.085$.}
 %\end{center}
 \label{fig:2c}
  \end{center}
  \end{figure}

\begin{figure}[!ht]
 \begin{center}
    \epsfxsize=.3
 \textwidth
  \epsffile{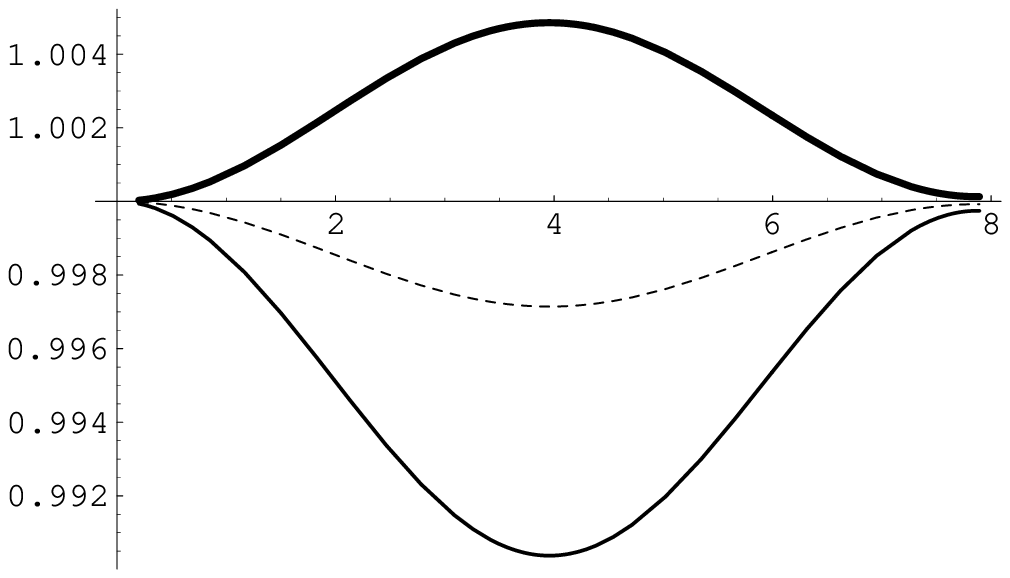}
  \epsfxsize=.3
 \textwidth
  \epsffile{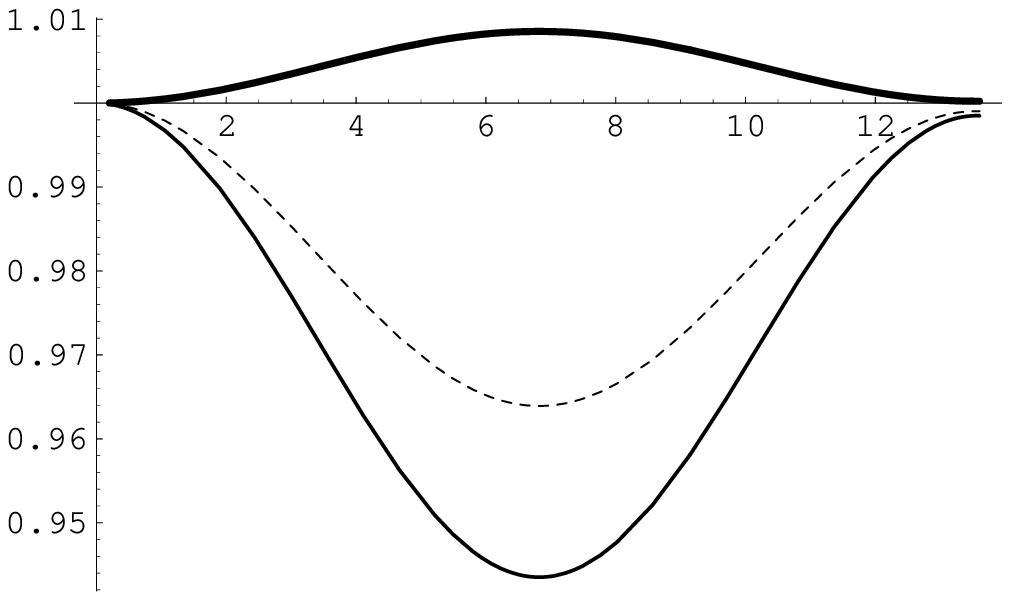}
  \epsfxsize=.3
 \textwidth
  \epsffile{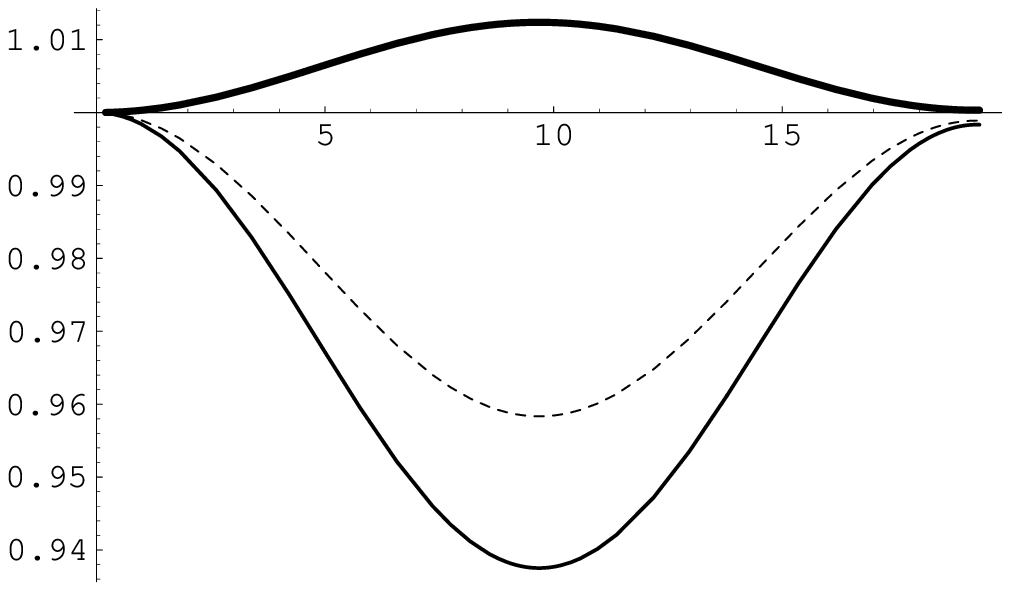}
 \caption{The same as in Fig. \ref{fig:1d} with $sin^2 2\theta_{13}=0.085$.}
 %\end{center}
 \label{fig:2d}
  \end{center}
  \end{figure}

\begin{figure}[!ht]
 \begin{center}
    \epsfxsize=.3
 \textwidth
  \epsffile{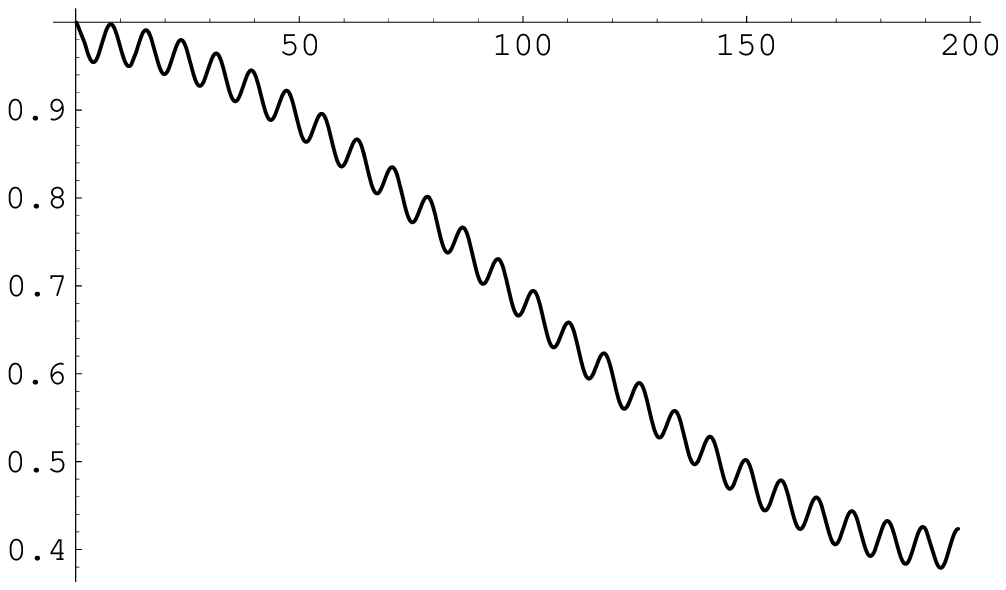}
  \epsfxsize=.3
 \textwidth
  \epsffile{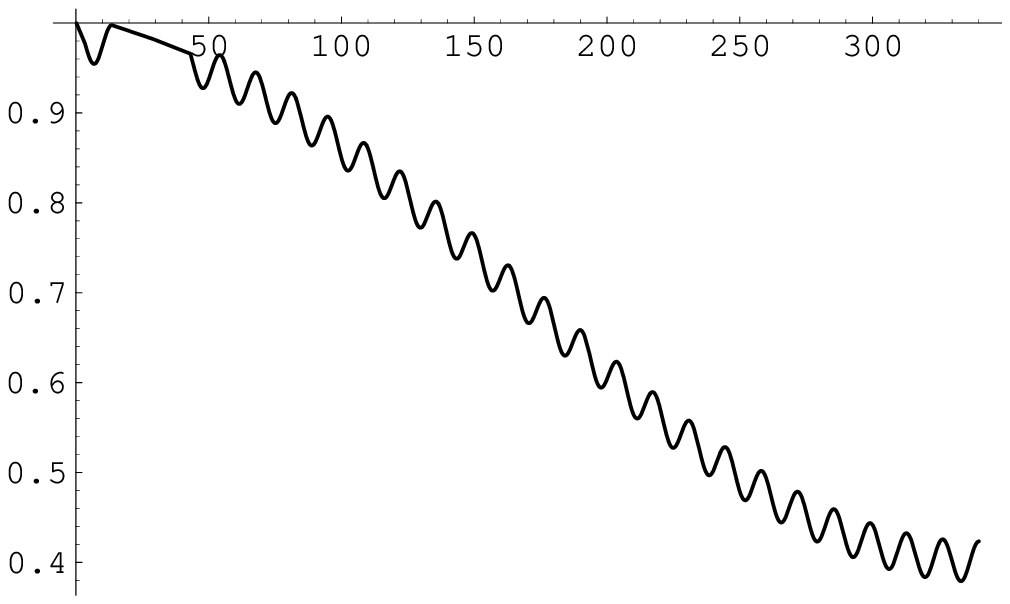}
  \epsfxsize=.3
 \textwidth
  \epsffile{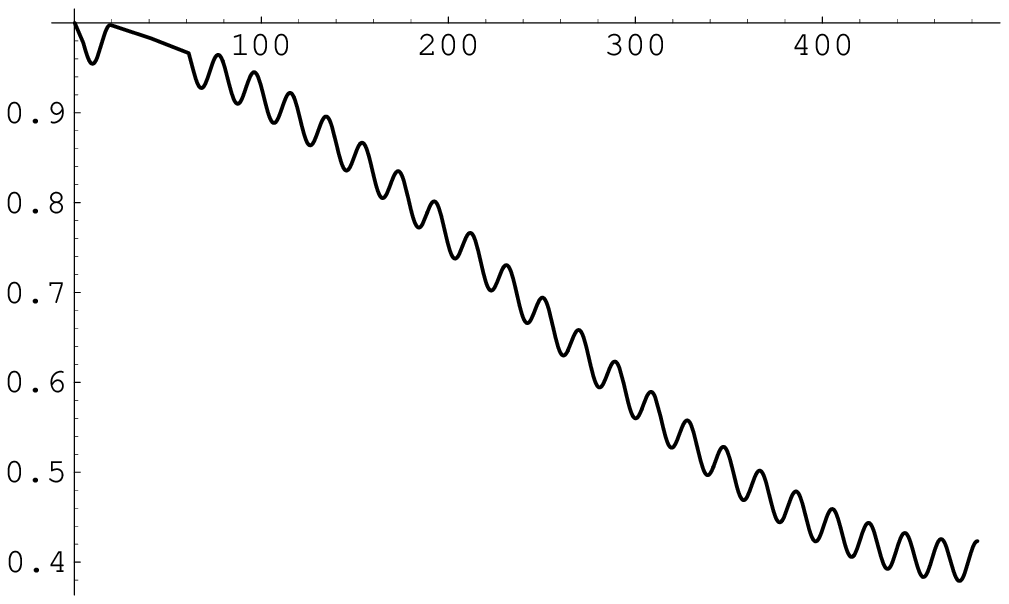}
 \caption{The same as in Fig. \ref{fig:1a} with $sin^2 2\theta_{13}=0.045$.}
 %\end{center}
 \label{fig:3a}
  \end{center}
  \end{figure}

\begin{figure}[!ht]
 \begin{center}
     \epsfxsize=.3
 \textwidth
  \epsffile{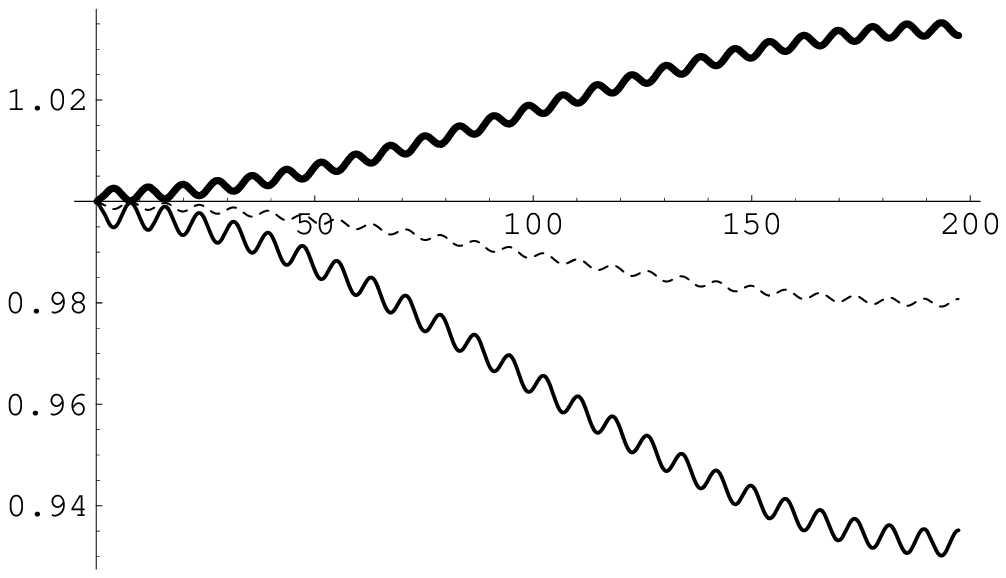}
  \epsfxsize=.3
 \textwidth
  \epsffile{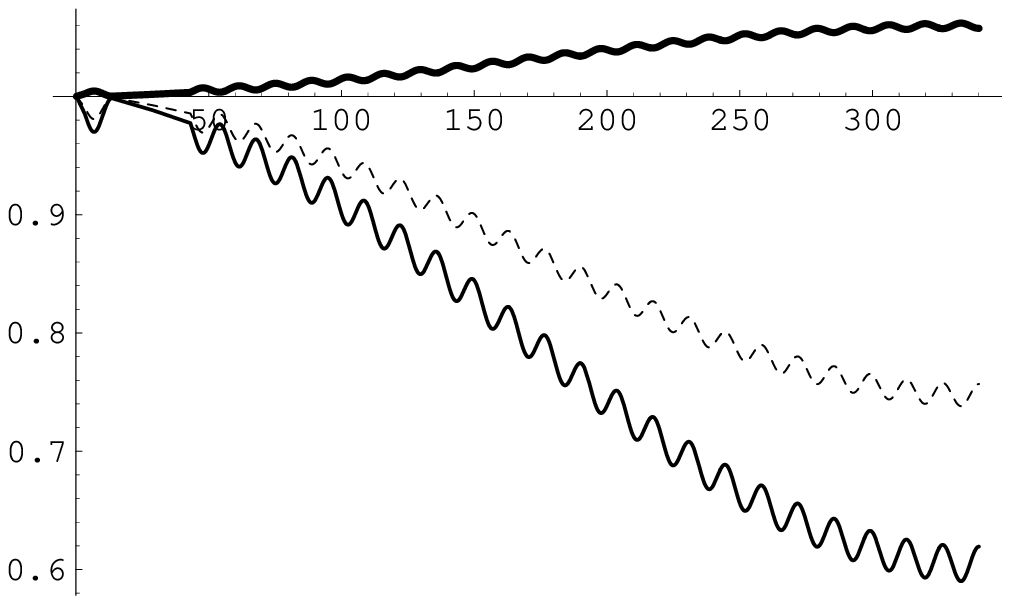}
  \epsfxsize=.3
 \textwidth
  \epsffile{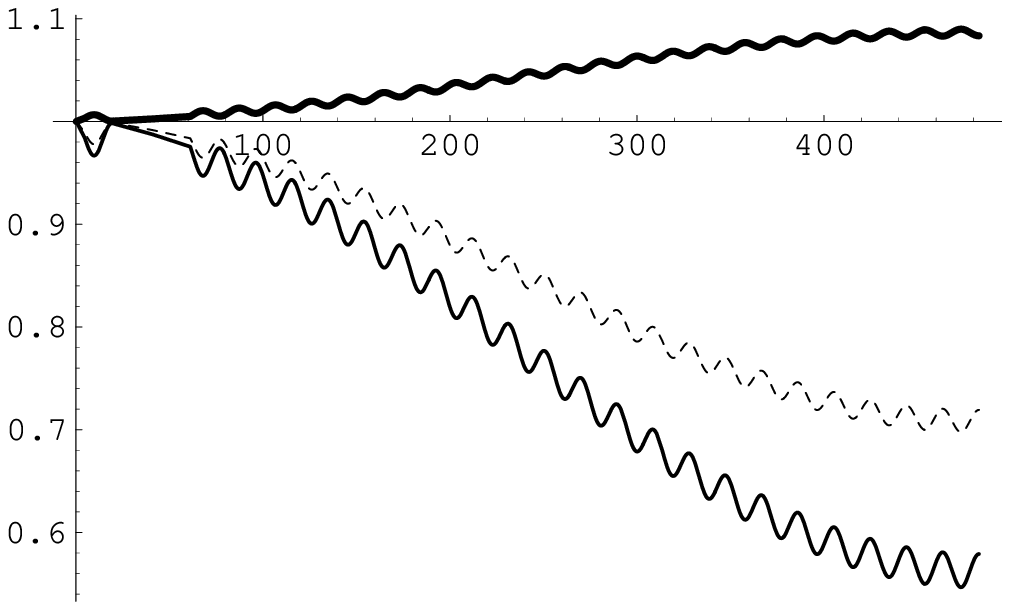}
%\includegraphics[scale=0.4]{bosc_045_1.eps}
%\includegraphics[scale=0.4]{bosc_045_2.eps}
%includegraphics[scale=0.4]{bosc_045_3.eps}
 \caption{The same as in Fig. \ref{fig:1b} with $sin^2 2\theta_{13}=0.045$.}
 %\end{center}
 \label{fig:3b}
  \end{center}
  \end{figure}

\begin{figure}[!ht]
 \begin{center}
     \epsfxsize=.3
 \textwidth
  \epsffile{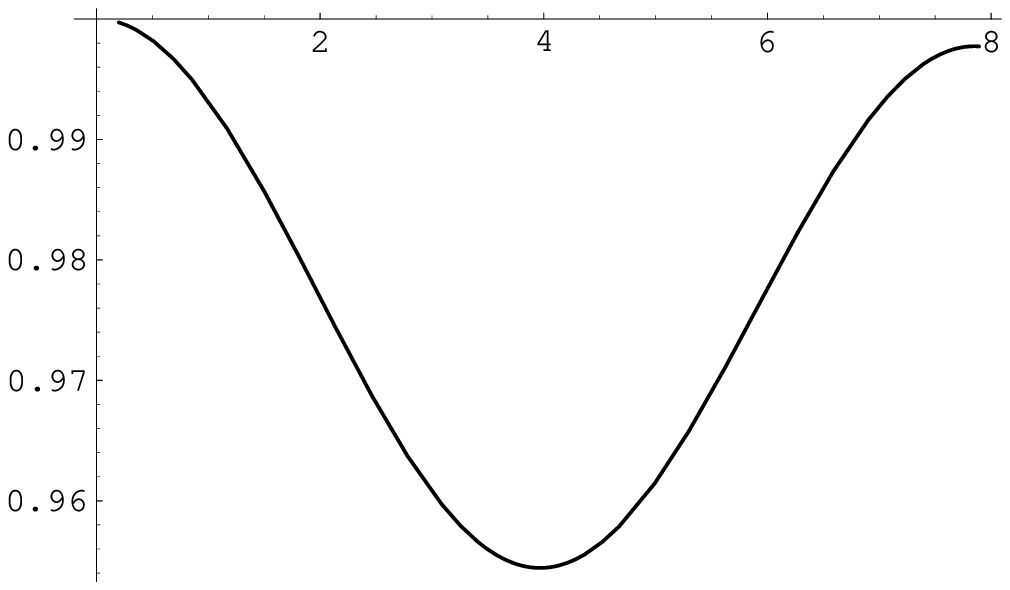}
  \epsfxsize=.3
 \textwidth
  \epsffile{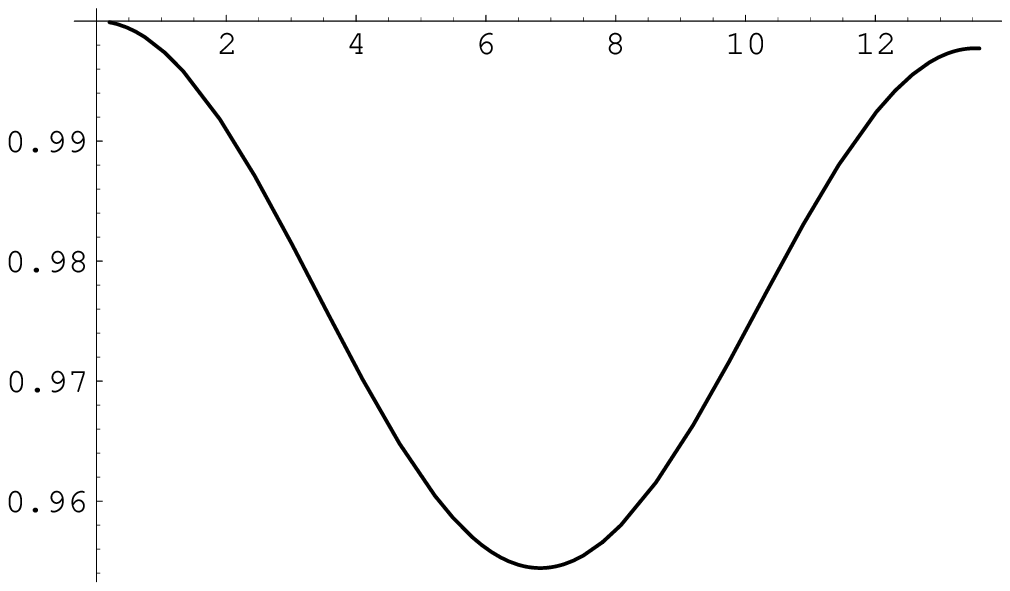}
  \epsfxsize=.3
 \textwidth
  \epsffile{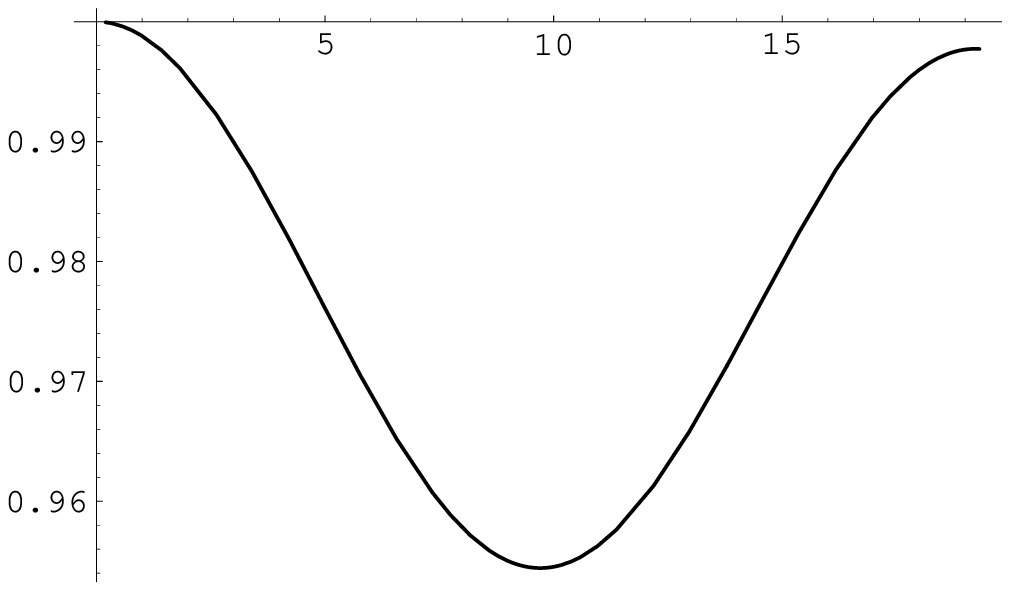}
 \caption{The same as in Fig. \ref{fig:1c} with $sin^2 2\theta_{13}=0.045$.}
 %\end{center}
 \label{fig:3c}
  \end{center}
  \end{figure}

\begin{figure}[!ht]
 \begin{center}
      \epsfxsize=.3
 \textwidth
  \epsffile{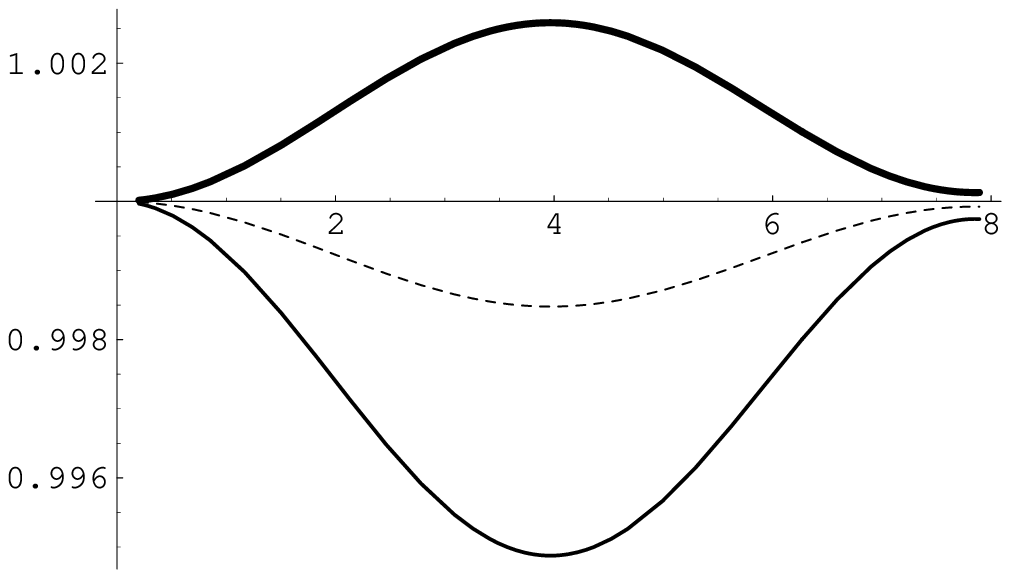}
  \epsfxsize=.3
 \textwidth
  \epsffile{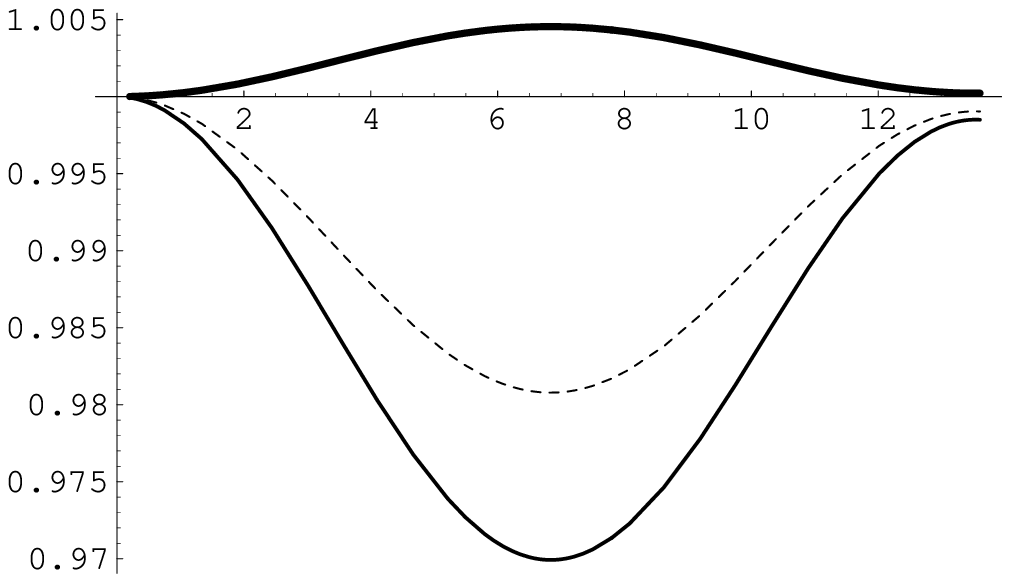}
  \epsfxsize=.3
 \textwidth
  \epsffile{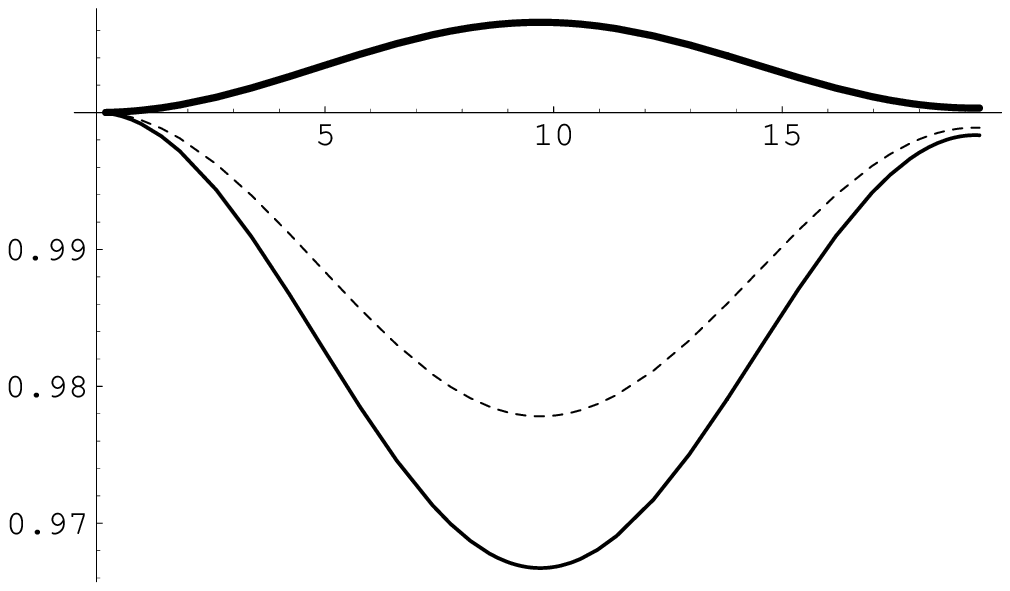}
%\includegraphics[scale=0.4]{sosc_045_1.eps}
%\includegraphics[scale=0.4]{sosc_045_2.eps}
%^\includegraphics[scale=0.4]{sosc_045_3.eps}
 \caption{The same as in Fig. \ref{fig:1d} with $sin^2 2\theta_{13}=0.045$.}
 %\end{center}
 \label{fig:3d}
  \end{center}
  \end{figure}

The oscillation probabilities for NOSTOS, i.e. near the source, as well as experiments further from the source for 
 $sin^2 2\theta_{13}=0.17$ with and without the function $\chi(E_{\nu},T)$ are shown in
Figs \ref{fig:1a}-\ref{fig:1d}. For  $sin^2 2\theta_{13}=0.085$ in Figs \ref{fig:2a}-\ref{fig:2d} and for
$sin^2 2\theta_{13}=0.045$ in Figs \ref{fig:3a}-\ref{fig:3d}
% \newpage
%  \end{slide*}
% \end{document}
 %\newpage
 %   \end{slide*}
 %\end{document}
  %\newpage
% \end{slide*}
 %\end{document}
%  \newpage
 %  \end{slide*}
 %\end{document}
% \newpage
 % \newpage
 % \newpage
% \newpage
 % \newpage
\section{AVERAGED NEUTRINO OSCILLATION PROBABILITIES}
With the above ingredients one can obtain an effective neutrino oscillation probability
 for various electron energies by averaging the expression of the
oscillation probability, Eq. (\ref{oscprob}), 
 over the weak differential cross section, Eq. \ref{elasw}, and the neutrino spectrum, see Fig. \ref{spectrum},
 i.e.
 \beq
\prec P(\nu_e \rightarrow \nu_e)\succ =\frac{\int\ P(\nu_e \rightarrow \nu_e) \left(\frac{d\sigma}{dT}\right)_{weak}
f_{\nu}(E_{\nu})dE_{\nu}}{\int\left(\frac{d\sigma}{dT}\right)_{weak}f_{\nu}(E_{\nu})dE_{\nu}}
\label{aveosc}
\eeq
 \begin{figure}[!ht]
 \begin{center}
      \epsfxsize=.3
 \textwidth
  \epsffile{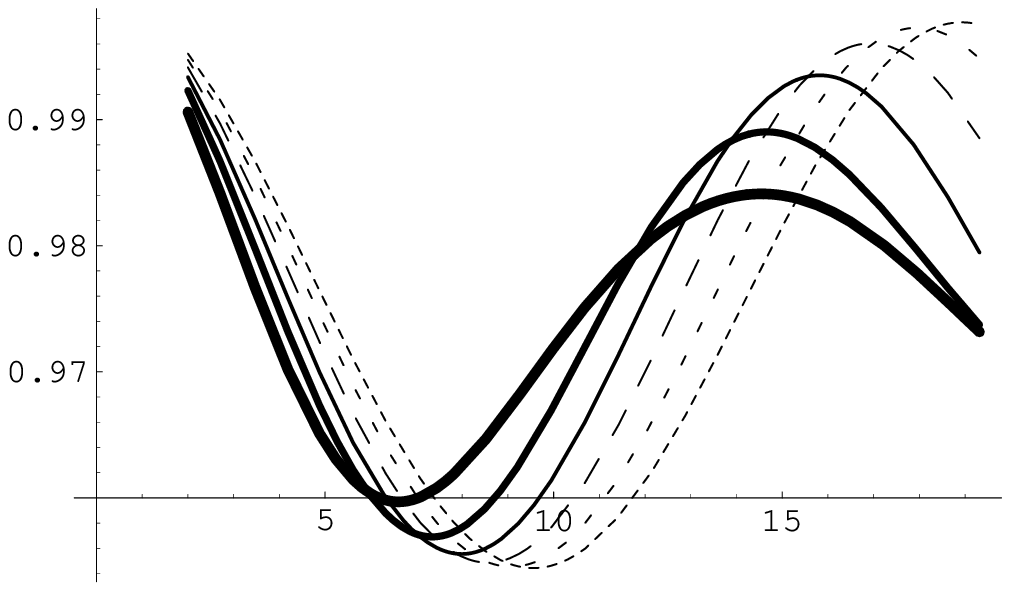}
  \epsfxsize=.3
 \textwidth
  \epsffile{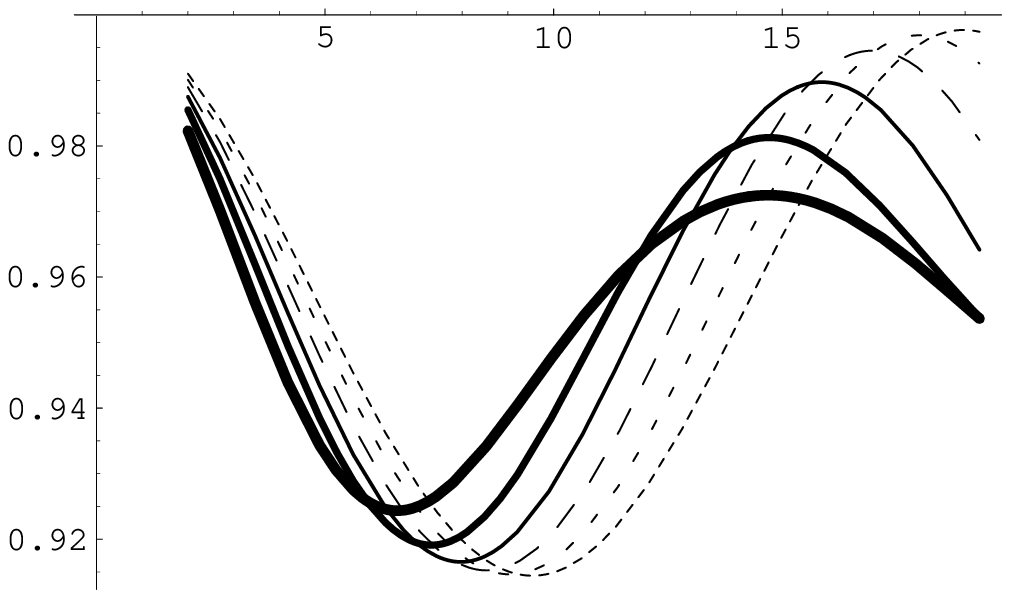}
  \epsfxsize=.3
 \textwidth
  \epsffile{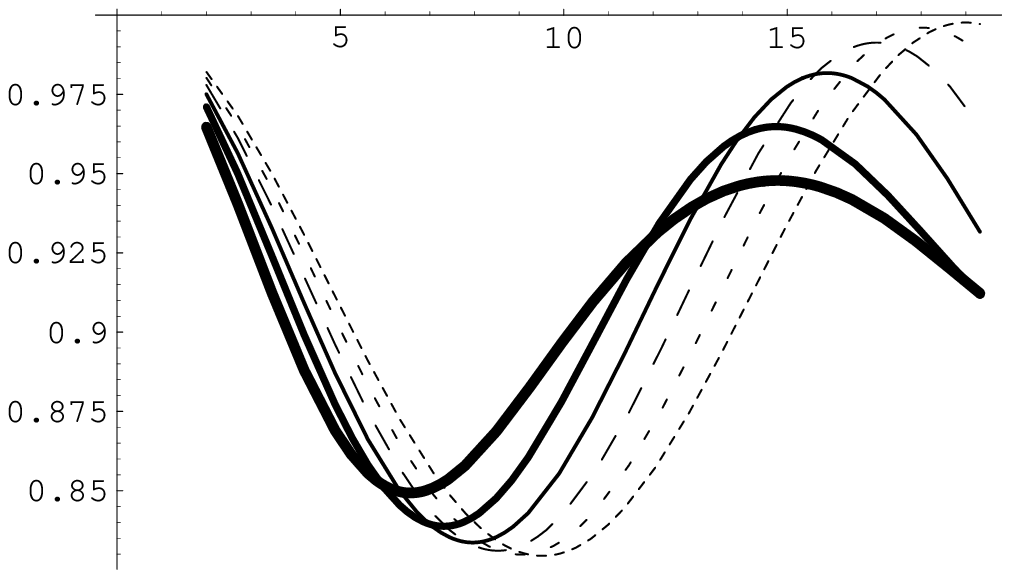}
 \caption{The average oscillation probability as a function of the radial distance $L$ (in meters).
 We have integrated over the neutrino spectrum, 
 but the  function $\chi(E_{\nu},T)$
%, which is the ratio of the charged current divided by the total $(\nu_e,e^-)$ cross section
 was not included. From left to right
$sin^2 2\theta_{13}=0.045,0.085,0.170$.  The depth is increasing with $T$}
 %\end{center}
 \label{fig:5a}
  \end{center}
  \end{figure}
Oscillations for $sin^2 2\theta_{13}=0.17$ as seen in distances far
greater than the NOSTOS experiment as well as in the NOSTOS experiment are shown 
in Figs \ref{fig:1a}-\ref{fig:1d} both with and without the function
$\chi(E_{\nu},T)$. Results for $sin^2 2\theta_{13}=0.085, 0.045$ are shown in Figs
 \ref{fig:2a}-\ref{fig:2d} and \ref{fig:3a}-\ref{fig:3d} respectively.  
\begin{figure}[!ht]
\begin{center}
      \epsfxsize=.3
 \textwidth
  \epsffile{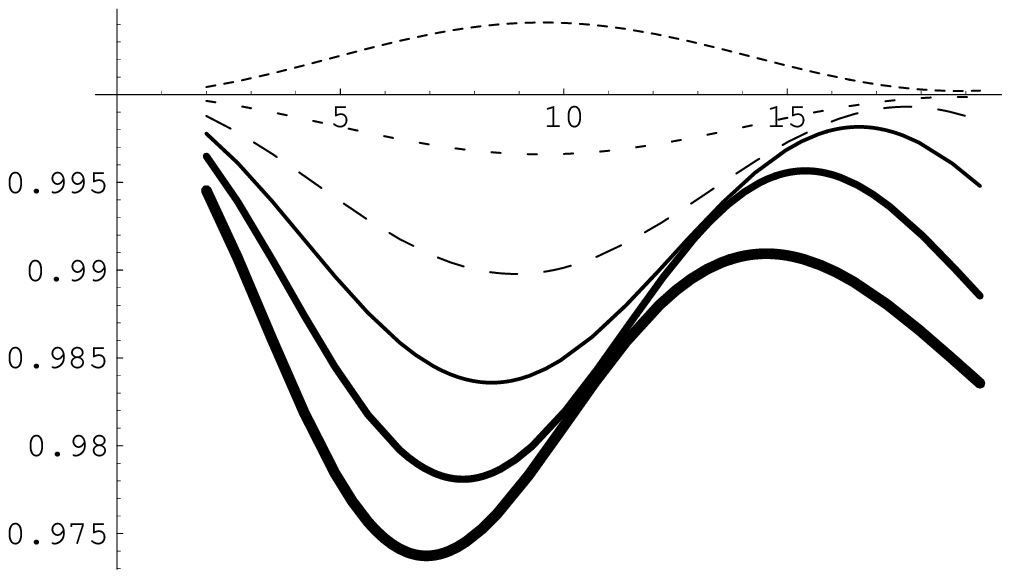}
  \epsfxsize=.3
 \textwidth
  \epsffile{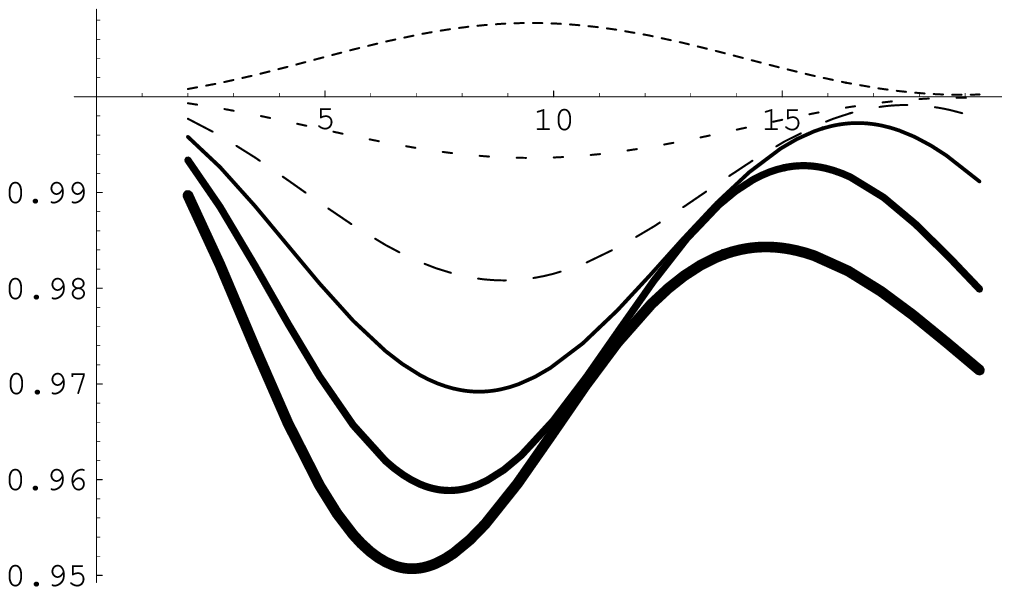}
  \epsfxsize=.3
 \textwidth
  \epsffile{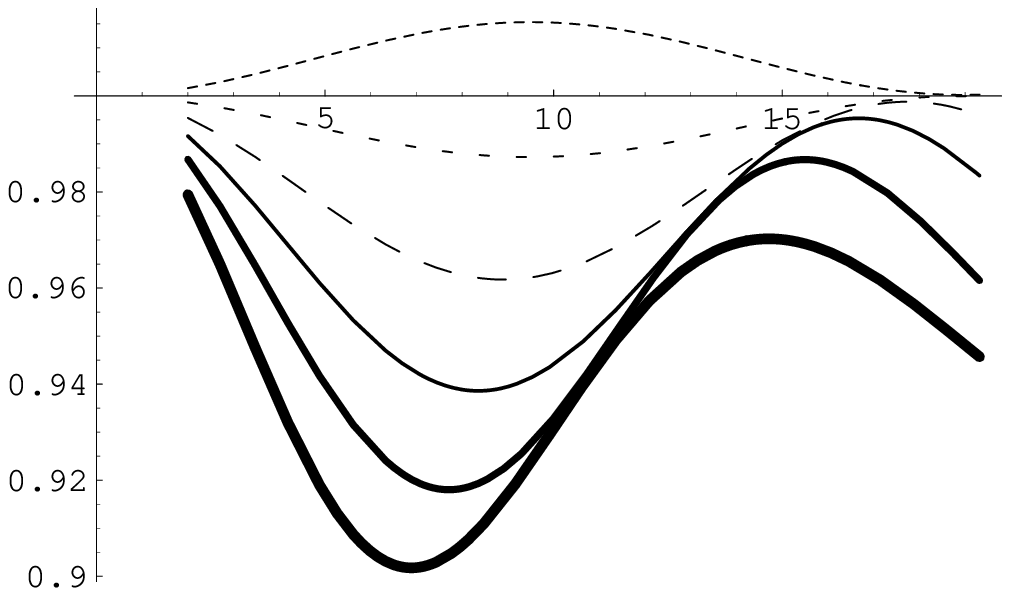}
 \caption{The average oscillation probability as a function of the radial distance $L$.
 We have integrated over the neutrino spectrum
 and included  the  function $\chi(E_{\nu},T)$.
%,  the ratio of the  $(\nu_e,e^-)$ cross sections, i.e. of the charged current divided by that of the  total, was included. 
From left to right
$sin^2 2\theta_{13}=0.045,0.085,0.170$. Labeling as in Fig. \ref{fig:5a}.}
 %\end{center}
 \label{fig:5b}
  \end{center}
  \end{figure}
\section{RESULTS}
\begin{figure}[!ht]
 \begin{center}
       \epsfxsize=.4
 \textwidth
  \epsffile{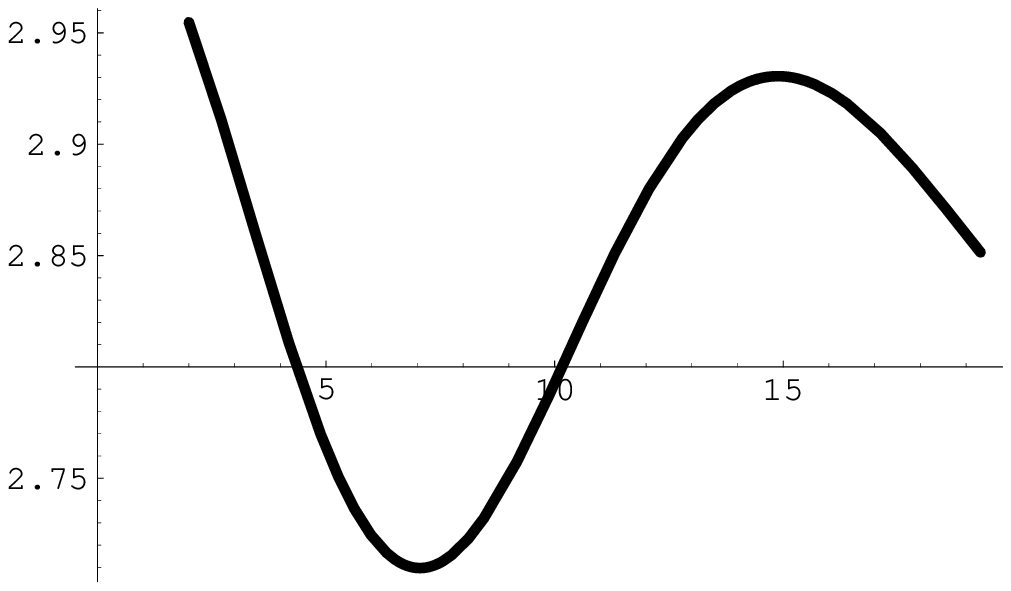}
    \epsfxsize=.4
 \textwidth
    \epsffile{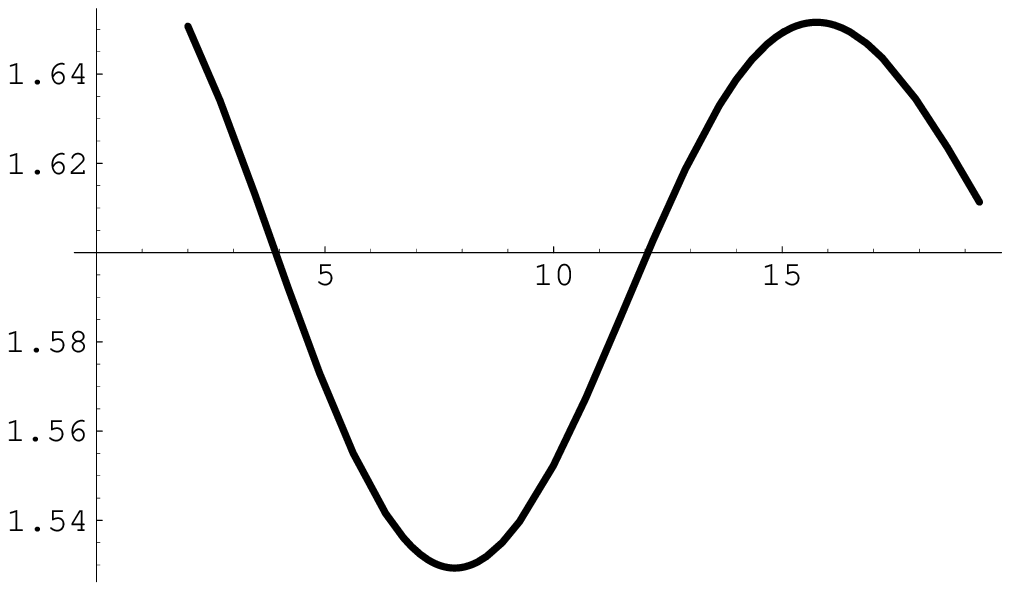}\\
  \epsfxsize=.4
 \textwidth
   \epsffile{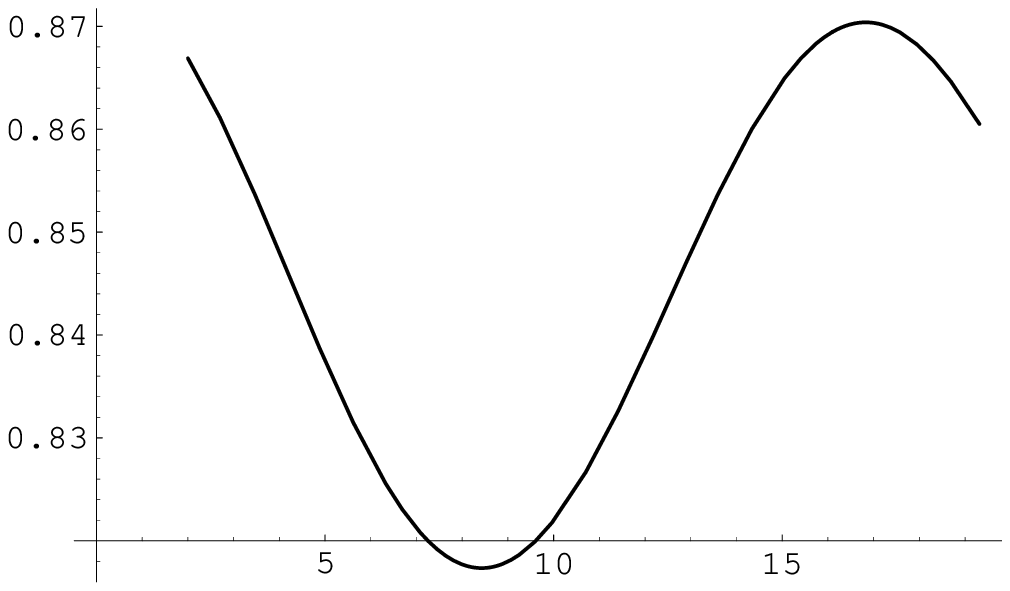}
     \epsfxsize=.4
 \textwidth
   \epsffile{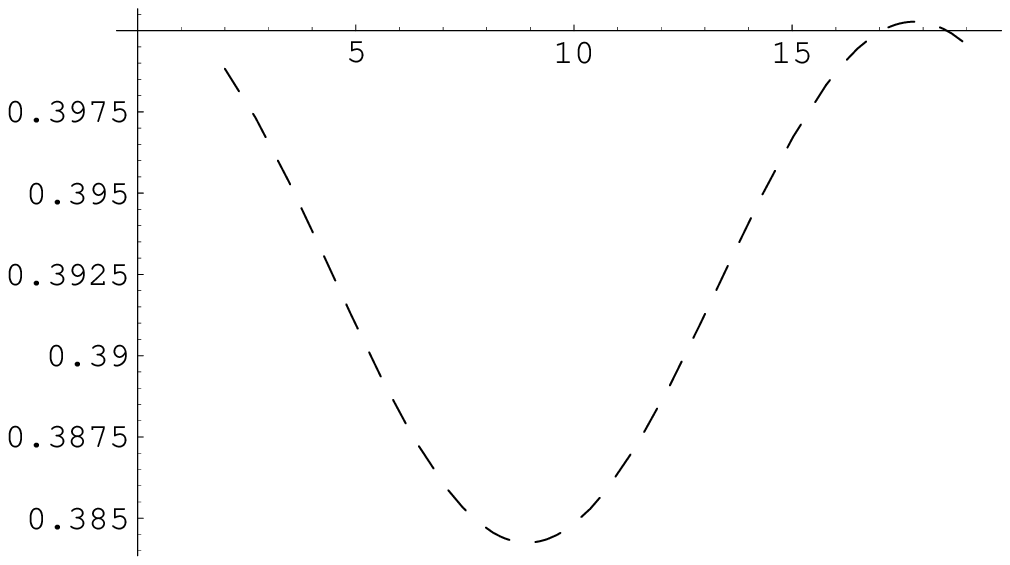}\\
    \epsfxsize=.4
    \textwidth
     \epsffile{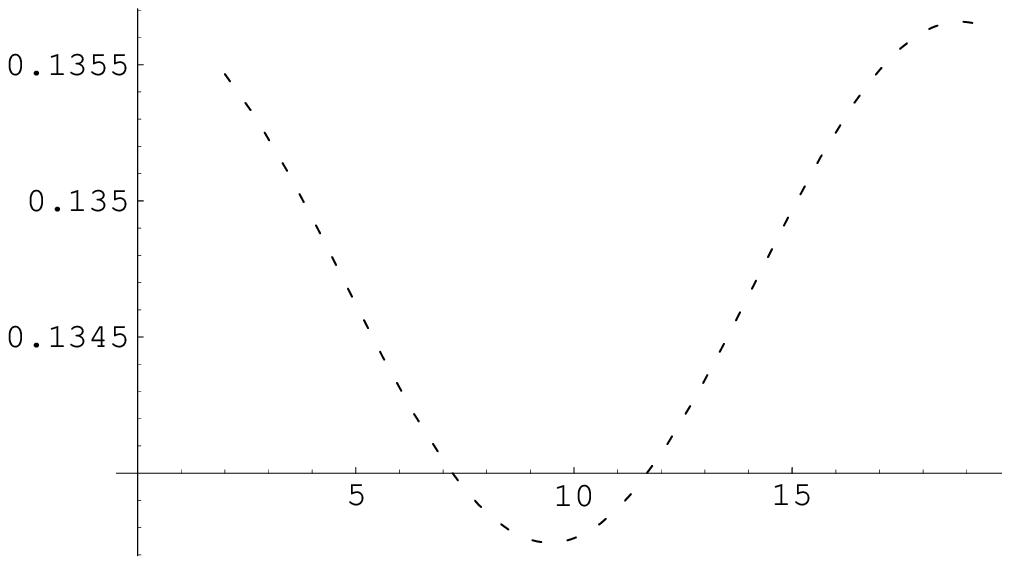} 
       \epsfxsize=.4
 \textwidth
    \epsffile{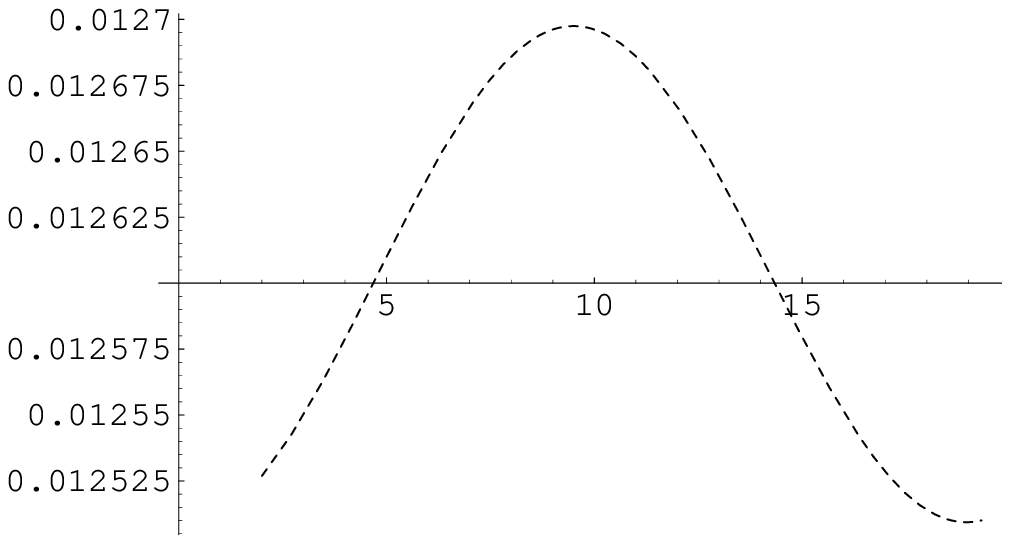}
%\includegraphics[scale=0.6]{fcrossex1_170.eps.}
%\includegraphics[scale=0.6]{fcrossex2_170.eps.}\\
%\vspace{0.5cm}
%\includegraphics[scale=0.6]{fcrossex3_170.eps.}
%\includegraphics[scale=0.6]{fcrossex4_170.eps.}\\
%\vspace{0.5cm}
%\includegraphics[scale=0.6]{fcrossex5_170.eps.}
%\includegraphics[scale=0.6]{fcrossex6_170.eps.}\\
%\vspace{0.5cm}
%\includegraphics[scale=0.8]{fcrossex_170.eps}
 \caption{The differential cross $(\nu_e,e^-)$ section,$\prec \frac{d\sigma}{dT}\succ$,
  in  units of $\frac{G^2_F m_e}{2 \pi}=4.5 \times 10^{-52}\frac{m^2}{keV}$,
 as a function of
 the source-detector distance averaged over the neutrino
energy for electron
energies from top to bottom and left to right 0.2, 0.4, 0.6, 0.8, 1.0 and 1.2 keV.
%At the bottom we plot all the cases in the same plot.
The results shown correspond to
$sin^2 2\theta_{13}=0.170$}
 %\end{center}
 \label{fcross:1}
  \end{center}
  \end{figure}

\begin{figure}[!ht]
 \begin{center}
        \epsfxsize=.4
 \textwidth
  \epsffile{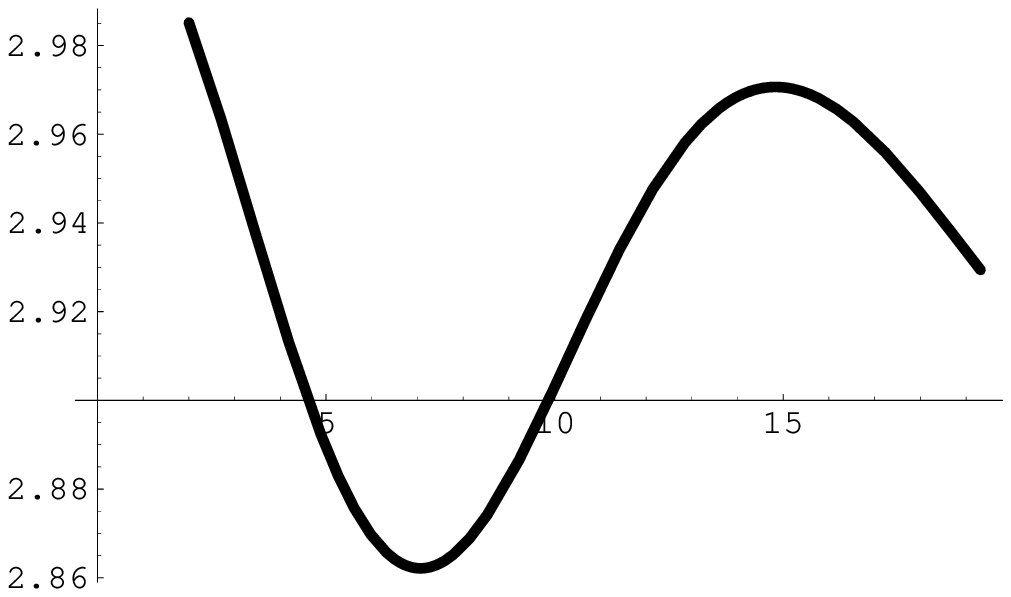}
    \epsfxsize=.4
 \textwidth
    \epsffile{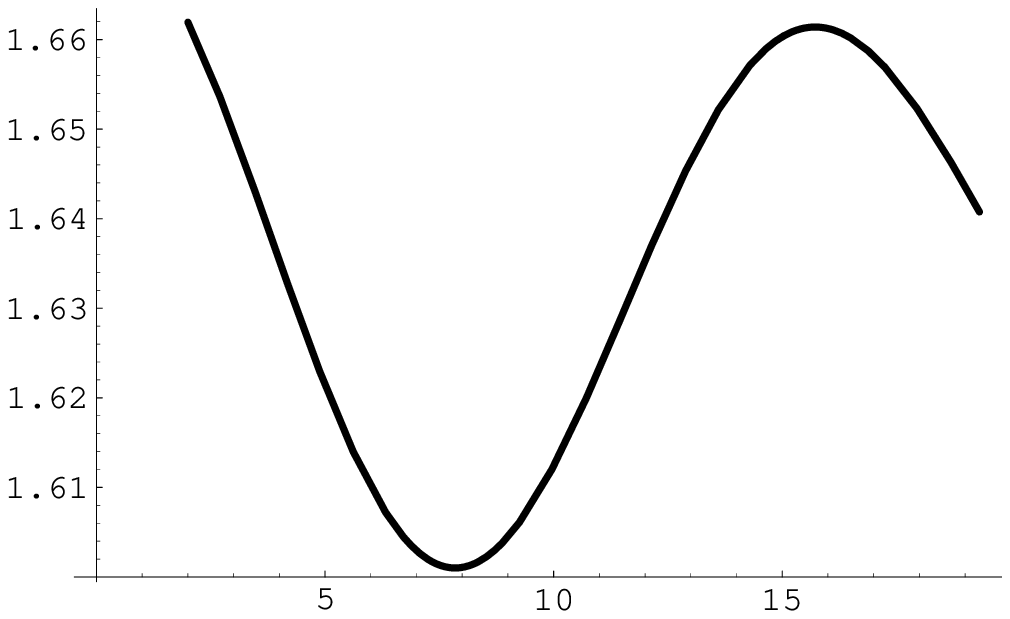}\\
  \epsfxsize=.4
 \textwidth
   \epsffile{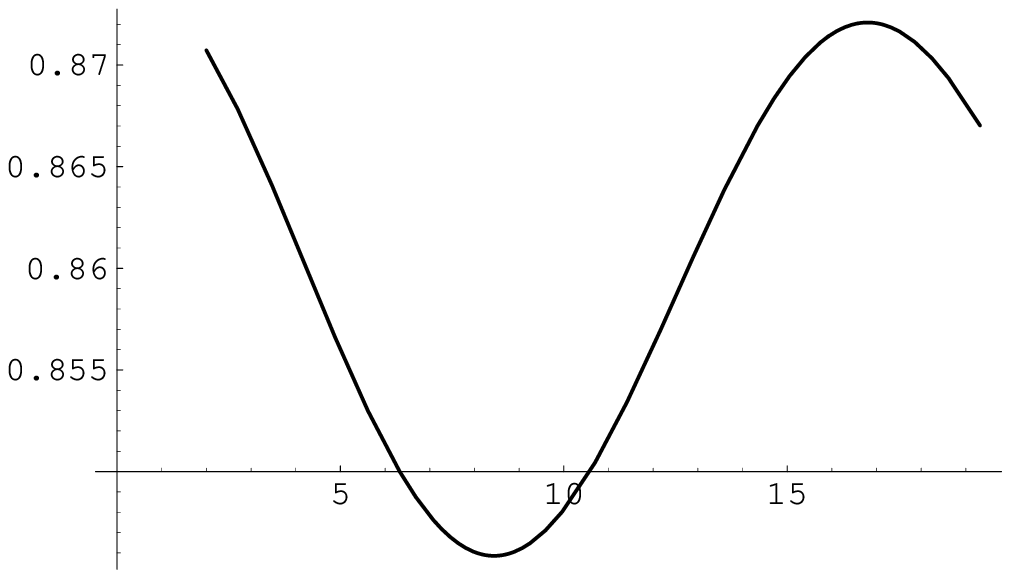}
     \epsfxsize=.4
 \textwidth
   \epsffile{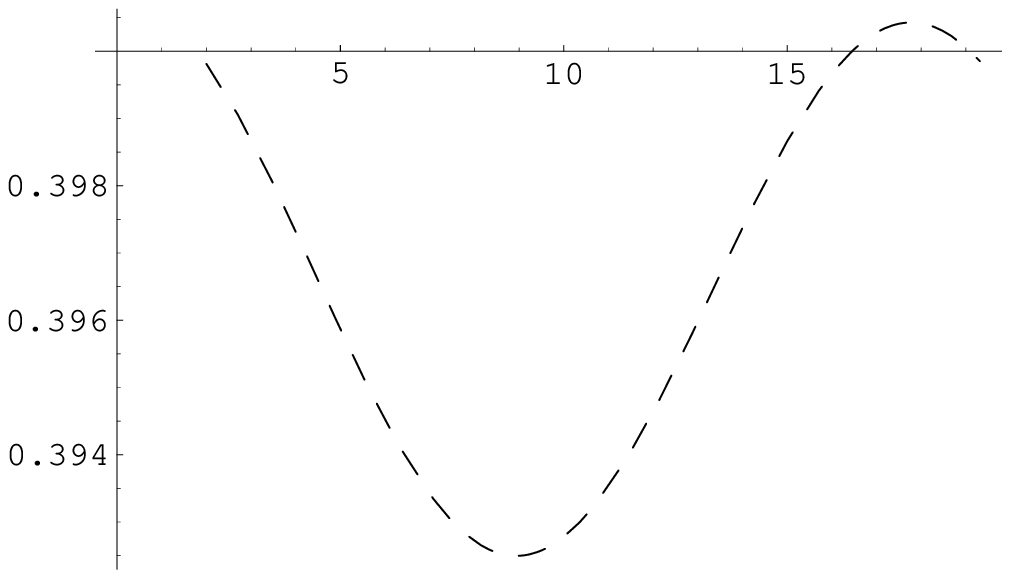}\\
    \epsfxsize=.4
    \textwidth
     \epsffile{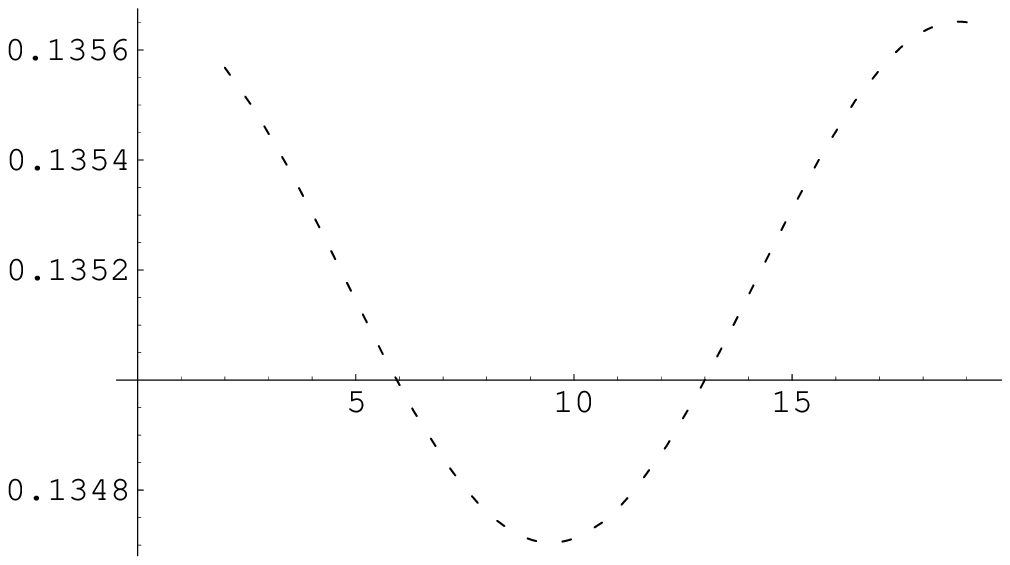} 
       \epsfxsize=.4
 \textwidth
    \epsffile{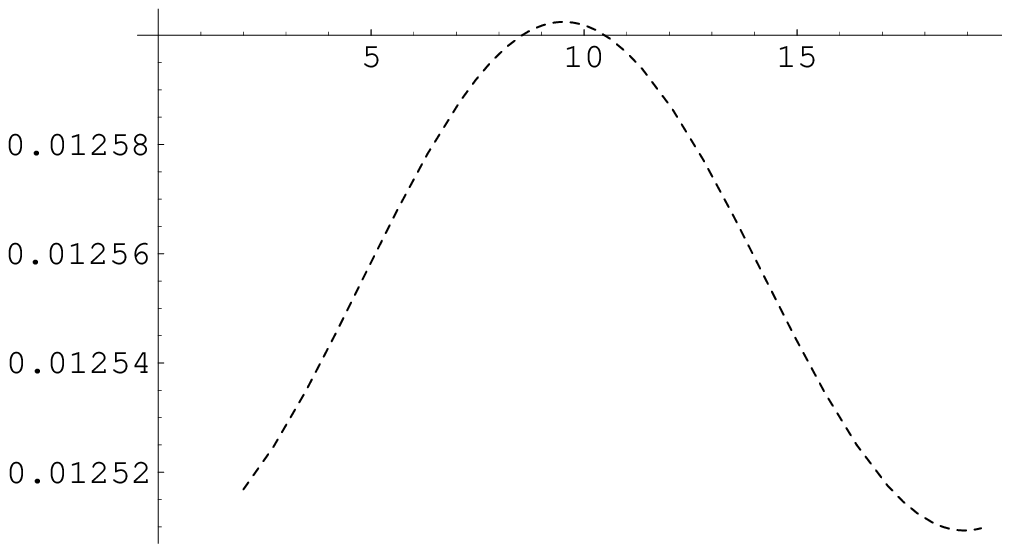}
%\includegraphics[scale=0.6]{fcrossex1_085.eps.}
%\includegraphics[scale=0.6]{fcrossex2_085.eps.}\\
%\vspace{0.5cm}
%\includegraphics[scale=0.6]{fcrossex3_085.eps.}
%\includegraphics[scale=0.6]{fcrossex4_085.eps.}\\
%\vspace{0.5cm}
%\includegraphics[scale=0.6]{fcrossex5_085.eps.}
%\includegraphics[scale=0.6]{fcrossex6_085.eps.}
%\vspace{0.5cm}
%\includegraphics[scale=0.8]{fcrossex_085.eps}
 \caption{The same as in Fig. \ref{fcross:1} for
$sin^2 2\theta_{13}=0.085$}
 %\end{center}
 \label{fcross:2}
  \end{center}
  \end{figure}

\begin{figure}[!ht]
 \begin{center}
 %\psfrag{y}{$\left <  g_e (T) \right >$}
        \epsfxsize=.6
 \textwidth
    \epsffile{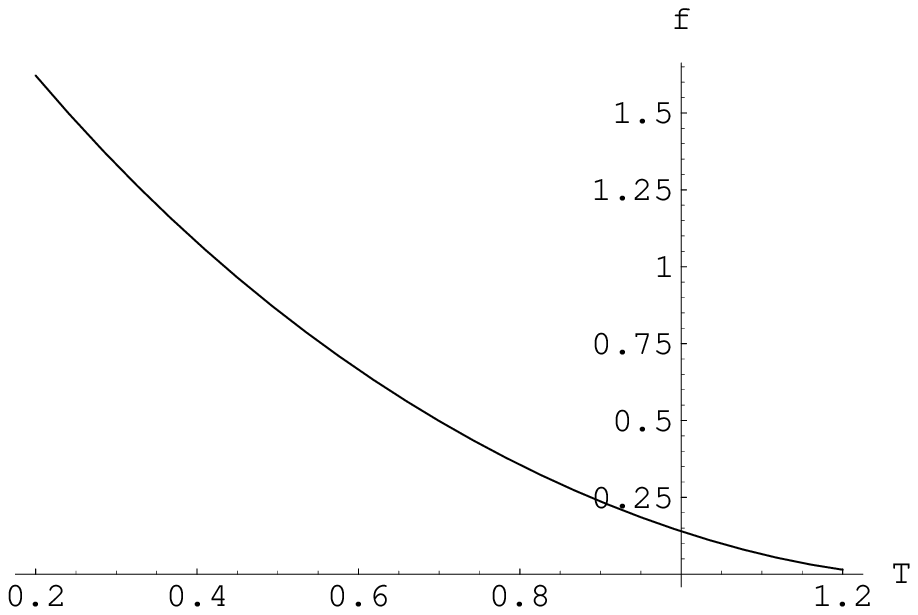}
 \caption{$ \left <f_e(T) \right >$, the differential cross $(\nu_e,e^-)$ section  without oscillations, in  units of
$\frac{G^2_F m_e}{2 \pi}=4.5 \times  10^{-52} \frac{m^2}{keV}$, averaged over the
neutrino spectrum, as a function of
 the outgoing electron energy in keV.}
 %\end{center}
 \label{figeg}
  \end{center}
  \end{figure}

\begin{figure}[!ht]
 \begin{center}
       \epsfxsize=.4
 \textwidth
  \epsffile{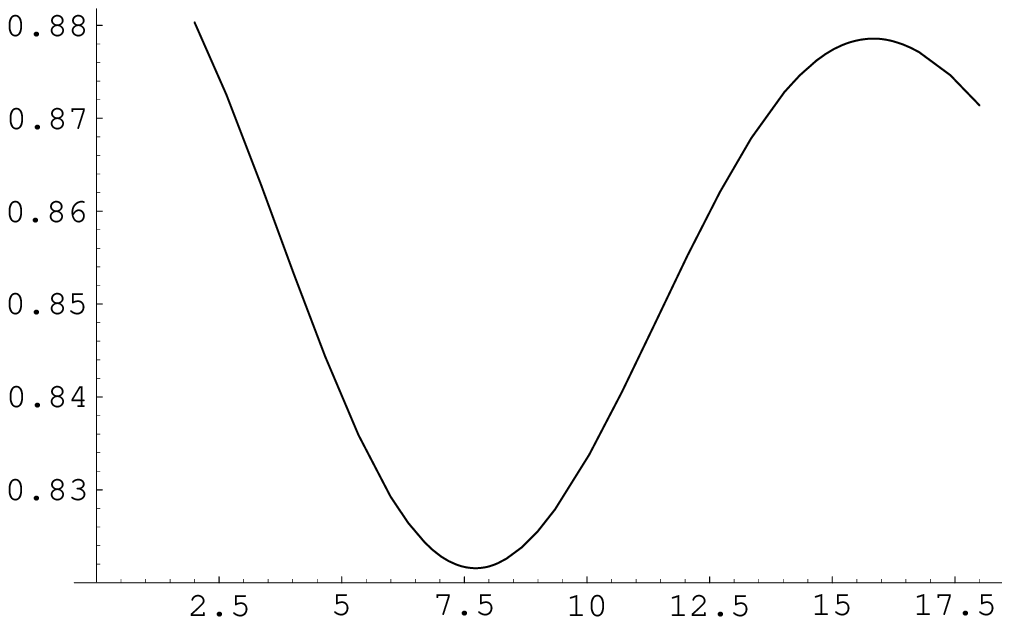}
  \epsfxsize=.4
 \textwidth
  \epsffile{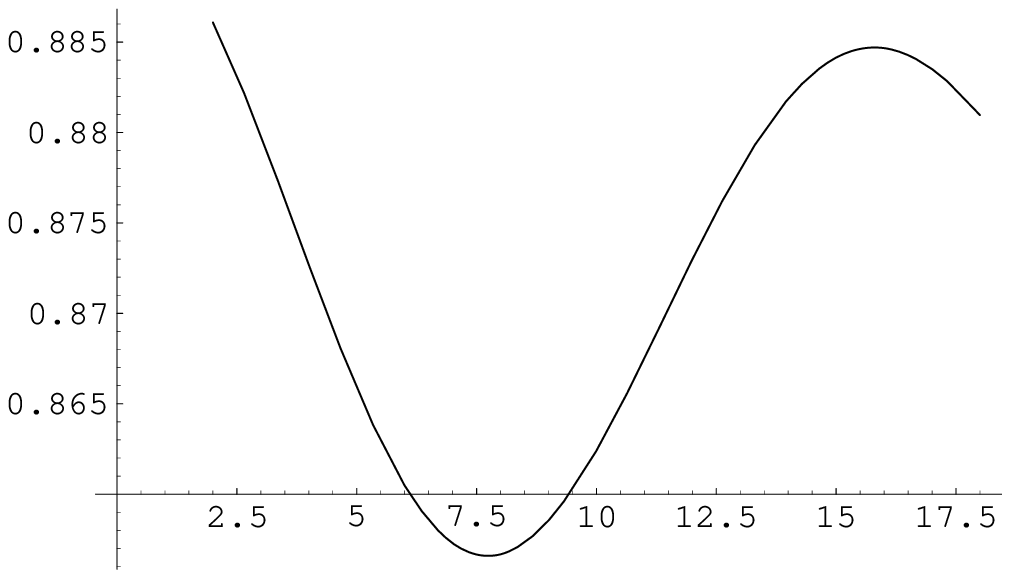}\\
  \epsfxsize=.4
 \textwidth
  \epsffile{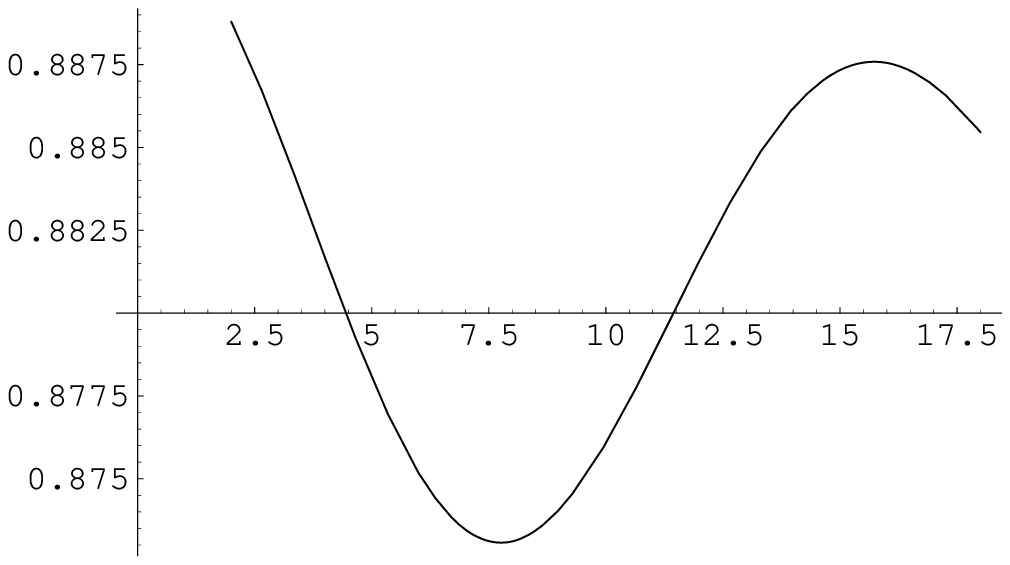}
%\includegraphics[scale=0.6]{totcross_170.eps.}
%\includegraphics[scale=0.6]{totcross_085.eps.}\\
%\vspace{0.5cm}
%\includegraphics[scale=1.0]{totcross_045.eps.}
%\vspace{0.5cm}
%\includegraphics[scale=0.8]{fcrossex_045.eps}
 \caption{The total cross section, in  units of
$\frac{G^2_F m_e}{2 \pi}=4.5 \times  10^{-52} cm^2$, as a function of the source detector distance. From top to bottom and
left to right for
$sin^2 2\theta_{13}=0.170,~0.85,~0.045$.}
 %\end{center}
 \label{totcross}
  \end{center}
  \end{figure}

We now are going to present what the NOSTOS experiment is going to measure, namely the differential cross
section averaged over the neutrino spectrum, i.e.
\begin{eqnarray}
\prec \frac{d\sigma}{dT}\succ  &=&\int\ P(\nu_e \rightarrow \nu_e)
\left(\frac{d\sigma}{dT}\right)_{weak} f_{\nu}(E_{\nu})dE_{\nu}
 \nonumber\\
  &=&\prec P(\nu_e \rightarrow
\nu_e) \succ \prec f_e(T) \succ \label{avecrossa}
 \end{eqnarray}
 Or \beq \prec
\frac{d\sigma}{dT}\succ  =\prec P(\nu_e \rightarrow \nu_e) \succ
\prec f_e(T) \succ \label{avecrossb} \eeq \beq \prec f_e(T) \succ
=\int
\left(\frac{d\sigma}{dT}\right)_{weak}f_{\nu}(E_{\nu})dE_{\nu}
\label{avecrossc} \eeq
 The results obtained are shown in Figs \ref{fcross:1}-\ref{fcross:2}.
 % \newpage
   \begin{figure}[!ht]
 \begin{center}
         \epsfxsize=.4
 \textwidth
  \epsffile{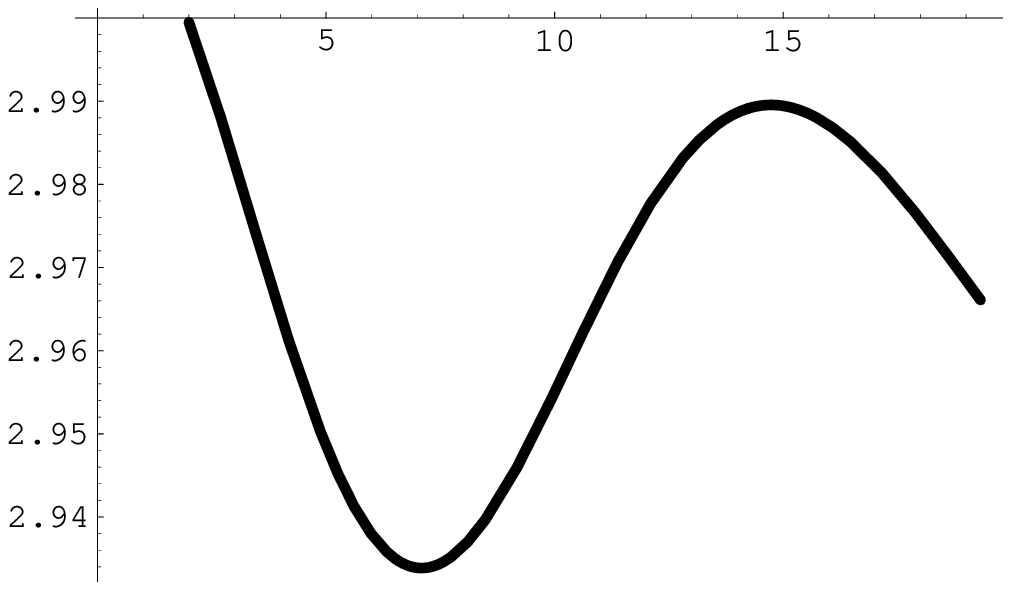}
    \epsfxsize=.4
 \textwidth
    \epsffile{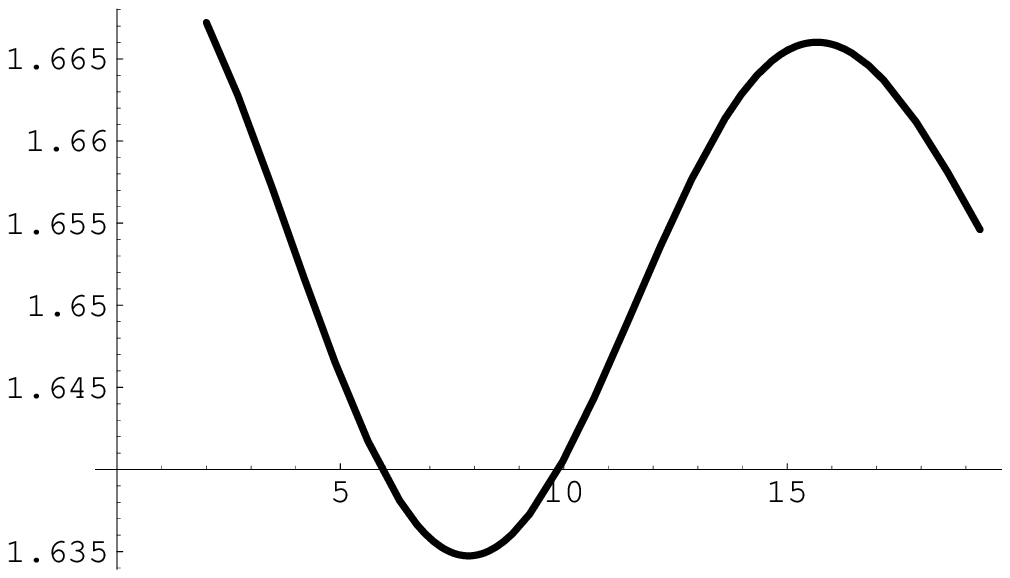}\\
  \epsfxsize=.4
 \textwidth
   \epsffile{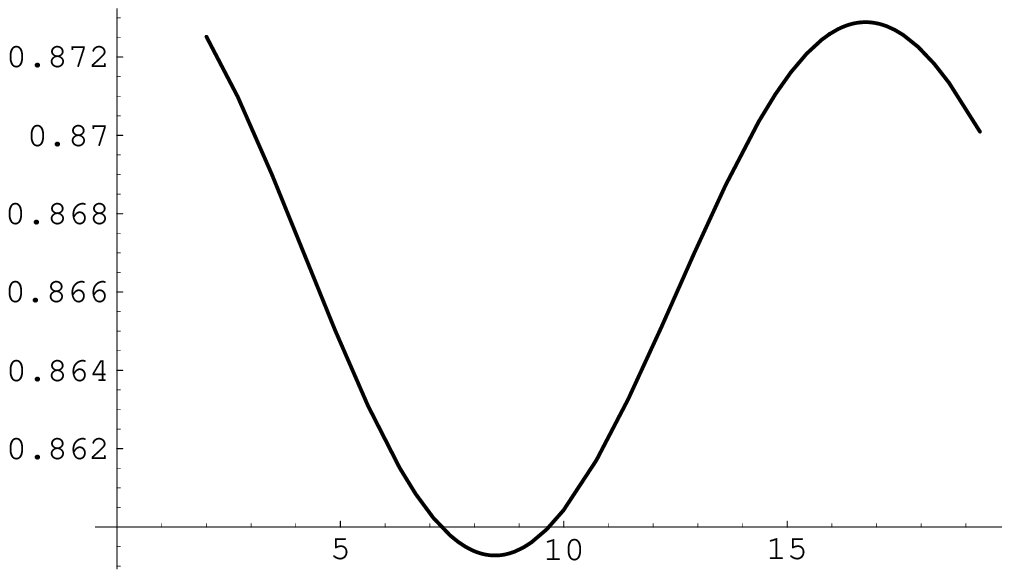}
     \epsfxsize=.4
 \textwidth
   \epsffile{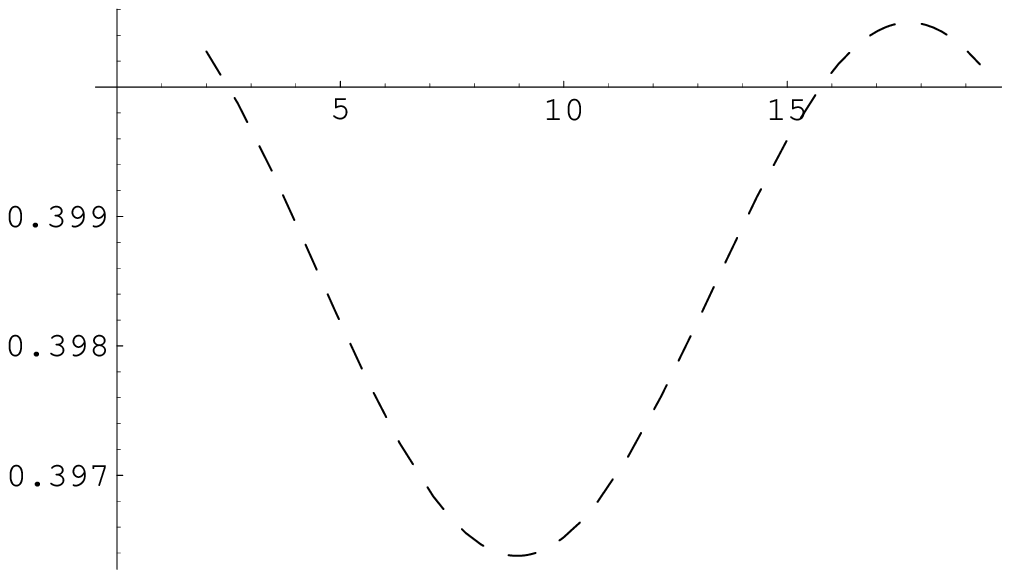}\\
    \epsfxsize=.4
    \textwidth
     \epsffile{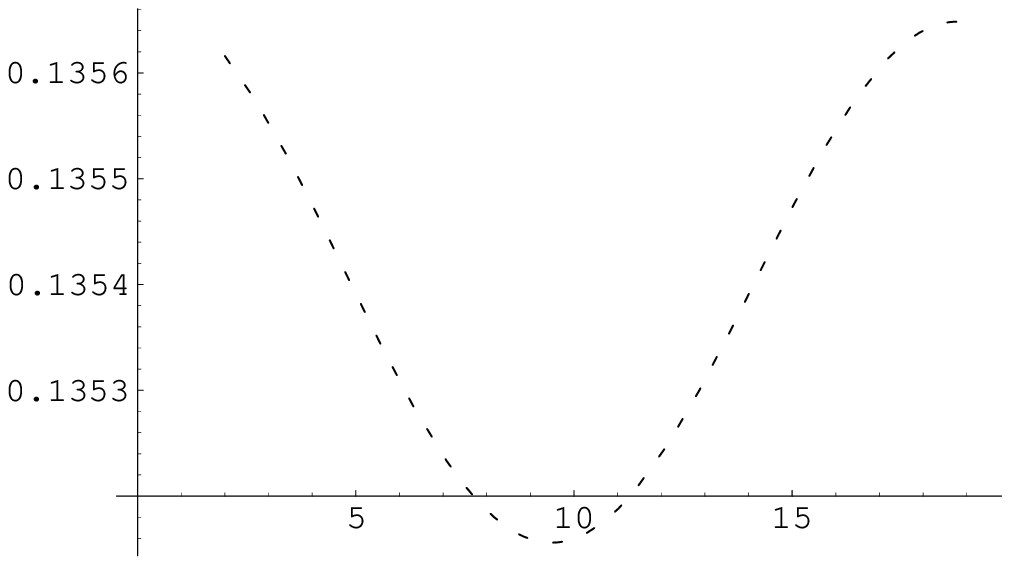} 
       \epsfxsize=.4
 \textwidth
    \epsffile{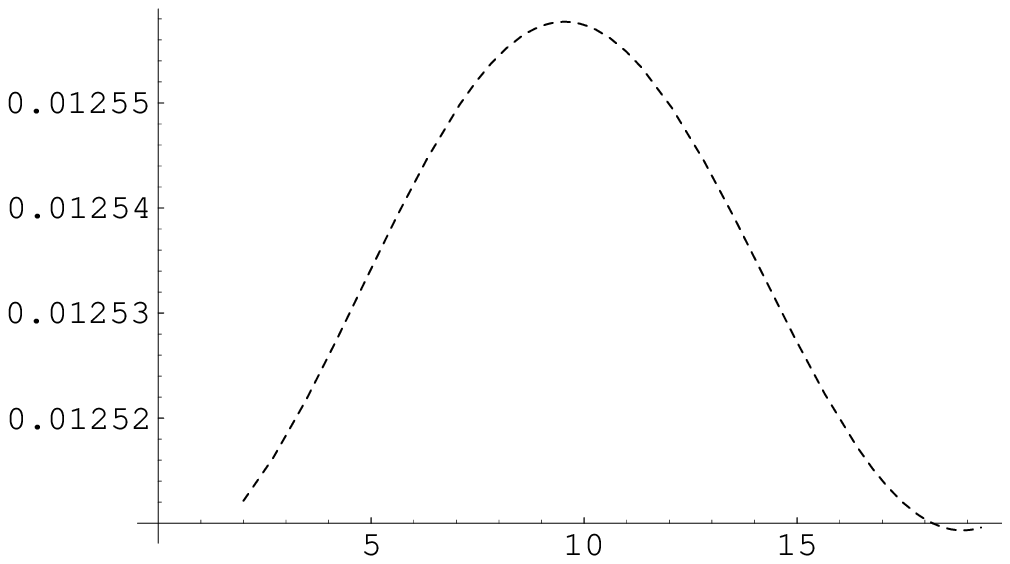}
%\includegraphics[scale=0.6]{fcrossex1_045.eps.}
%\includegraphics[scale=0.6]{fcrossex2_045.eps.}\\
%\vspace{0.5cm}
%\includegraphics[scale=0.6]{fcrossex3_045.eps.}
%\includegraphics[scale=0.6]{fcrossex4_045.eps.}\\
%\includegraphics[scale=0.6]{fcrossex5_045.eps.}
%\includegraphics[scale=0.6]{fcrossex6_045.eps.}\\
%\vspace{0.5cm}
%\includegraphics[scale=0.8]{fcrossex_045.eps}
 \caption{The same as in Fig. \ref{fcross:1} for
$sin^2 2\theta_{13}=0.045$}
 %\end{center}
 \label{fcross:3}
  \end{center}
  \end{figure}
   We see that, unfortunately, the  unexpected
 oscillation features are washed out. They stood out mainly due to the fact that we normalized the amplitude in
the usual way, i.e. so that the oscillation length independent
part of the amplitude is equal to unity. They happen, however, to occur in a kinematically unfavored region, 
where the cross sections is small. Indeed
 the differential cross section, averaged over the neutrino
spectrum, as we have seen can also be obtained by multiplying the average oscillation probability with the
function  given by Eq. \ref{avecrossc} (see Eqs \ref{avecrossb} and  \ref{aveosc}).
The latter quantity is shown in Fig. \ref{figeg}. We thus see that the contribution  of the high energy electrons
becomes very small.
  One can, of course, obtain the total cross section by integrating the differential cross section over the electron 
energy as a function of the distance from the source. The thus obtained results are shown in Fig. \ref{totcross}. 
  
\begin{figure}[!ht]
 \begin{center}
          \epsfxsize=.4
 \textwidth
  \epsffile{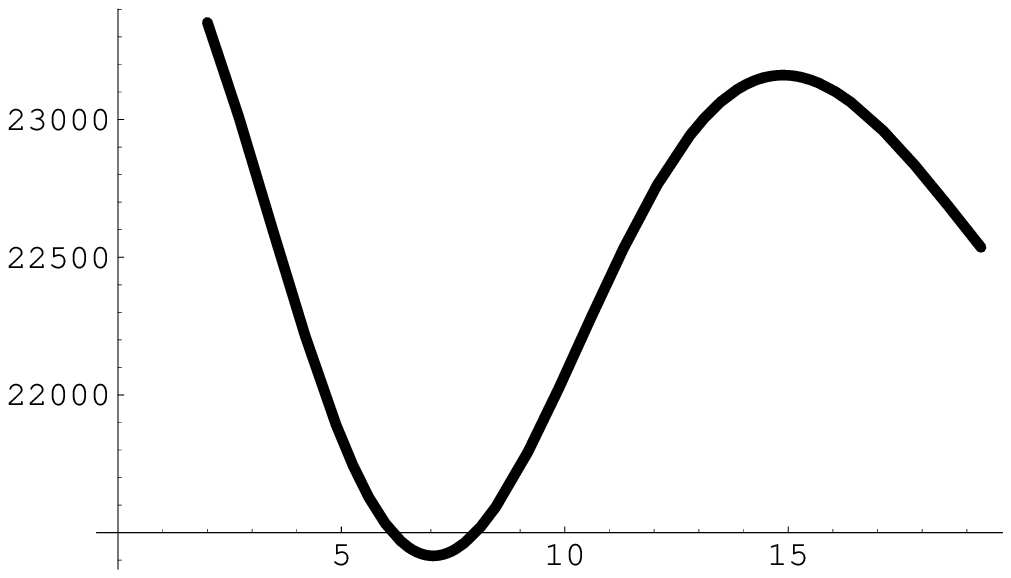}
    \epsfxsize=.4
 \textwidth
    \epsffile{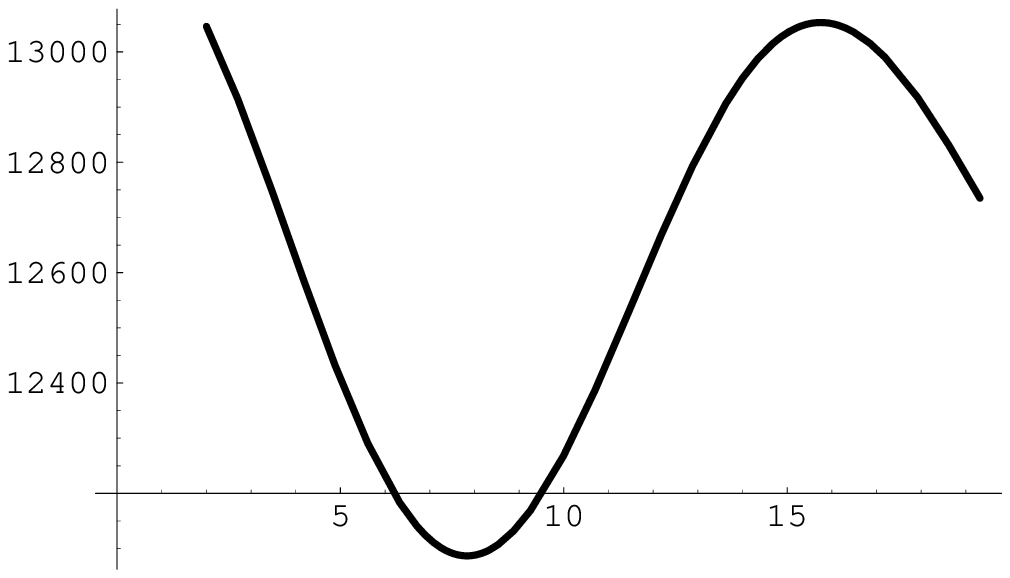}\\
  \epsfxsize=.4
 \textwidth
   \epsffile{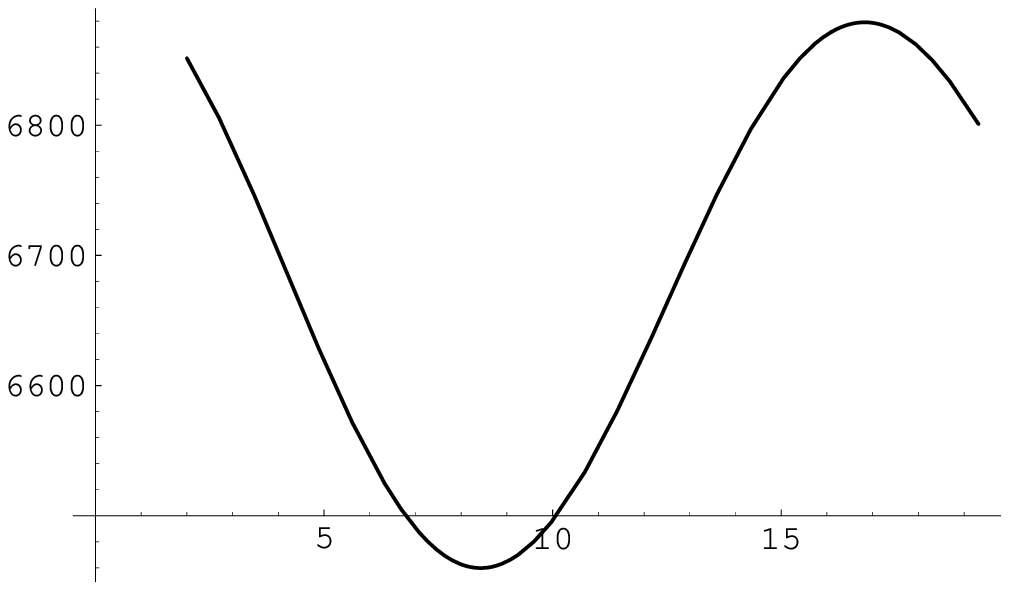}
     \epsfxsize=.4
 \textwidth
   \epsffile{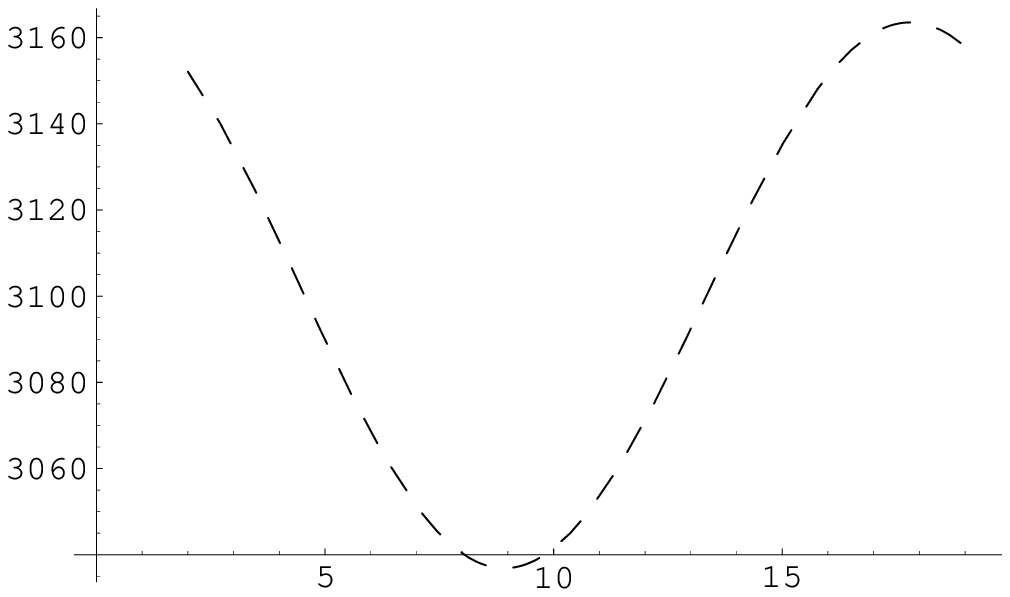}\\
    \epsfxsize=.4
    \textwidth
     \epsffile{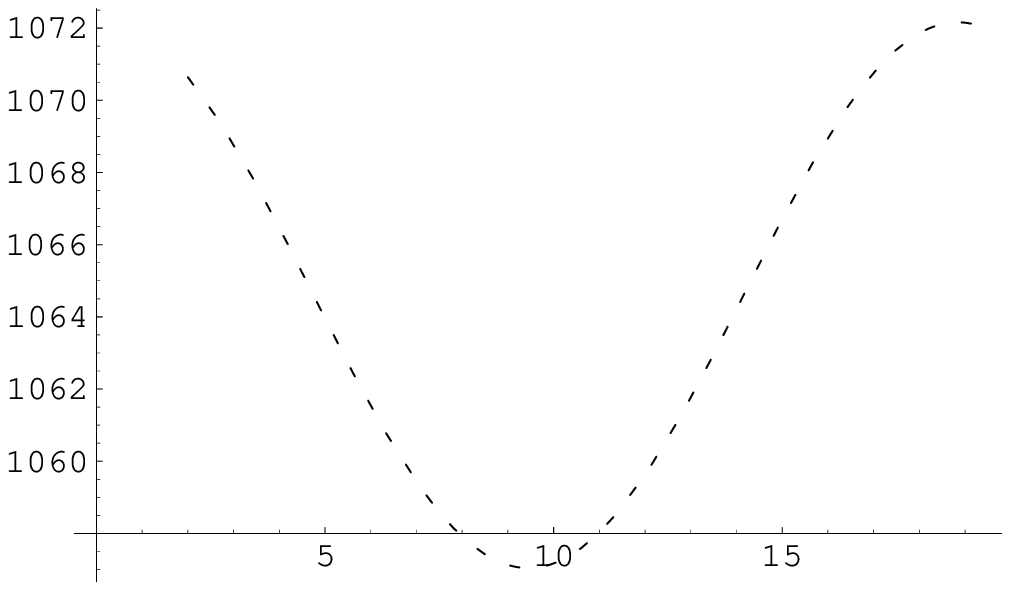} 
       \epsfxsize=.4
 \textwidth
    \epsffile{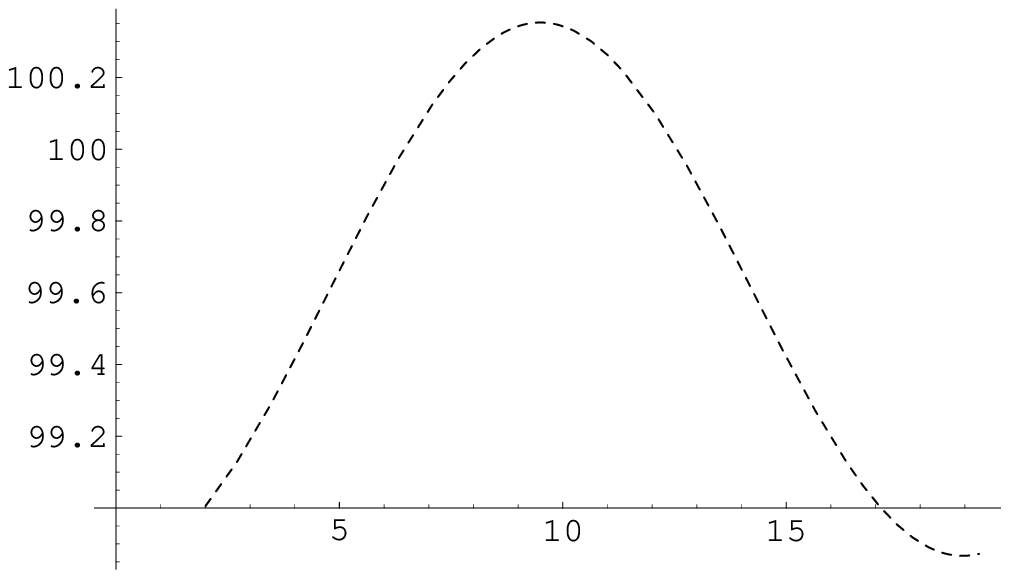}
%\includegraphics[scale=0.6]{frate1_170.eps.}
%\includegraphics[scale=0.6]{frate2_170.eps.}\\
%\vspace{0.5cm}
%\includegraphics[scale=0.6]{frate3_170.eps.}
%\includegraphics[scale=0.6]{frate4_170.eps.}\\
%\vspace{0.5cm}
%\includegraphics[scale=0.6]{frate5_170.eps.}
%\includegraphics[scale=0.6]{frate6_170.eps.}
%\vspace{0.5cm}
%\includegraphics[scale=0.8]{fcrossex_170.eps}
 \caption{The differential rate $\frac{dN}{dT~dL}$ (per keV-meter) for Ar at $10$ Atm with $20$ Kg of tritium 
 as a function of
 the source-detector distance (in m ), averaged over the neutrino
energy, for electron
energies from top to bottom and left to right 0.2, 0.4, 0.6, 0.8, 1.0 and 1.2 keV.
%At the bottom we plot all the cases in the same plot.
The results shown correspond to
$sin^2 2\theta_{13}=0.170$. This rate must be multiplied by $1-e^{-t/\tau}$ to get the number of events after
running time $t$. Thus only $5\%$ of these are expected during the first year of running.}
 %\end{center}
 \label{difrates1}
  \end{center}
  \end{figure}

\begin{figure}[!ht]
 \begin{center}
           \epsfxsize=.4
 \textwidth
  \epsffile{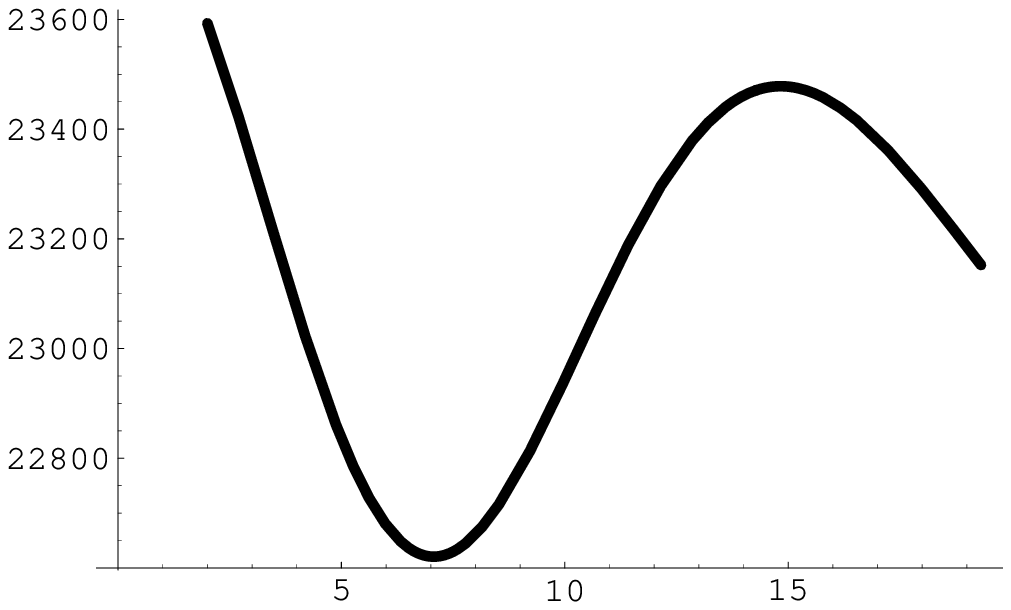}
    \epsfxsize=.4
 \textwidth
    \epsffile{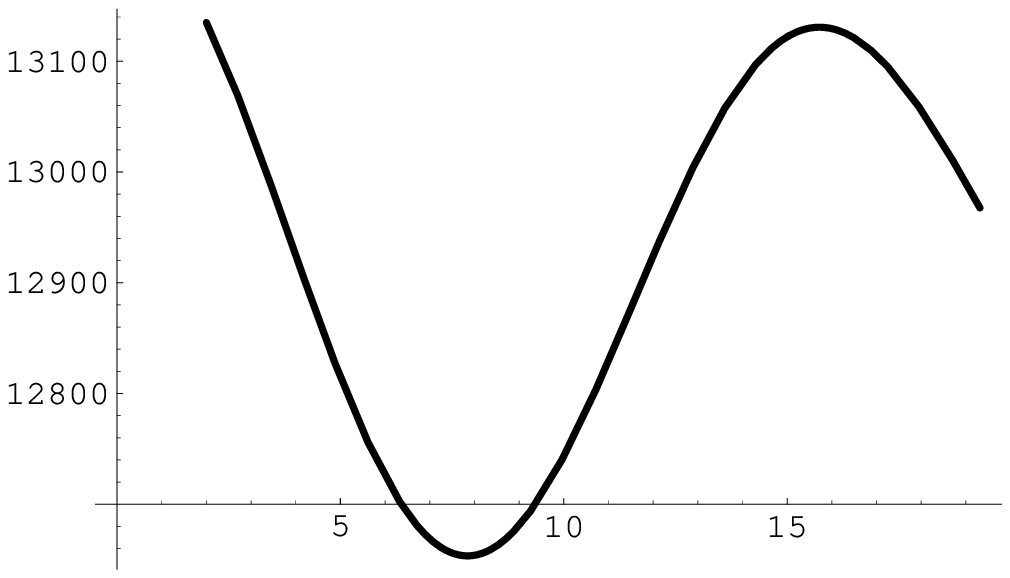}\\
  \epsfxsize=.4
 \textwidth
   \epsffile{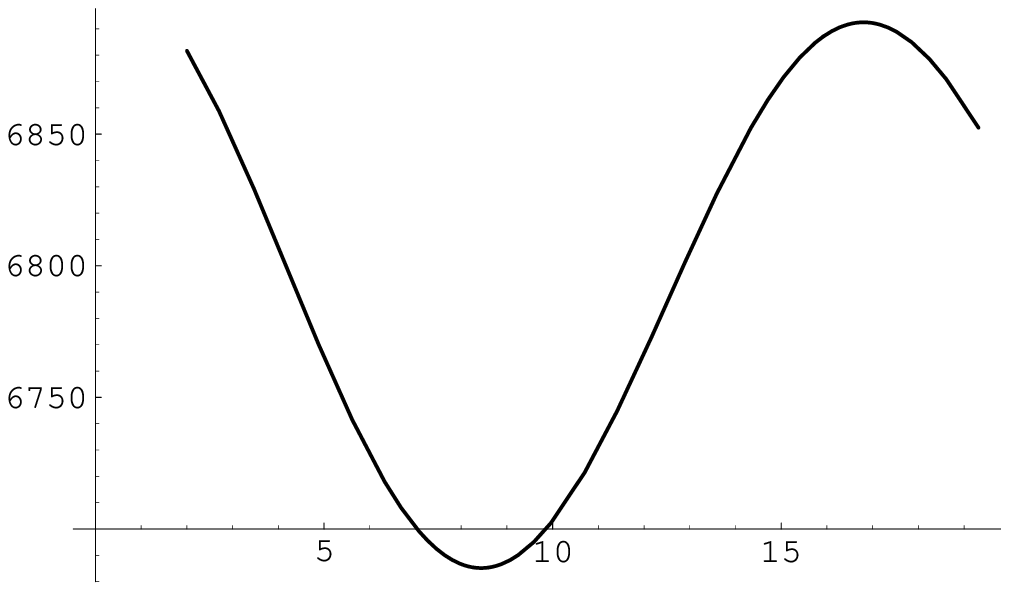}
     \epsfxsize=.4
 \textwidth
   \epsffile{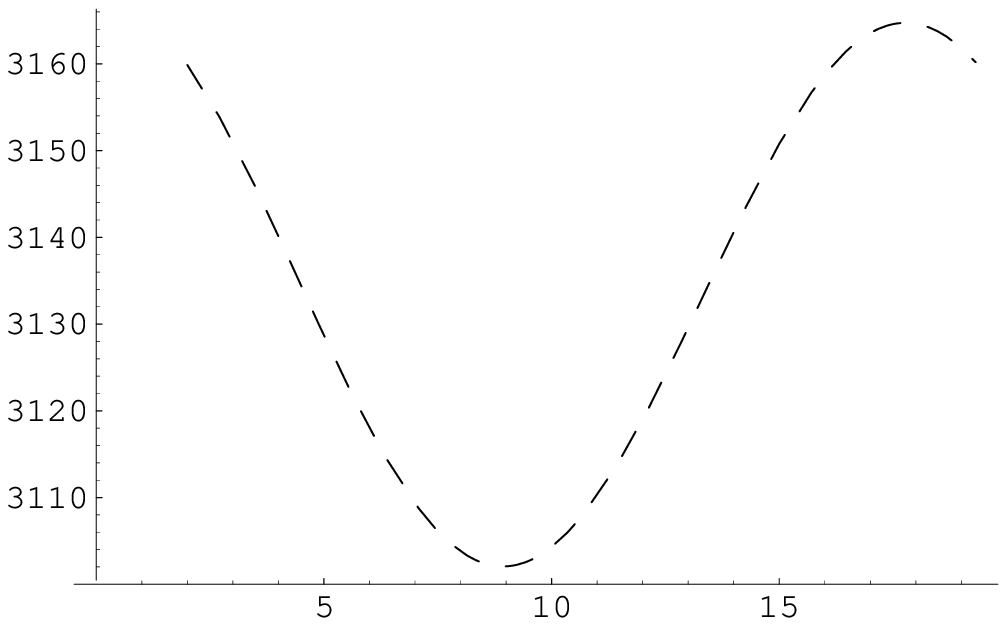}\\
    \epsfxsize=.4
    \textwidth
     \epsffile{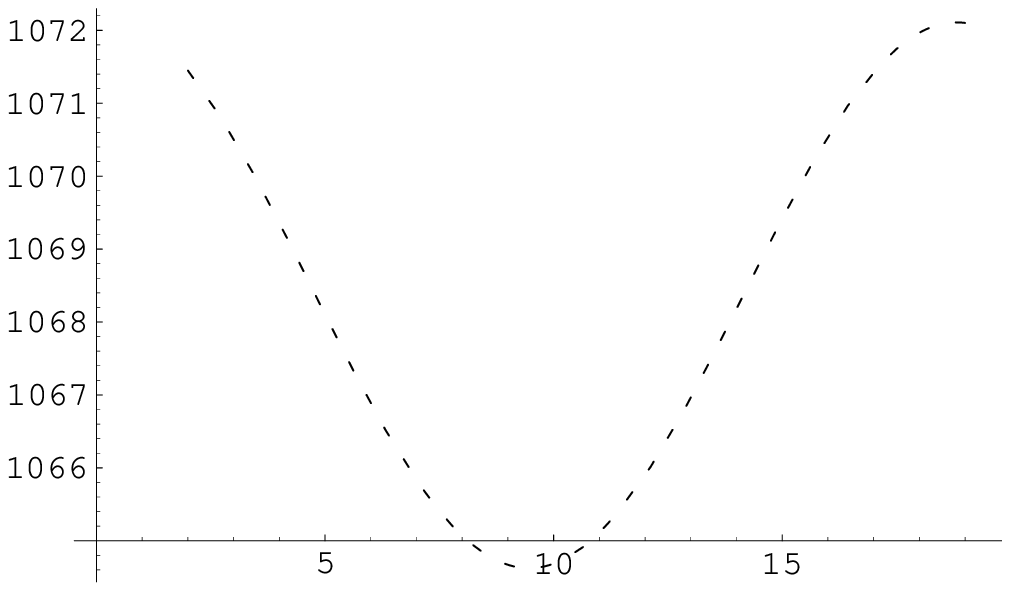} 
       \epsfxsize=.4
 \textwidth
    \epsffile{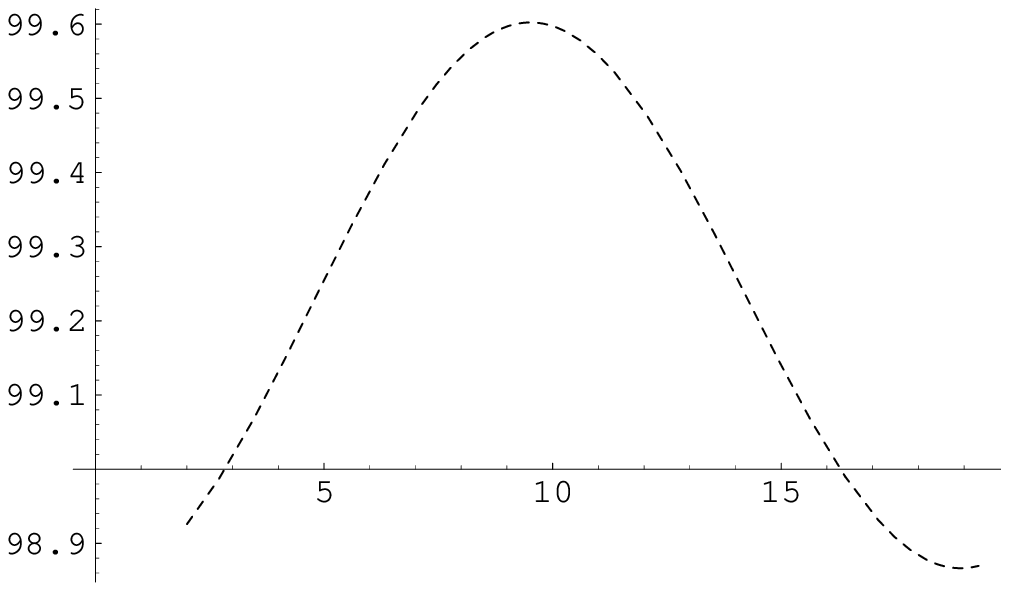}
%\includegraphics[scale=0.6]{frate1_085.eps.}
%\includegraphics[scale=0.6]{frate2_085.eps.}\\
%\vspace{0.5cm}
%includegraphics[scale=0.6]{frate3_085.eps.}
%\includegraphics[scale=0.6]{frate4_085.eps.}\\
%\vspace{0.5cm}
%\includegraphics[scale=0.6]{frate5_085.eps.}
%\includegraphics[scale=0.6]{frate6_085.eps.}
%\vspace{0.5cm}
%\includegraphics[scale=0.8]{fcrossex_170.eps}
 \caption{The same as in Fig. \ref{difrates1} for
$sin^2 2\theta_{13}=0.085$.}
 %\end{center}
 \label{difrates2}
  \end{center}
  \end{figure}

\begin{figure}[!ht]
 \begin{center}
            \epsfxsize=.4
 \textwidth
  \epsffile{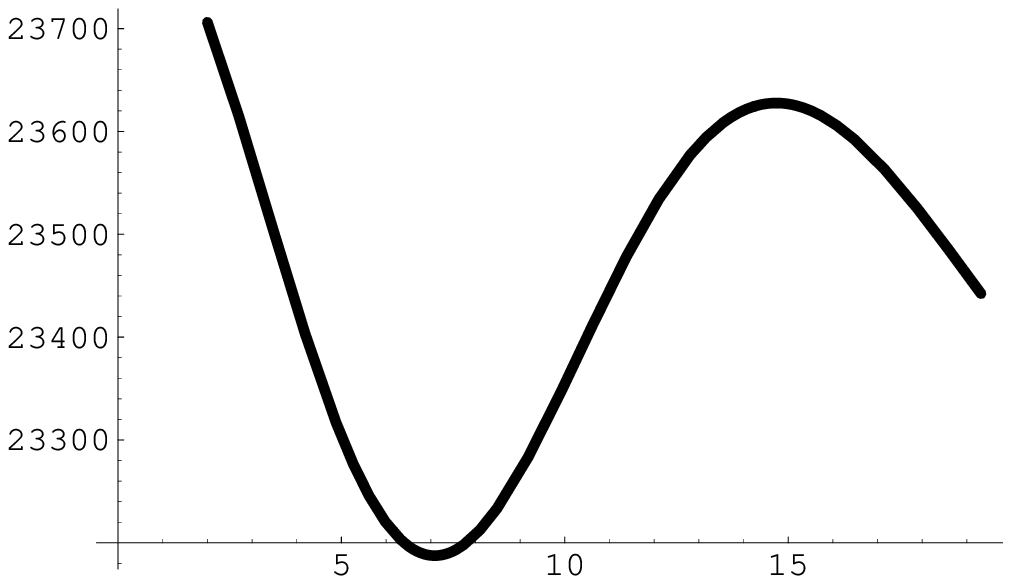}
    \epsfxsize=.4
 \textwidth
    \epsffile{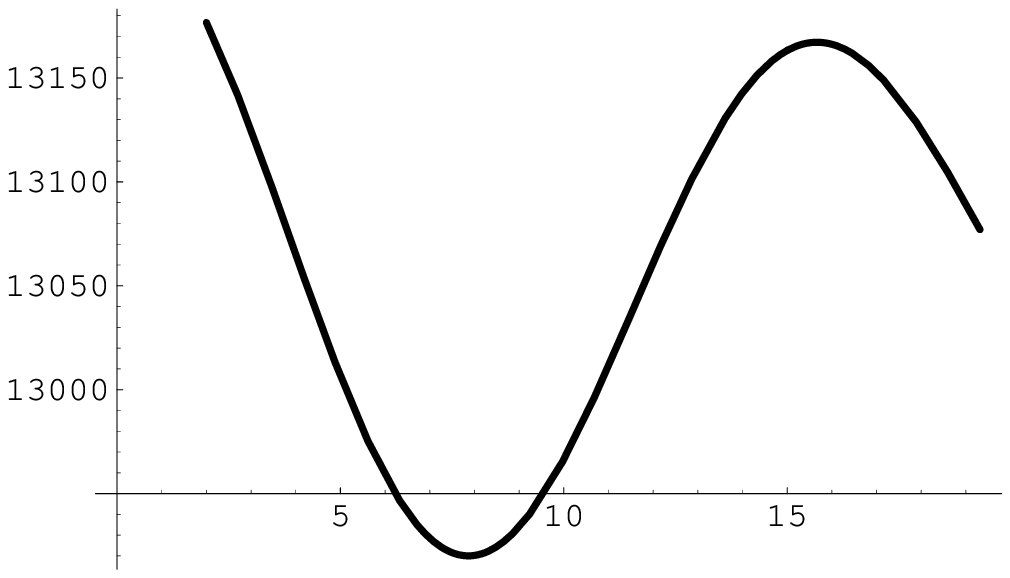}\\
  \epsfxsize=.4
 \textwidth
   \epsffile{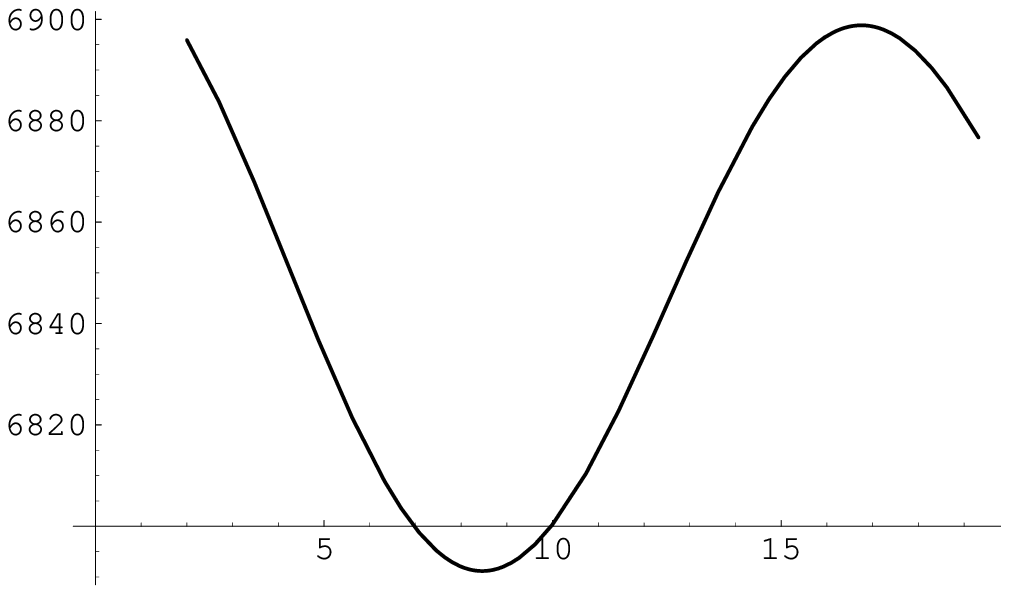}
     \epsfxsize=.4
 \textwidth
   \epsffile{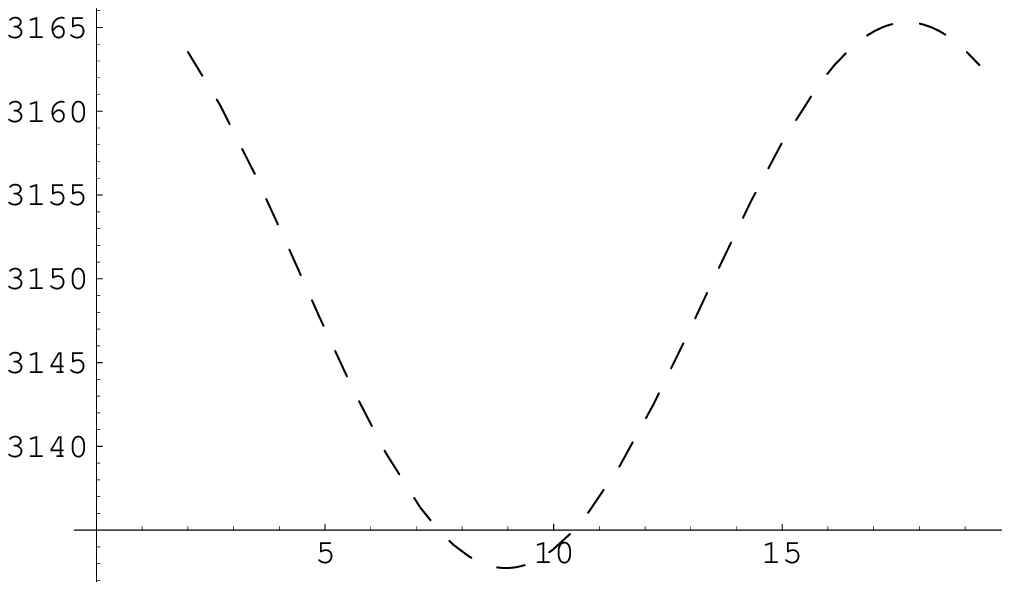}\\
    \epsfxsize=.4
    \textwidth
     \epsffile{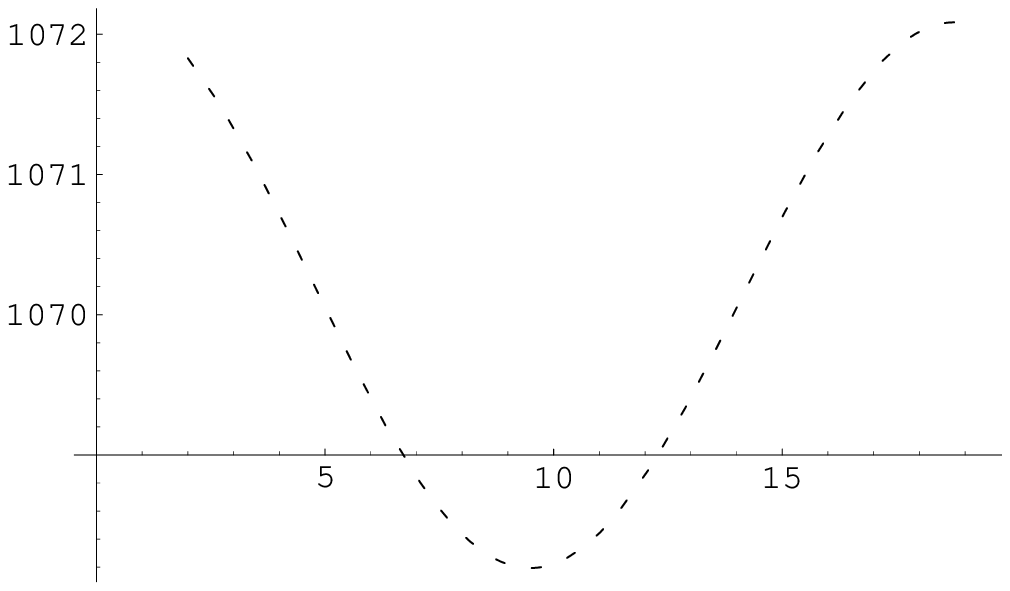} 
       \epsfxsize=.4
 \textwidth
    \epsffile{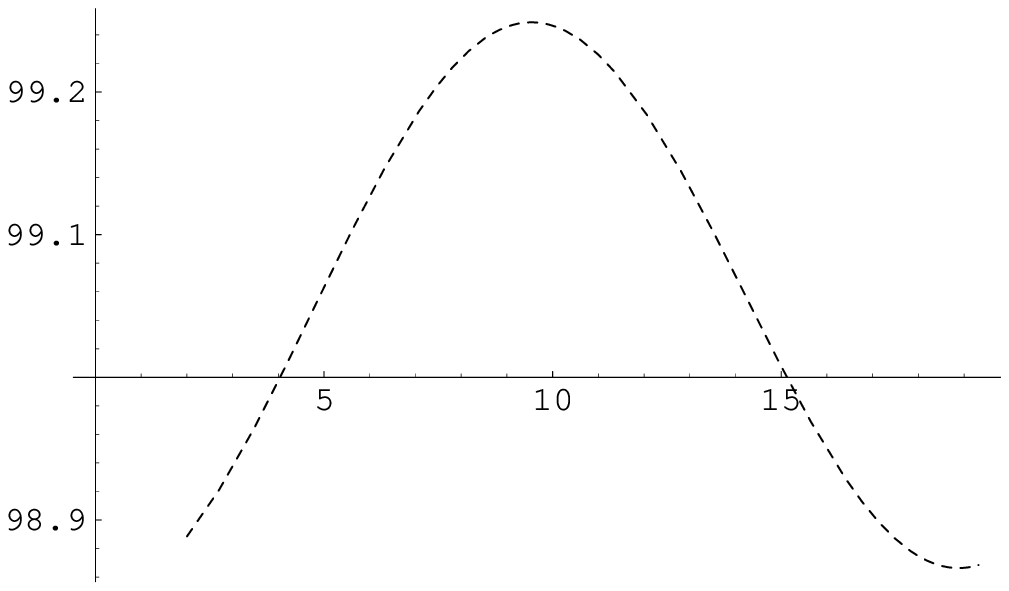}
%\includegraphics[scale=0.6]{frate1_045.eps.}
%\includegraphics[scale=0.6]{frate2_045.eps.}\\
%\vspace{0.5cm}
%\includegraphics[scale=0.6]{frate3_045.eps.}
%\includegraphics[scale=0.6]{frate4_045.eps.}\\
%\vspace{0.5cm}
%\includegraphics[scale=0.6]{frate5_045.eps.}
%\includegraphics[scale=0.6]{frate6_045.eps.}
%\vspace{0.5cm}
%\includegraphics[scale=0.8]{fcrossex_170.eps}
 \caption{The same as in Fig. \ref{difrates1} for
$sin^2 2\theta_{13}=0.045$.}
 %\end{center}
 \label{difrates3}
  \end{center}
  \end{figure}

\begin{figure}[!ht]
 \begin{center}
             \epsfxsize=.4
 \textwidth
  \epsffile{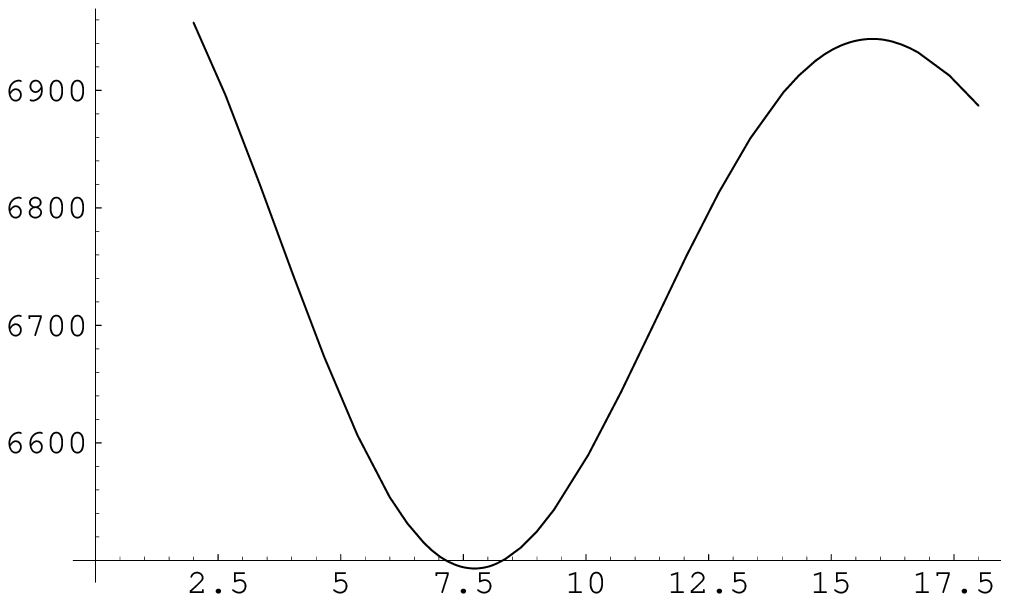}
    \epsfxsize=.4
 \textwidth
    \epsffile{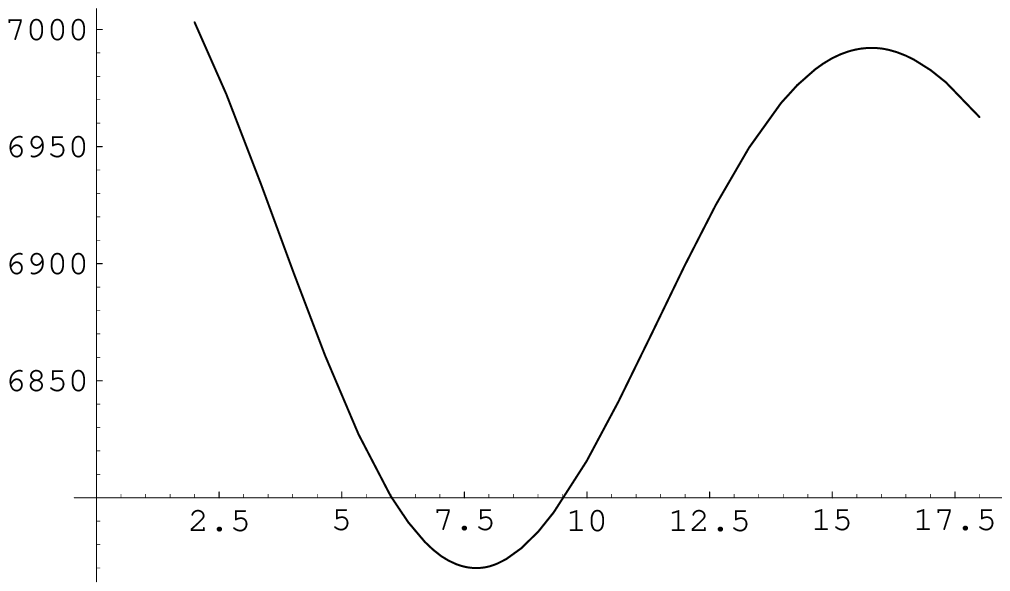}\\
  \epsfxsize=.4
 \textwidth
   \epsffile{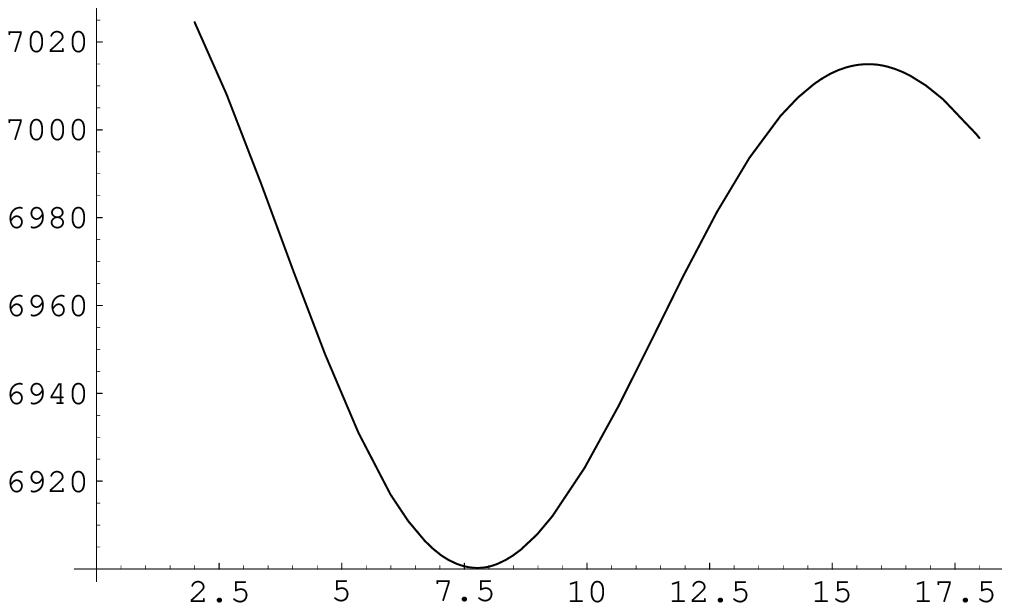}
%\includegraphics[scale=0.6]{totrate_170.eps.}
%\includegraphics[scale=0.6]{totrate_085.eps.}\\
%\includegraphics[scale=1.0]{totrate_045.eps.}
%\vspace{0.5cm}
%\includegraphics[scale=0.8]{fcrossex_045.eps}
 \caption{The total event rate as a function of the source detector distance. The notation is the same
as in Fig. \ref{totcross}. The input is the same as in Fig. \ref{difrates1}.}
 %\end{center}
 \label{totrates}
  \end{center}
  \end{figure}

 Experimentally, of course, one measures events. The number of events is obtained  by:
 \begin{itemize}
\item Inserting in the relevant cross sections the scale of week interaction
$$\frac{G^2_F m_e}{2 \pi} =4.50\times 10^{-52}\frac{m^2}{keV}$$
%\nonumber
%\eeq
\item Considering the total number of neutrinos emitted by the source of 20 Kg of tritium
$$N_{\nu}= \frac{20}{3 \times 1.67\times 10^{-27}}=4.00\times 10^{27}$$
for any given time $t$ the number of neutrinos emitted must be multiplied by the fraction 
$1-e^{-t/\tau}$ with $\tau=\frac{T_{1/2}}{\ell n 2}\simeq 17.7y$ ($T_{1/2}=12.3y$).
%\nonumber
%\eeq
This yields the neutrino flux:
$$\Phi_{\nu}=\frac{N_{\nu}}{4\pi L^2}(e^{-t/\tau}/\tau)$$
\item Assuming that in the target the number of electrons per unit volume is
%\beq
$$n_e= Z\frac{P}{kT_0}=4.4\times 10^{27}m^{-3} \frac{P}{10~Atm}\frac{Z}{18}\frac{300}{T_0}$$
($Z$ the atomic number, $P$ is the pressure and $T_0$ the temperature).
\item The total number of events with energies between $T$ and $T+dT$ in the length region between $L$
and $L+dL$ is
$$dN=\Phi_{\nu}~\frac{d\sigma}{dT}~ n_e~ 4\pi L^2~ dTdL=N_{\nu}~\frac{d\sigma}{dT} ~n_e~dTdL.$$
For any given time $t$ we get only the fraction $1-e^{-t/\tau}$ of these events, e.g $5\%$ the first year, $11\%$ for
 two years of running, $16\%$ after 3 years and $43\%$ after 10 years.
%\nonumber
%\eeq
\end{itemize}
The thus obtained results for the differential rate $\frac{dN}{dT~dL}$, during the life time of the source ($20$ Kg of tritium),
 employing $^{40}_{18}$Ar at $P=10$ Atm,
 are shown in Fig. \ref{difrates1}-\ref{difrates3}.
  The total rate per unit length, obtained by integrating the differential rate of Fig. \ref{difrates1}-\ref{difrates3} 
with
respect to the electron energy, is shown in Fig.
  \ref{totrates}.
  \section{Discussion and Conclusions}
  In the experiment one will measure the differential rate, i.e. the number of electrons detected per unit energy
and per unit length, as
  a function of the distance $L$ from the source, i.e. the radius of the sphere. Integrating the differential rate one
  obtains the number of electrons per unit length as a function of $L$. From Figs \ref{difrates1}-\ref{difrates3}
 and \ref{totrates},
obtained with a source of $20$ Kg of tritium over its lifetime and using Ar at $10$ Atm, we see that the 
  $L$-dependence of the event rates resembles closely that of the corresponding cross sections shown in Figs 
\ref{fcross:1}-\ref{fcross:3} and \ref{totcross}. To get the event rate after some time $t$ of running, we must multiply
by the factor $1-e^{-t/\tau},~\tau=17.7$y, e.g. $5\%$ during the first year etc. The event rates scale with
the atomic number $Z$. From these plots we see that:
\begin{itemize}
\item In the case of the differential rate:
\begin{enumerate}
\item At high electron energies 
it increases as a function of L, i.e. it behaves like appearance oscillation. The electrons are mainly
produced by the other two appearing neutrino flavors. This is a novel feature of the neutrino electron scattering.
\item At low electron energies, where the cross section is large, one
 recovers the usual features of electron flavor disappearance.
\item The difference between the maximum and the 
minimum is at the level of $8,4$ and $2$ $\%$ for $sin^2 2\theta_{13}=0.170,~0.085$ and $0.045$ respectively.
\item The location
of the maximum depends on the electron energy. 
\end{enumerate}
\item In the case of the total cross section:
\begin{enumerate}
\item The percentage of the difference
between the maximum and the minimum is about the same as before. 
\item The maximum of the oscillation probability occurs at $L=7.5$ m.
\end{enumerate}
\end{itemize}
 It is clear that with such a large number of events the above features of neutrino oscillation can be exploited to measure 
 $sin^2 2\theta_{13}$ and perhaps to accurately
 determine the oscillation length.
 
acknowledgments: This work was supported in part by the
European Union under the contract
MRTN-CT-2004-503369 and by the program PYTHAGORAS-1 of the 
Operational Program for Education and Initial Vocational Training of the 
Hellenic Ministry of Education under the 3rd Community Support Framework 
and the European Social Fund.


\begin{thebibliography}{8.}
\addcontentsline{toc}{section}{References}
 %\overline
%\bibitem{VERGADOS}
% See, e.g.\\
% J.D. Vergados, {\it  Phys. Rep.} {\bf 361} (2002) 1;\\
% J.D. Vergados, {\it  Phys. Rep.} {\bf 133} (1986) 1.
 \bibitem{NOSTOS1}
 Y. Giomataris and J.D. Vergados, Nucl. Instr. Meth. A {\bf 53} (2004) 330.
%\bibitem{VOGBEAC}
%P. Vogel and J.F. Beacom, {\it Phys. Rev. D} {\bf 60}
%(1999) 053003
\bibitem{SUPERKAMIOKANDE}
Y. Fukuda {\it et al}, The Super-Kamiokande  Collaboration,  {\it
Phys. Rev. Lett.} {\bf 86},  (2001) 5651; {\it ibid} {\bf 81}
(1998) 1562 $\&$ 1158; {\it ibid} {\bf 82}  (1999) 1810 ;{\it
ibid} {\bf 85} (2000) 3999.
\bibitem{SOLAROSC}
Q.R. Ahmad {\it et al}, The SNO Collaboration, {\it Phys. Rev.
Lett.} {\bf 89}  (2002) 011302; {\it ibid} {\bf 89}  (2002) 011301
;
{\it ibid} {\bf 87} (2001) 071301.\\
K. Lande {\it et al}, Homestake Collaboration, {\it Astrophys, J}
{\bf 496}, (1998) 505\\
W. Hampel {\it et al}, The Gallex Collaboration, {\it Phys. Lett.
B} {\bf 447}, (1999) 127;\\
J.N. Abdurashitov {\it al}, Sage Collaboration, {\it Phys. Rev. C}
{\bf 80} (1999) 056801;\\
G.L Fogli {\it et al}, {\it Phys. Rev. D} {\bf 66} (2002) 053010.
\bibitem{KAMLAND}
K. Eguchi {\it et al}, The KamLAND Collaboration, Phys. Rev. Lett.
90 (2003) 021802,  hep-exp/0212021.
\bibitem{BAHCALL02}
J,N. Bahcall, M.C. Gonzalez-Garcia, C. Pe$\tilde{n}$a-Garay, {\it
hep-ph/0212147}
\bibitem{NUNOKAWA}
H Nunokawa {\it et al}, {\it hep-ph/0212202}.
\bibitem{ALIANI}
P. Aliani {\it et al}, {\it hep-ph/0212212}.
\bibitem{MSW}
M. Maltoni, T. Schwetz and J.F. Valle, hep-ph/0212129;
S. Pakvasa and J.F. Valle, hep-ph/0301061
\bibitem{BARGER02}
V. Barger and D. Marfatia, {\it hep-ph/0212126}.
\bibitem{HOOFT}
G. 't Hooft, Phys. Lett. B  {\bf 37} (1971) 195.
 \bibitem{REINES}
F. Reines, H.S. Gurr and H.W. Sobel, Phys. Rev. Lett. {\bf 6}
(1976) 315.
\bibitem{VogEng}
P. Vogel and J. Engel, {\it Phys. Rev. D} {\bf 39} (1989) 3378.
\end{thebibliography}
\end{document}